\documentclass{emulateapj}

\begin{document}

\newcommand{\etal}{{\em et~al.\,}}

\title{Evolution in the Color--Magnitude Relation of Early--Type
  Galaxies in Clusters of Galaxies at z$\simeq$1 \altaffilmark{1}}
\altaffiltext{1}{Based on observations with the NASA/ESA Hubble Space
Telescope, obtained at the Space Telescope Science Institute, which is
operated by the Association of Universities for Research in Astronomy,
Inc. under NASA contract No. NAS5-26555.}

\author{B. P. Holden\altaffilmark{2,3}, S. A. Stanford\altaffilmark{2,4}}
\affil{Department of Physics, University of California, Davis}
\altaffiltext{2}{Participating Guest, Institute of Geophysics and
Planetary Physics, Lawrence Livermore National Laboratory}
\altaffiltext{3}{Presently at Lick Observatory, University of California}
\altaffiltext{4}{Visiting Astronomer, Kitt Peak National Observatory, 
National Optical Astronomy Observatory, which is operated by the
Association of Universities for Research in Astronomy, Inc. (AURA)
under cooperative agreement with the National Science Foundation.}
\email{holden@ucolick.org,adam@igpp.ucllnl.org}

\author{P. Eisenhardt\altaffilmark{4}}
\affil{Jet Propulsion Laboratory, California Institute of Technology,
MS 169-327, 4800 Oak Grove Drive, Pasadena, CA 91109}
\email{prme@kromos.jpl.nasa.gov}

\and

\author{M. Dickinson\altaffilmark{4}}
\affil{Space Telescope Science Institute\altaffilmark{5}, 3700 San Martin Dr.,
Baltimore, Maryland 21218}
\email{med@stsci.edu}
\altaffiltext{5}{The Space Telescope Science Institute is operated by
the AURA, Inc., under National Aeronautics and Space Administration
(NASA) Contract NAS 5-26555.}

\begin{abstract}

We present a study of the color evolution of elliptical and S0
galaxies in six clusters of galaxies inside the redshift range $0.78 <
z < 1.27$.  For each cluster, we used imaging from the Hubble Space
Telescope to determine morphological types of the galaxies.  These
types were determined both by an automated technique and from visual
inspection.  We performed simulations to determine the accuracy of the
automated classifications and found a success rate of $\sim$75\% at
$m(L_{\star})$ or brighter magnitudes for most of our HST imaging data
with the fraction of late--type galaxies identified as early--type
galaxies to be $\sim$10\% at $m(L_{\star})$ to $\sim$20\% at
$m(L_{\star}+2)$. From ground based optical and near-infrared imaging,
we measured the zero-point and scatter in the color--magnitude
relation of the elliptical and S0 galaxy populations, which we combine
with the previous sample of \citet{stanford98}, yielding a sample of
cluster early--type galaxies that span a lookback time of almost 9
gigayears from the present. We see the colors of the early--type
cluster members become bluer with increasing redshift as expected from
a passively evolving stellar populations.  We fit a set of models to
the change in the color as a function of redshift with the best
fitting values ranging from a formation redshift of $3^{+2}_{-1}$ to
$5_{-3}$, depending on the specific model used, though we find no
dependence for the formation epoch on the metallicity of the
populations.  The large scatter in resulting formation epochs, which
depends on the details of the models used, implies that we can
conclude that the oldest stars in the elliptical galaxies appear to
have formed at redshifts of $z>3$.  We find possible evolution in the
scatter of the colors, with some high redshift clusters showing
scatter as small as the Coma cluster but others showing much larger
scatter.  Those clusters with a small scatter imply either a formation
redshift of at least $z \sim 3$ or a smaller spread in the range of
formation redshifts at lower redshifts, assuming a Gaussian
distribution of star-formation around the mean epoch.

\end{abstract}
\section{Introduction}

The color--magnitude relation of galaxies has long been used to study
the evolution and formation of elliptical and S0, or early--type,
galaxies in clusters \citep[for
example]{faber73,visvanathan77,sandage78}.  The canonical picture of
the formation of early--type galaxies was the monolithic collapse model
of \citet{els62}.  In this scenario, elliptical galaxies form through
the gravitational collapse of large clouds of gas.  As the system
forms, it goes through a burst of intense star formation which quickly
ends as the majority of the gas in the galaxy is either locked up in
stars or heated to extreme temperatures, preventing further star
formation.  One of the crucial pieces of evidence for this model was
the work of \citet{bower92b}, since updated with
\citet{terlevich2001}.  The authors found a small scatter in the
colors of early--type galaxies in the Coma cluster.  Such uniform
colors for the early--type galaxies implies an early epoch of star
formation for the galaxies and small range in the epochs from galaxy
to galaxy.  \citet{aragon93}, \citet[SED95]{stanford95}
\citet{ellis97} and \citet[SED98]{stanford98} extended this work to
moderate redshifts.  In those papers the authors found no evolution in
scatter or slope of the early--type galaxy color--magnitude relation.
Further, in SED98 and \citet{dePropris99}, the authors found that the
evolution of both the colors of early--type cluster galaxies and the
luminosities of all cluster members agrees with passively evolving
stellar populations.

The monolithic collapse model was called into question with the work
of \citet{kauffmann98}.  In their picture, early--type cluster galaxies
build up wards from the accretion of smaller systems, a process often
referred to as hierarchical merging.  \citet{kauffmann98} were able to
recreate the observed color--magnitude relation of cluster early--type
galaxies at low redshift with a merging formation model.  However,
their model alone does not answer the question of how the small
scatter in cluster member colors comes about.  \citet{pvd_mf2001}
suggest a solution to that problem in the hierarchical merger scenario
with the ``progenitor bias.''  In this model, early--type galaxies form
continuously through merging but are only identified as early--type
galaxies when star formation has ceased.  The selection of an ensemble
of early--type galaxies will, therefore, only find the most extreme
members of the distribution of star-formation histories in the whole
of the galaxy population.  Anytime the most extreme members of a
random distribution are chosen, the distribution of properties among
those members is quite small.  Nonetheless, the uniformity of the
population of stars in early--type galaxies is such that
\citet{pvd_mf2001} finds that early--type galaxies have more than half
of their stars formed at or near the beginning of the star formation
epoch. \citet{bower98} find a slightly stronger result, requiring that
the majority, $\ge50$\%, of stars be formed within the first four or
so gigayears with only residual star formation still occurring at later
epochs.

In this paper, we investigate the color--magnitude relation of
early--type galaxies in the redshift range $0.58 < z < 1.27$.  We focus
on six clusters with $0.78 < z < 1.27$, to probe the highest
observable redshifts in order to have the most sensitivity to the
formation epoch of the stars in early--type galaxies.  We present the
early--type galaxy color--magnitude relation for these six clusters and
quantify the observed evolution in the slope and scatter.  In
addition, we discuss the evolution in the zero-point, or average
color, of the relation.  We rely on the data of \citet{stanford2002}
in this work, but review our data reduction procedures in \S 2 with
particular attention paid to how we constructed and merged catalogs of
Hubble Space Telescope (HST) imaging and ground based imaging.  One of
the most difficult problems is reliably morphologically typing high
redshift galaxies, even when using HST images.  We used two different
methods and performed a number of checks to ensure we understand how
well we can identify early--type cluster members, all of which is
discussed in \S 3.  We compared the observed color--magnitude relations
with that of the Coma cluster using the data described in
\citet{dePropris98}.  The comparison, outlined in \S 4, was done in a
similar manner as SED98.  We report our measurements in \S 4 and then
discuss our results in \S 5.

\section{Data}

The eight clusters in this sample represent an extension of the SED98
sample, though only six clusters are used in the final analysis.  The
aim of this paper is to increase the sample of SED98 at the highest
redshifts. All clusters in this paper are in the redshift range $0.58
< z < 1.27$, and with this sample we triple the number of clusters at
redshifts $z > 0.75$ from three in SED98 to a total of nine with ones
in this paper.  The photometric catalogs are published, along with the
data for SED98, in \citet{stanford2002}.  Below we will briefly review
how we prepared our data and highlight any differences with SED95 and
SED98.

\subsection{Cluster Selection}

Each of the clusters in this paper were selected based on the
availability of archival HST imaging and photometry from
\citet{stanford2002}.  The original sample of \citet{stanford2002} was
constructed to encompass all of the high redshift clusters known at
the time of the original survey, 1994 through 1996, regardless of the
method used to find the cluster.  The clusters in this paper's sample
are from the catalog of \citet{gunn86}, the Einstein Medium
Sensitivity Survey \citep{gioia90a,henry92,gioia94} or from clusters
found in the follow-up of the 3C catalog discussed in
\citet{spinrad85}.  The one exception is RDCS~0848+4453
\citep{stanford97}, which was found over the course of the SPICES
survey \citep{eisenhardt_spices} but is part of the {\tt ROSAT} Deep
Cluster Survey discussed in \citet{rosati98}.

The heterogeneous selection of clusters for this paper means that
there is a variety of follow-up work.  In \citet{oke98}, four of the
clusters in this paper were surveyed to determine cluster membership
and velocity dispersions.  Two of them, GHO~0229+0035 and
GHO~2155+0321 are found to be marginal over-densities at best. This
implies that they are unlikely to have a significant number of
early--type galaxies above the number expected from just the field
population.  GHO~0021+0406, discussed in \citet{lubin98} as CL
0023+0423, appears to have a number of cluster members, but also a
large number of the galaxies in the field appear to be projected along
the line of sight.  The systems appear to be two groups merging, one
with a velocity dispersion of $152^{+42}_{-33}\ {\rm km\ s^{-1}}$ and
the second with a velocity dispersion of $415^{+102}_{-63}\ {\rm km\
s^{-1}}$.  A side effect of this is that there will appear to be a
large number of galaxies in this system, but as the fraction of
early--type galaxies is lower in low density environments, there will
be a lower than expected early--type fraction given the number of
total galaxies.  The final cluster in the sample of \citet{oke98}
included in our sample is GHO~1604+4329, discussed in
\citet{postman2001} as CL 1603+4321.  This cluster has 41
spectroscopically confirmed members and velocity dispersion of
$935^{+126}_{-91}\ {\rm km\ s^{-1}}$, though no detectable X-ray
emission according to \citet{castander94}.

The remaining clusters of our sample were discovered originally from
X-ray selected catalogs or radio surveys.  MS1137.5+6625 has X-ray
imaging as well as X-ray and optical spectroscopy which is discussed
in \citep{donahue99b}.  The gravitational arcs and high X-ray
temperature both point to this being the most massive system in this
paper.  3C184 was spectroscopically confirmed in \citet{deltorn97},
with 11 redshifts and a measured velocity dispersion of
$634^{+206}_{-102}\ {\rm km\ s^{-1}}$.  Finally, in addition to a deep
Chandra observation discussed in \citet{stanford00}, RDCS~0848+4453
has fifteen spectroscopically confirmed cluster members.  The
color--magnitude relation and the fundamental plane of a subset of
these galaxies are discussed in \citet{vandokkum2001} and
\citet{vandokkum2003}.  It is unlikely that GHO~0229+0035 and
GHO~2155+0321 will yield enough early--type galaxies to provide a good
statistical sample, and that 3C184 and GHO~0021+0406 may have a large
number of interloping galaxies because of their small binding masses
and, thus, small implied richnesses.

\subsection{Ground based imaging}
\label{grounddata}
We obtained optical and near-IR images of the clusters using CCD and
HgCdTe array cameras on NOAO telescopes at Kitt Peak in 1993-1996.
Our imaging provides multi-wavelength photometric data of uniform
quality through a standard set of filters, and we used well-understood
photometric systems.  These features are particularly advantageous
when attempting to systematically investigate the evolution of large
samples of faint galaxies over a broad redshift range.  The sample of
clusters studied here, along with the band passes used, is summarized
in Table \ref{groundsummary}.  In the near infrared, atmospheric
transmission windows require us to use fixed band passes, so we have
imaged through standard $J$, $H$ and $K_s$ filters.  We have not
obtained all three filters for every cluster in the sample in this
paper, but we do have $J$ and either $H$ or $K_S$ for every cluster in
our sample.

The optical imaging was generally obtained in two bands which we have
adjusted according to the cluster redshift in order to ensure that
they span the $\lambda_0 \sim 4000$\AA\ region in the cluster rest
frame.  As in SED98, the clusters were divided into redshift ranges
for this purpose: $0.45 < z < 0.7$ ($V$ and $I$ bands); and $0.7 < z <
0.9$ ($R$ and $I$ bands) while for our clusters at $z=0.920$,
$z=0.996$ $z=1.169$ and $z=1.273$ the combination of the $I$ and $J$
filters straddle the 4000\AA\ region.  In this paper, we will
generally refer to the two pass bands that straddle the 4000\AA\
region as {\it blue} and {\it red}.  The former measures rest-frame
emission at wavelengths similar to or somewhat bluer than the
rest-frame $U$-band, while the latter corresponds to rest-frame
wavelengths roughly from $B$ to $V$.

We chose the exposure times in all band passes to provide galaxy
photometry with $S/N > 5$ for galaxies with the spectral energy
distributions of present-day ellipticals, down to $\sim$2 magnitudes
fainter than the apparent magnitude of a $L_{\star}$, or
$m(L_{\star})$ galaxy at the cluster redshift.  This magnitude also
includes the same aperture correction as in SED98 between the total
magnitude as measured by FOCAS and the true total magnitude of an
elliptical galaxy.  This permits us to study galaxy properties over a
similar range of luminosities for all clusters in our sample,
regardless of their redshift.  Our ground based images typically cover
a field size of $\sim$1 Mpc at the cluster redshift, which is
generally larger than that covered by the WFPC2 data used to select
the early--type galaxy subsamples (see below).  The optical data were
calibrated onto the Landolt system wherein Vega has $m_V = +0.03$, and
the IR images onto the CIT system wherein Vega has $m = 0$.  The
typical root mean square error of the transformations is 0.02 in the
optical and 0.03 in the near-IR.  The effective angular resolution of
the images is $\sim$1.2 \arcsec for our sample.  The last column in
Table \ref{groundsummary}, which is labeled as Table Number, refers to
which table in \citet{stanford2002} that contains the photometric
catalog for the cluster.  We use the same identifiers for the galaxies
in this paper as used in \citet{stanford2002}.

\subsection{HST Imaging}

The HST imaging we used in this work came from a combination of
archival data and pointing observations by some of the authors.  Most
of the clusters were observed in the F814W or F702W bands using the
WFPC2 instrument.  One cluster, GHO~0229+0035, was observed in F606W
and two, RDCS~0848+4453 and 3C210, were observed with the NICMOS
Camera 3 in the F160W (roughly an $H$ band) as well as with WFPC2 in
the F814W filter.  In Table \ref{hstsummary}, we list the total
exposure time and the coordinates of the pointings on a filter by
filter basis.  The NICMOS F160W observations are listed for each
central dither position, with the number of the dither position listed
next to the cluster name.  Three were necessary to adequately cover
the WFPC2 fields for those two clusters.  The last column in Table
\ref{hstsummary} lists the table in this paper where the HST imaging
results are listed.

\subsubsection{WFPC2 Image Preparation}

For a given set of images towards a cluster, all images with centers
within 0\farcs 05 of each other were stacked together.  This is less
than half a pixel in the WF images.  Each of these groupings of images
was then combined with {\tt crreject}.  The sum of each grouping of images
was then registered.  We shifted the registered images and then
combined them with crreject.  When at all possible, we used integer
shifts.  Integer shifts, instead of combining the data with drizzle,
allow us to use GIM2D as a method of identifying early--type galaxies,
which we shall discuss in more detail in \S \ref{class}.  

We have, after all of the above, an image for each chip.  We then
combined the data quality files into one final data quality file.  Any
pixel that was marked in one of the data quality files, even if only
partially covered by a shifted pixel, was flagged in the final data
quality file.  We used this final data quality file as a bad pixel
mask in later processing with GIM2D.  More recent versions of the
GIM2D software allow the use of weight maps, instead of forcing us to
flag a pixel as bad in all of the data when it is bad in only one of
the images.  

\begin{figure}[tbp]
\begin{center}
See 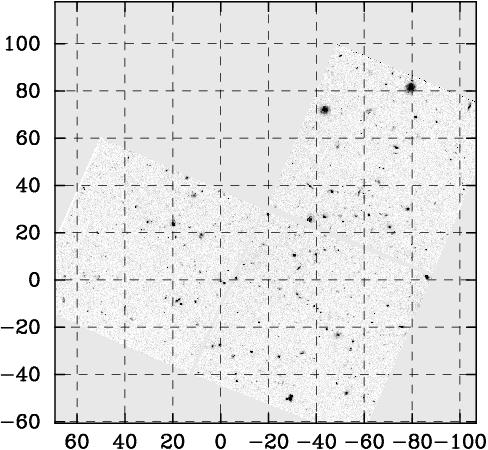
\end{center}
\caption[holden.fig1.ps]{Mosaic of WFPC2 frames for GHO~0229+0035 at
z=0.607.  North is up and East is to the left.  The dashed lines
represent offsets as measured from the brightest object near the
center of the field and are the same lines as used in Figure 29 of
\citet{stanford2002}.}
\label{cl0231_mos}
\end{figure}

\begin{figure}[tbp]
\begin{center}
See 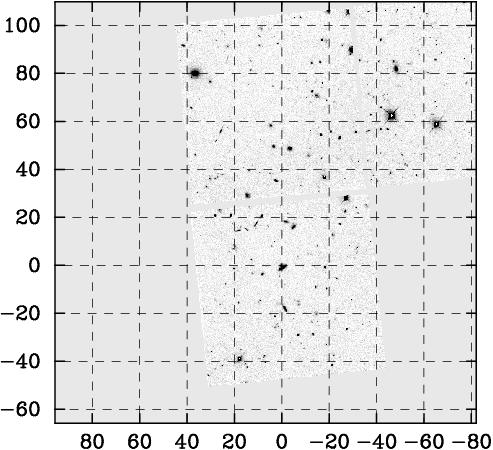
\end{center}
\caption[holden.fig2.ps]{Same as \ref{cl0231_mos} but for
GHO~2155+0321 z=0.7.  The corresponding image in \citet{stanford2002}
is Figure 31.
\label{cl2157_mos}}
\end{figure}

\begin{figure}[tbp]
\begin{center}
See 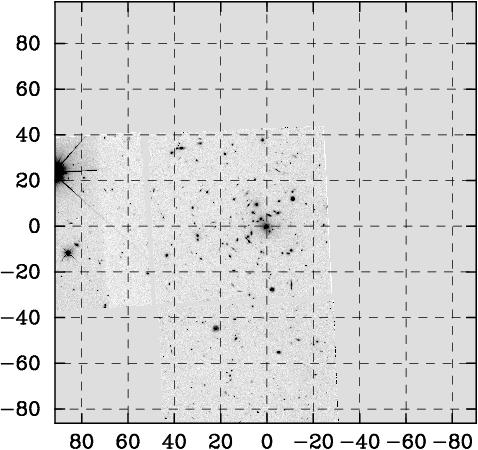
\end{center}
\caption[holden.fig3.ps]{Same as \ref{cl0231_mos} but for
  MS~1137.5+6625 at z=0.782.  The corresponding image in
  \citet{stanford2002} is Figure 35.
\label{ms1137_mos}}
\end{figure}

\begin{figure}[tbp]
\begin{center}
See 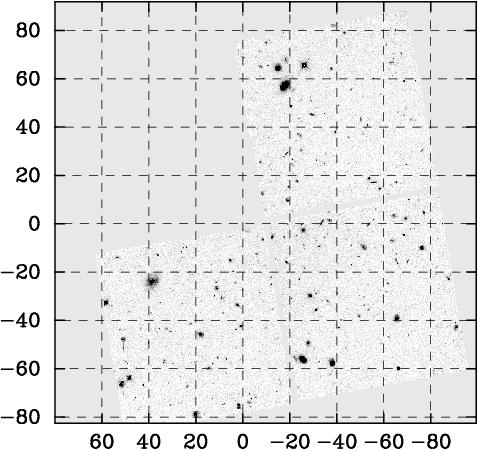
\end{center}
\caption[holden.fig4.ps]{Same as \ref{cl0231_mos} but for
GHO~0021+0406 at z=0.832.  The corresponding image in \citet{stanford2002} is
Figure 37.
\label{cl0023_mos}}
\end{figure}

\begin{figure}[tbp]
\begin{center}
See 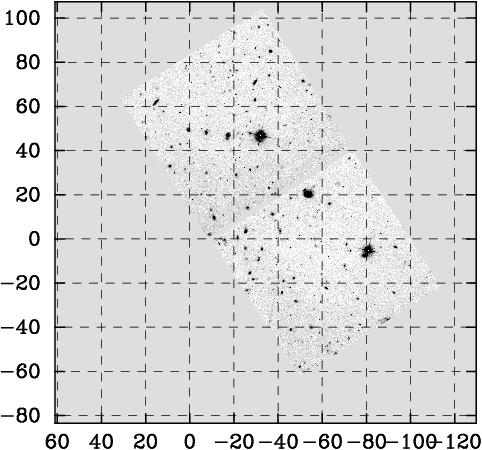
\end{center}
\caption[holden.fig5.ps]{Same as \ref{cl0231_mos} but for
GHO~1604+4329 at z=0.920.  The corresponding image in
\citet{stanford2002} is Figure 40.  The third WF chips does not
overlap with much of the ground based data and so was included.
\label{cl1604_mos}}
\end{figure}

\begin{figure}[tbp]
\begin{center}
See 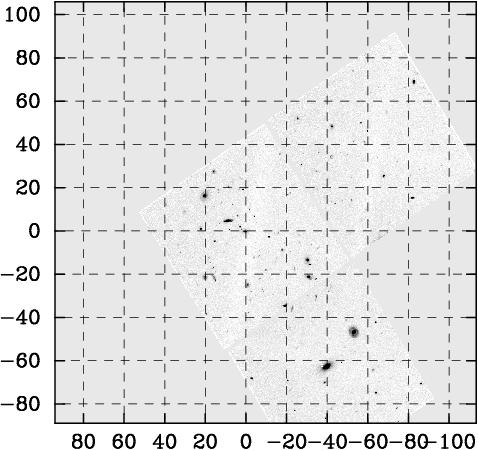
\end{center}
\caption[holden.fig6.ps]{Same as \ref{cl0231_mos} but for 3C184 at
z=0.996.  The corresponding image in \citet{stanford2002} is Figure
41.
\label{3c184_mos}}
\end{figure}

\begin{figure}[tbp]
\begin{center}
See 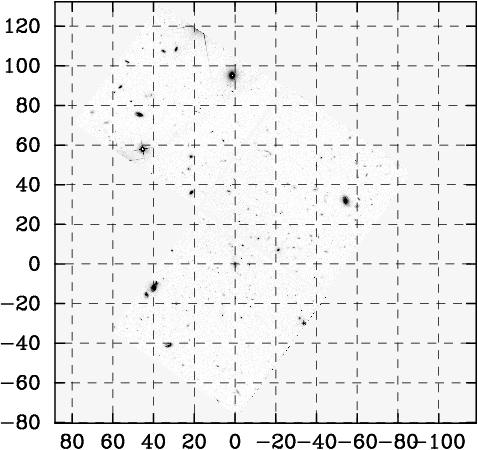
\end{center}
\caption[holden.fig7.ps]{Same as \ref{cl0231_mos} but for 3C210 at
z=1.169.  The corresponding image in \citet{stanford2002} is Figure
42.
\label{3c210_wfpcmos}}
\end{figure}

\begin{figure}[tbp]
\begin{center}
See 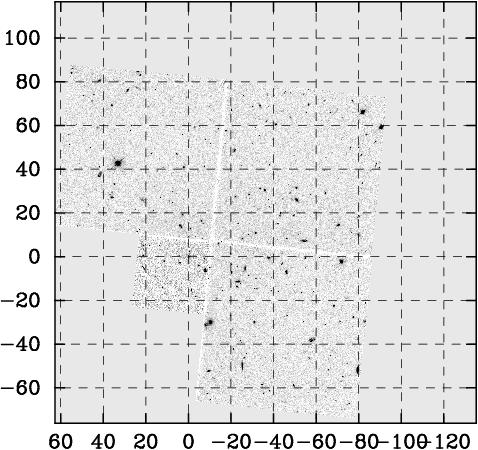
\end{center}
\caption[holden.fig8.ps]{Same as \ref{cl0231_mos} but for
RX~J0848+4453 at z=1.273.  The corresponding image in
\citet{stanford2002} is Figure 43.
\label{cl0848_wfpcmos}}
\end{figure}

In Figures \ref{cl0231_mos} through \ref{cl0848_wfpcmos}, we plot the
mosaic images of the WFPC2 imaging.  These mosaics have been
geometrically transformed into the coordinate system of the data
published in \citet{stanford2002}.  In most cases, the PC chip was not
transformed.  The small area generally meant that very few objects
with ground based photometry were observed in the PC chip. For
RDCS~0848+4453, however, the deeper imaging ground based data allowed
the PC chip to be included in the mosaic .  For GHO~1604+4329, one of
the three WF chips overlaps with little of the ground based data, and
so was not displayed in Figure \ref{cl1604_mos}.  In each figure,
North is up and East is to the left.  The dashed grid lines are the
same as lines for the images in \citet{stanford2002}.

\subsubsection{NIC3 Imaging}
\label{nic_imaging}

RDCS~0848+4453 was observed with the NICMOS Camera 3 in three different
pointings.  Each pointing was observed with a dither pattern of eight
exposures, one of 1344 seconds and seven of 1408 seconds.  In Table
\ref{hstsummary} we list the center position of each dither pattern
and the total exposure time.  The data were processed using a
combination of STScI pipeline routines and custom software.  All 24
images were combined into a single mosaic using the ``drizzling''
method of \citep{fruchter2002}.  The NICMOS images have a point-spread
function (PSF) with full width at half maximum (FWHM) of 0\farcs 22,
primarily limited by the pixel scale (0\farcs 2) of Camera 3.  The
final ``drizzled'' data were sampled at the resolution of the PC
camera, 0\farcs 04 per pixel.

3C210 was also observed with NICMOS Camera 3 in three different
pointings.  Each pointing was observed six times for a total of 5248
seconds, four exposures of 896 seconds and two of 832 seconds.  The
data were reduced using the same procedure as for RDCS~0848+4453.  Some
of the observations, however, were not usable as it appears that the
HST lost lock during the observation.  We list the total usable
exposure time for each of the pointings in Table \ref{hstsummary}.

\begin{figure}[tbp]
\begin{center}
See 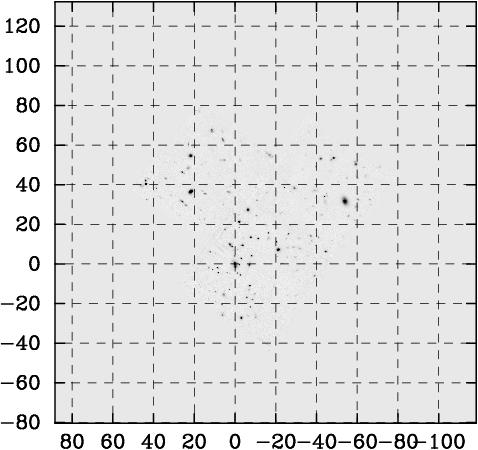
\end{center}
\caption[holden.fig9.ps]{Mosaic of NICMOS F160W pointings for 3C210,
  at z=1.169.  The corresponding image in \citet{stanford2002} is
  Figure 42.
\label{3c210_nicmos}}
\end{figure}

\begin{figure}[tbp]
\begin{center}
See 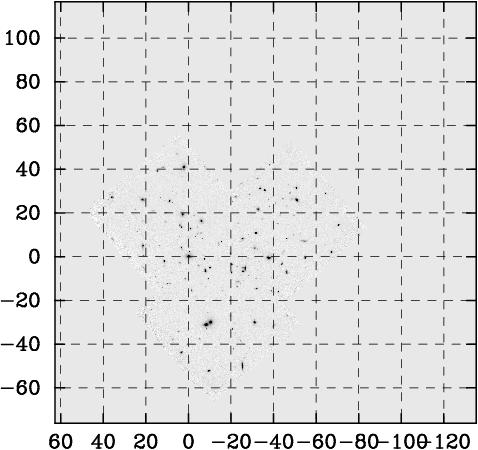
\end{center}
\caption[holden.fig10.ps]{Mosaic of NICMOS F160W pointings for
  RX~J0848+4453 at z=1.273.  The corresponding image in
  \citet{stanford2002} is Figure 43.
\label{cl0848_nicmos}}
\end{figure}

We plot mosaics of the NICMOS images, similar to the WFPC2 mosaics, in
Figures \ref{3c210_nicmos} and \ref{cl0848_nicmos}.  The coordinate
system and the dashed lines are the same as in the WFPC2 mosaics
as shown in Figures \ref{cl0231_mos} through \ref{cl0848_wfpcmos}.

\subsection{Galaxy Catalogs}

For each set of imaging data, we constructed different galaxy
catalogs.  Each catalog had a different scientific goal.  With the
ground-based imaging data, we measured the total magnitudes and colors
of galaxies.  The purpose of the catalogs of the HST data was to
identify morphological types.

\subsubsection{Ground Based Imaging Catalog}
\label{gcat}

Images in all band passes were co-aligned and convolved to matching
point spread functions (PSFs) in order to ensure uniform photometry at
all wavelengths through fixed apertures.  We detected objects on the
$K$ images using a modified version (Adelberger, personal
communication) of FOCAS (Valdes 1982).  We used the same software to
obtain photometry in each band through circular apertures with a
diameter equal to the twice the PSF FWHM.  This aperture was the same
angular size for all of the images as we convolved all of our data for
each cluster to match the largest PSF for that cluster.  These
apertures will be used for our color measurements while the total
magnitude was measured in the longest available infrared band, $K_s$
or $H$.  These total magnitudes are from the
modified version of FOCAS used to detect and measure the fluxes of
galaxies.  The definition in FOCAS is that the total magnitude is the
magnitude within an aperture that encompasses twice the detection
area.  Simulations with FOCAS, see SED98, show that these total
magnitudes are, in general, 0.3 magnitudes fainter than the true total
magnitude of the early--type galaxies.  We do not, however, correct the
total magnitudes in our data for this bias.

The observing methods, data reduction techniques, and photometric
methods for our ground-based data set are described in detail in SED95
and \citet{stanford2002}.  All photometry has been corrected for
reddening using the interstellar extinction curve given in
\citet{mathis90}, with values for $E(B-V)$ taken from \citet{bh82}.
The final photometric catalogs are in \citet{stanford2002}.

\subsubsection{WFPC2 Catalog}
\label{deblend}

We used SExtractor v2.1.1 \citep{bertin96} to construct a catalog of each of
the HST images.  For each image, we used the same threshold for
detection (1.5 $\sigma$ of the background and a minimum area of ten
pixels) and the same filter, a Gaussian with a FWHM of 1.5 pixels.
For SExtractor, the other significant parameter that controls the
number of detected objects is the minimum contrast parameter for
de-blending.  We choose a rather small value of 0.00175 (this is the
minimum fraction of total light within the detection isophote that the
program requires to form a new sub-object), as compared to the value
of 0.005 recommended in \citep{bertin96}.  The combination of the
de-blending contrast chosen and the small convolution kernel meant, in
general, irregular looking galaxies will be broken up into
sub-components.  However, close pairs or trios of nearby galaxies will
almost always be broken up into separate objects.

The main purpose of using SExtractor was to create a catalog of
galaxies for further post-processing.  Therefore, we did not try
optimize the detection threshold or de-blending parameters to ensure
we found every galaxy in the field.  Rather, we are interested in the
brighter galaxies that can be reliably morphologically typed and used
SExtractor to define those galaxies.  We used the {\tt flux\_auto}
parameter in SExtractor, a flux measured within 2.5 Kron radii
\citep{kron80}, as a total flux measurement.  We then trimmed our
galaxy catalog to a no-evolution $m(L_{\star}) + 3$ in the observed
HST band after converting the fluxes to magnitudes using the
zero-points from \citet{holtzman95}

\subsubsection{NICMOS Catalog}

As with the WFPC2 imaging, we used SExtractor to generate an image
catalog for each position.  Unlike with the WFPC2 data, we did not
combine each pointing for a dither position into one image.  Instead
we generated a catalog for each of the eight images separately.  The
catalogs were than joined so that an object found in the reference
image, the first image in the dither sequence, was matched with the
same object in each subsequent image.  To create this catalog, we used
SExtractor v2.1.1 with a 2.0 $\sigma$ detection limit and a minimum
area of 15 pixels.  We choose the fiducial value of 0.005 of the
contrast parameter.  We tested a large number of values and found
little change in the output catalog.  We chose these numbers as they
maximize the number of detected objects in the resulting catalog.

\subsubsection{Merging of catalogs}

The ground based catalog and the WFPC2 catalogs had to be merged into
a final list of galaxies.  The main problem is the very different
resolution in the ground based imaging as opposed to the HST imaging.
We transformed the coordinates in the HST catalog to those in the
ground based catalog.  We chose a square with five pixels in the HST,
or 0\farcs 5 for the WFPC2 data and 1\farcs 0 for the NICMOS, on a
side to match objects in the two catalogs.  The large size of this
box, especially when compared to the resolution of the HST, meant we
often had multiple matches in the HST catalog for one object in the
ground based catalog.  These duplicate matches were culled based on
visual inspection of ground and HST images.  In Tables
\ref{cl0231data} through \ref{cl0848nic} we list the positions of the
objects detected in the final merged catalogs.  Included are the
positions and identifiers from \citet{stanford2002}.  In that paper,
we list the magnitudes and colors used in this paper.  We list the
information used to determined the morphological types as well, which
will be discussed in the next section.

In Table \ref{numsummary}, we summarize the collected catalogs.  We
list, in addition to the cluster name and redshift, the value of
$m(L_{star})$.  We use the absolute magnitude for a $L_{\star}$ galaxy
from \citet{dePropris98}.  Table \ref{numsummary} lists the no
evolution $m(L_{\star})$ magnitude ($K$ or $H$) for each cluster
assuming a $\Omega_m = 0.1, \Omega_{\lambda} = 0.0, {\rm H_{o} = 65\
km\ s^{-1}\ Mpc^{-1}}$ cosmology, to be consistent with the values
from SED98.  In parenthesis, we also list the $\Omega_m = 0.3,
\Omega_{\lambda} = 0.7, {\rm H_{o} = 65\ km\ s^{-1}\ Mpc^{-1}}$
cosmology values for comparison.  We list the total number of galaxies
brighter than $m(L_{\star})$ in the merged catalogs.  We list the
total number of early--type galaxies brighter than $m(L_{\star})$ as
classified by our two different classification techniques discussed
next, with a slash separating the two values.  For 3C210 and
RDCS~0848+4453 our total number of galaxies and our number of
early--type galaxies are estimated from the NICMOS catalogs.  Finally,
we list an estimate of the number of field galaxies.  We discuss below
how that estimate is arrived at.

\section{Morphological Types}

\subsection{Classifications}
\label{class}

We used the high-angular resolution imaging from the HST to determine the
morphological types of the galaxies in two ways, by visual typing and
using an automated machined-based approach.  Two methods were used in
order to compensate for the difficulties of consistently
morphologically identifying galaxies at high redshift.

One of us (SAS) typed, by eye, all galaxies within the merged
HST/ground imaging catalogs.  Small images extracted from the
SExtractor catalogs of the WFPC2 or NICMOS data were used.  Each
postage stamp size was based on the isophotal size of the image
measured by SExtractor.  For RDCS~0848+4453 and 3C210, the images from
the drizzled NICMOS mosaic were used in the visual typing when an
object fell within the field of view of the data, else the WFPC2
images were used.

For an automated approach, we used the GIM2D package \citep{simard02}
in conjunction with the classification scheme of \citet{im00}.  The
package, described in \citet{simard02}, fits a two component model to
the two-dimensional galaxy images.  GIM2D uses as a galaxy model a de
Vaucouleurs law with a fixed ellipticity and an optically thin
exponential disk inclined to the line of sight.  The only constraints
shared between the bulge and disk components of the model is that both
have the same center and the same background value.  The model is then
convolved with a point spread function.  The best fitting model is
found for the two-dimensional image data by minimizing the $\chi^2$
statistic.

After fitting a galaxy model, we used the classification scheme of
\citet{im00}.  In this scheme, galaxies are classified with a T-Type
of less than 0 if the fraction of light in the model coming from the
de Vaucouleurs component is at least 40\% of the total light, and fit
residuals must be less than 8\% of the light in the galaxy, as
determined by the galaxy model normalization.  The residuals are
quantified using the two asymmetry parameters from \citet{schade95}.
One parameter measures the fraction of the total symmetric flux while
the other measures the fraction of the total asymmetric flux in the
residuals.  \citet{im00} add the two which yields the fraction of
flux in the residuals compared with the total amount of light in the
object.  This fraction is measured within two half light radii.
Generally, if the residual fraction is smaller than 8\% for a galaxy
with a bulge fraction greater than or equal to 40\%, the galaxy is
classified as an early--type.  There is an adjustment for the
half-light radius of the model, where for small half-light radii, less
than three pixels, the fraction of flux allowed in the residuals
becomes 7\%; see \citet{im00} for details on the classification and
the simulations used to calibrate these criteria.

Because the classification scheme relies on the amount of residuals,
after removing a model which assumes a constant background, this
approach is sensitive to close neighbors or galaxies with extended
envelopes.  For MS~1137.5+6625, we removed the brightest cluster
member first, using the best fitting model from GIM2D, and then reran
SExtractor and GIM2D to process the remainder of the cluster.
Removing other bright galaxies from MS~1137.5+6625 did not change the
results of the GIM2D fits, because only the brightest galaxy has the
kind of extended envelope found in the most luminous of cluster elliptical
galaxies.

For the RDCS~0848+4453 and 3C210 NICMOS imaging, along with the F814W PC
data for RDCS~0848+4453, we used a different routine.  A specially
modified version of GIM2D was used that simultaneously fits a model to
the different images for a given pointing, eight for RDCS~0848+4453 and
up to six for 3C210.  The fitting program then minimizes the $\chi^2$
statistic for all of the images simultaneously.  For the NICMOS data,
the variance maps, constructed for the drizzled images, were used to
construct variance maps for the model fitting.  To determine the
early--type galaxies, the same parameters were used, the bulge to total
ratio, half-light radius and the sum of the residuals, as described
above.  The only difference was that the residuals were computed on an
image by image basis, so we compared the average residuals with
maximum values given in \citet{im00}.

For Table \ref{numsummary}, we list an expected number of field
early--type galaxies.  One of the difficulties is that there does not
yet exist a large published sample of early--type members with high
resolution infrared imaging at high redshift.  Instead, we use the
sample of \citet{im00}.  This has two advantages.  First, it uses the
same automated identification process that our paper uses.  Second, it
contains a large sample, 145 objects in total.  The only disadvantage
is that the sample does not go faint enough to reach $m(L_{\star})$
for the highest redshift clusters in our sample.  We augment the
sample using the rest of the Groth Strip Survey from \citet{simard02}.
To select potential early type galaxies, we use the same parameters as
\citet{im00} did for the faintest magnitude bin.  This selection
process will yield an unknown number of contaminants.  However, we go
only $\simeq 1$ magnitude below the limit of \citet{im00} at the
highest redshift.  In addition, we use a color cut.  We select
galaxies red ward of the color of the bluest early--type galaxy in
each cluster field.  This should eliminate many of the contaminants in
our morphological selection.  In general we compute the number of
early--type galaxies expected for three WFPC2 chips. For both 3C210
and RDCS~0848+4453, however, we provide the expectation value for the
NICMOS pointings, which are 0.42 the size of the WFPC2 pointings.  For
GHO~1604+4329, we rescale by two-thirds the expected number because we
only use two of the three WF chips.

The morphological types are listed in Tables \ref{cl0231data} through
\ref{cl0848nic}.  Each table lists in the first column an
identification number which corresponds to the identifier used in the
photometric catalogs presented in \citet{stanford2002}.  The second
and third columns list the offsets, in arc-seconds, in both right
ascension and declination from the central cluster galaxy.  These
positions are the same as used in \citet{stanford2002}.  The fourth
column corresponds to the morphological class assigned by visual
inspection.  For Tables \ref{cl0231data} through \ref{3c184data}, the
morphological types are T Types \citep{devauc1991} where an early--type
galaxy is classified as -5 through -1, with 0 being classified as
S0/Sa and not included as an early--type galaxy for this paper.  Values
smaller than -5 are classified as stars.  Tables \ref{3c210data},
\ref{3c210nicdata}, \ref{cl0848wfpc} and \ref{cl0848nic} list a number
from zero to five instead of a T Type.  A zero represents a star, one
represents an early--type galaxy and larger values are for steadily
less symmetric systems.  We made this change because we could not
classify the galaxies reliably on the T Type scale.  Instead, we use
broad bins for classification.  The numbers can be translated, using a
a value between -5 and -1 for galaxies labeled with a 1.  

The next three columns in all of these tables represent the values
from GIM2D used to classify the galaxy; the bulge to total ratio, the
fraction of the total galaxy light in the residuals, and the
half-light radius in arc-seconds.  For some objects, the fraction of
galaxy light is strongly negative, such as -199.98.  This comes from
bright stars where the cores have been removed by the cosmic ray
rejection algorithm.  Fainter stars can usually be identified by a
half light radius with a size of less than one pixel, or 0\farcs 1 for
the WFPC2 data.  The final columns contain the identification of the
galaxy in the HST imaging.  For the WFPC2 data, this is the chip the
galaxy was in and the position on the chip in pixels.  For NICMOS
data, we list the pointing number along with the position.  The
pointings are identified as one, two or three and the numbers
correspond to those listed in Table \ref{hstsummary}.

Two entries in Table \ref{g0023data} have a -1 in every GIM2D
column.  These galaxies were too close to the edge of the WFPC2 frame
for GIM2D to fit a model but we could determine visual
classifications.  Table \ref{g2157data} lists object \#8 twice.  It is
clearly separated into two objects in the HST frame but is a blend in
the ground based image.  Table \ref{cl0848wfpc} lists object \#6 three
times as it was de-blended into three separate objects in the WFPC2
data.  In the NICMOS data, however, it remains one object.  Object \#6
is a confirmed cluster member and possible merger candidate, see
\citet{vandokkum2001} for further discussion.  

\subsection{Comparing visual types with GIM2D}

\begin{figure}[tbp]
\begin{center}
\includegraphics[width=3.0in]{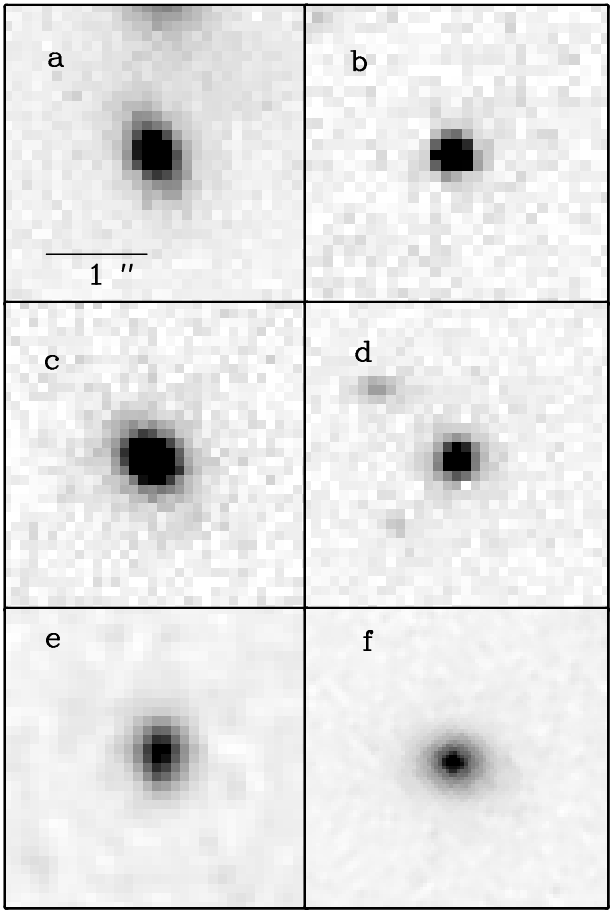}
\end{center}
\caption[holden.fig11.ps]{ 

Six early--type galaxies, all within $\pm 0.2$ magnitudes of
$m(L_{\star})$, are plotted.  Each galaxy is classified as an
early--type visually and by using GIM2D.  Each galaxy image is the
same angular size, a solid line in part a is provided for scale.  a)
Object \# 57 (194,270) from MS1137.5+6625, classified as -1.  b)
Object \# 57 from GHO~0021+0406, classified as a -5.  c) Object \# 62
from GHO~1604+4329, a -3.  d) Object \# 19, a -2 from 3C184.  e)
Object \# 50 from the NICMOS imaging of 3C210.  f) Object \# 43 from
the NICMOS imaging of RDCS 0848+4453.  The images for e and f were
both based on drizzled data, though the GIM2D models were not fit to
that data.

}
\label{normal}
\end{figure}

We plot in Figure \ref{normal} examples of early--type galaxies.  We
have plotted six galaxies, with a magnitude within 0.2 magnitudes of
$m(L_{\star})$ that was classified as an early--type by both visual
inspection and by GIM2D.  Each postage stamp has the same field of
view, 3\farcs 2 , and we plot in one figure a 1\farcs 0 bar for
scale.  Two of the galaxies are from the NICMOS imaging, one from
3C210 and one from RDCS~0848+4453.  Both stamps are drizzled images,
and therefore not the images we used for GIM2D fits.  The 3C210 image
has pixels of 0\farcs 1, as do all of the WFPC2 images.  The
RDCS~0848+4453 data was drizzled to a scale of 0\farcs 04 so as to
match up with the pixel scale of the PC chip.  

For a comparison between the approaches to morphological
classification, we used our results for MS~1137.5+6625.  This cluster
is a rich, compact system at a redshift of $z=0.78$
\citep{gioia94,clowe98,donahue99b} with a large number of early--type
galaxies in the cluster for which we can compare classifications.  The
visual classifications for MS~1137.5+6625 yielded 37 early--type
galaxies while the automated classification system yielded 35
early--types.  Of the 37 visually classified galaxies, 10 are not
classified as early--type galaxies by GIM2D.

\begin{figure}[tbp]
\begin{center}
\includegraphics[width=3.0in]{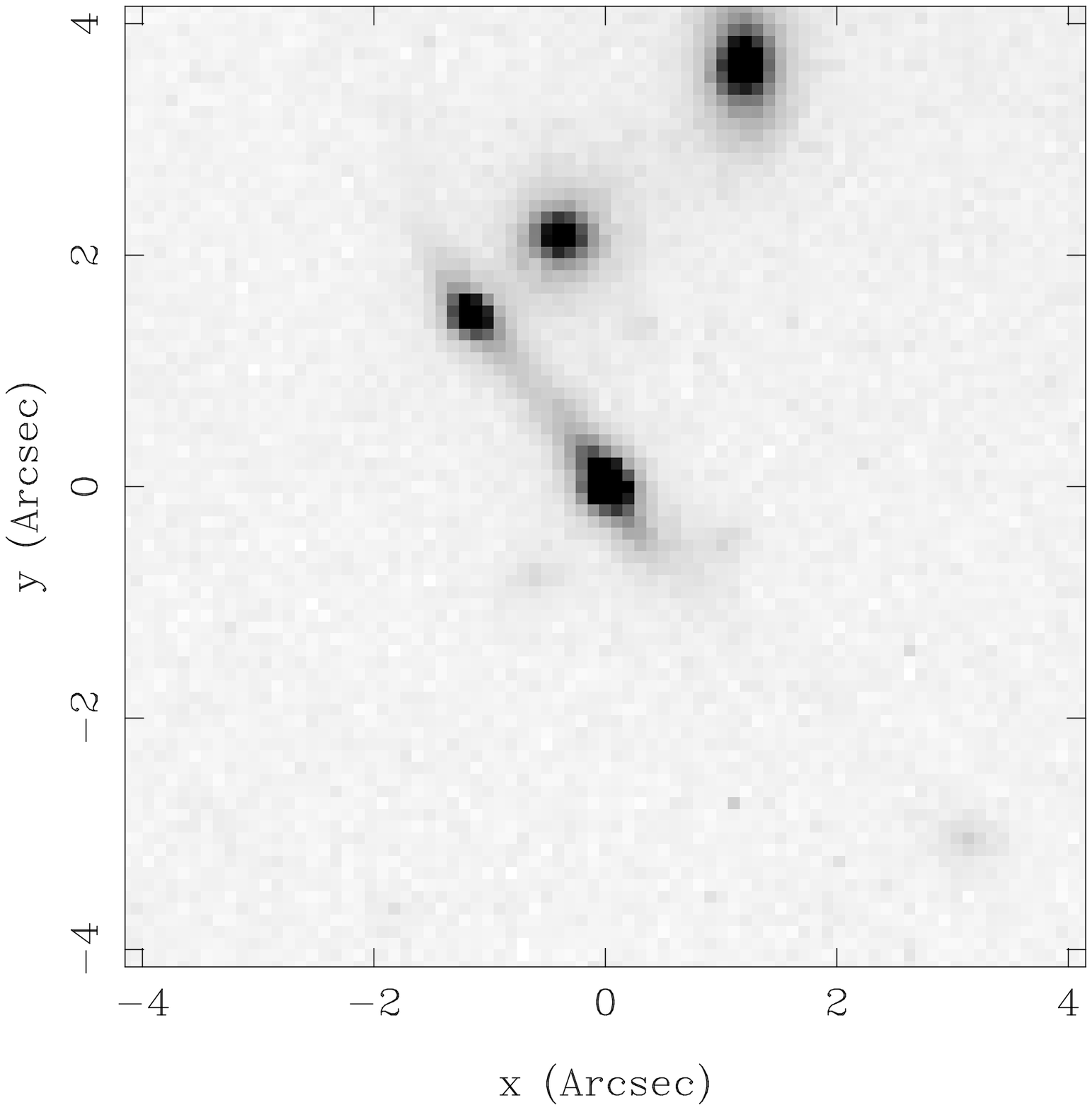}
\end{center}
\caption[holden.fig12.ps]{ The brightest early--type galaxy, object \#
11 in Table \ref{m1137data}, as classified by SAS but not by GIM2D in
MS~1137.5+6625.  In the ground-based image the galaxy in the center is
blended with the galaxies to the north.}
\label{diff_a}
\end{figure}

\begin{figure}[tbp]
\begin{center}
\includegraphics[width=3.0in]{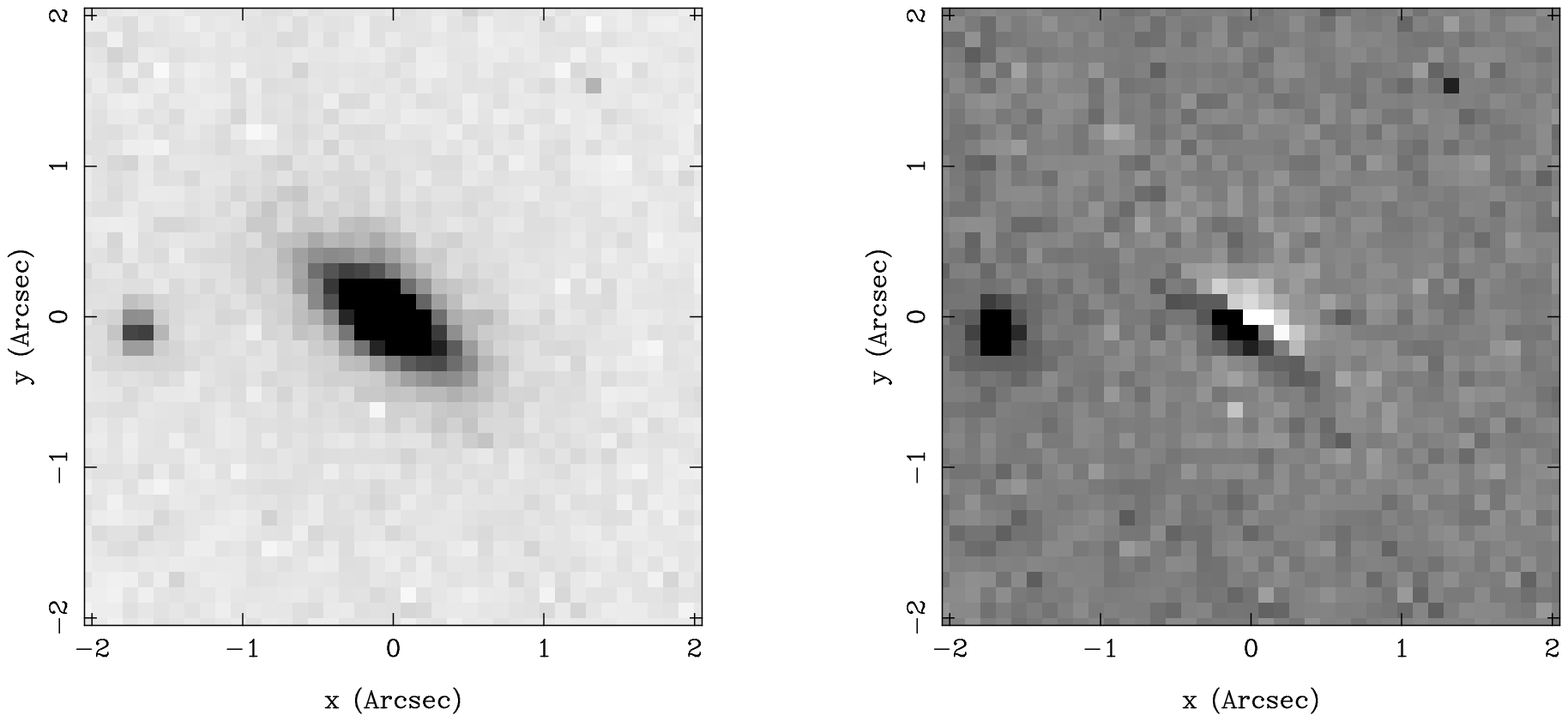}
\end{center}
\caption[holden.fig13.ps]{ In the left panel is object \# 40 in Table
\ref{m1137data}, a S0 according to SAS, with a GIM2D bulge to total
ratio of 0.74 but with large model residuals.  The residuals are
plotted in the right panel.}
\label{diff_b}
\end{figure}

The galaxies classified visually as early--type galaxies but not
classified as such by GIM2D fall into three categories: galaxies
classified as early--type spirals by GIM2D (two), galaxies with close
neighbors (five) and edge-on galaxies (three).  The brightest galaxy
classified by SAS as an early--type but not classified as such by
GIM2D is in Figure \ref{diff_a} (\#11 in Table \ref{m1137data}).  This
is a complicated system with three galaxies very close on the sky and
is possibly lensed.  Nonetheless, this makes a good example of the
first type of discrepant object, one with a bright, nearby
companion. Most of the clusters in this sample are much less rich and
compact when compared with MS~1137.5+6625. So, the errors in the
automated classification caused by image crowding should be less
pronounced and there should be fewer edge-on galaxies, as they only
make up small percentage of the over-galaxy population.  The second
kind of discrepant object can be seen in Figure \ref{diff_b} (\#40 in
Table \ref{m1137data}).  The galaxy is an isolated object classified
as a S0 by SAS.  The best fitting model is bulge dominated, with a
bulge to total fraction of 0.74, but with residuals that make up 11\%
of the total flux.  The residuals are anti-symmetric, having large
positive values in one half of the galaxy and large negative values
for the second half.  Such residuals are an artifact of an absorbing
material, such as dust, which would depress the flux from the galaxy
center.  This cause the best fitting model to be offset in the
centroid from the center of the galaxy, causing this particular set of
residuals.  Therefore, the visual classification is likely correct as
there are S0 galaxies with small dust lanes observed at the present
epoch.

There are eight galaxies not classified as an early--type by visual
classification but meet the criteria from GIM2D.  Of these eight,
three are classified as T Type 0, or S0/Sa, three are T Type 1 or a Sa
and two are T Type 2 for Sa/b.  As all are visually classified as
early--type spirals, we conclude that the discrepancy comes from the
scatter in galaxy classification technique, inherent in any
measurement.  A further confirmation of this comes from examining
those galaxies which have colors in agreement with galaxies classified
as early--types but have T Types of greater than 0 according to the
automated classification.  We use the 27 early--type as classified by
GIM2D to determine a color--magnitude sequence.  Ten late-type
galaxies have colors similar to this color--magnitude sequence.  Of
these ten late-type galaxies, none has been visually classified to
have a T Type greater than 2.

\subsection{Comparing GIM2D with the Medium Deep Survey}
\label{mds_compare}

\citet{lubin98} use the automatic object detection and classification
methods of the Medium Deep Survey \citep[MDS]{rog99} to quantitatively
analyze the morphology of GHO~0021+0406.  The approach of the MDS is
similar to what we have done in that the pipeline consists of a two
step process, an object detection step and an object classification
step.  For the MDS, object detection is done by looking for contiguous
pixels with a signal to noise of $1\sigma$ in the WFPC2 image.  The
MDS team then makes an object mask and uses the masked image to
determine the pixel associated with a given object.  The object
classification step begins with a simple moment analysis of the pixels
in a given objects mask, then proceeds with a maximum likelihood fit
to the object pixels of a model.

Though the methods used by the MDS are quite similar to the SExtractor
and GIM2D analysis we have performed, there are two main differences.
First, we chose, in SExtractor, a very small number for the de-blending
parameter, knowing that we would separate galaxies with complicated
morphologies into several smaller objects (see \S \ref{deblend}).  We
chose this approach because we wanted to use the asymmetry parameters
and we wanted to ensure that objects projected into the line of sight
of early--type galaxies would be de-blended in most cases.  In
contrast, the MDS team de-blends in a much less stringent manner, to
prevent flocculent spirals or other galaxies with internal structure
from being de-blended into multiple objects.  The second major
difference is that the MDS team did not fit a full bulge plus disk
model to all of the objects as we did.  Only those objects with
a total signal to noise above a certain threshold were fit
\citep{rog99}.  Also, if a galaxy was almost as well fit by a pure
bulge or pure disk model, that model was used in place of the combined
model.

\begin{figure}[tbp]
\begin{center}
\includegraphics[width=3.0in]{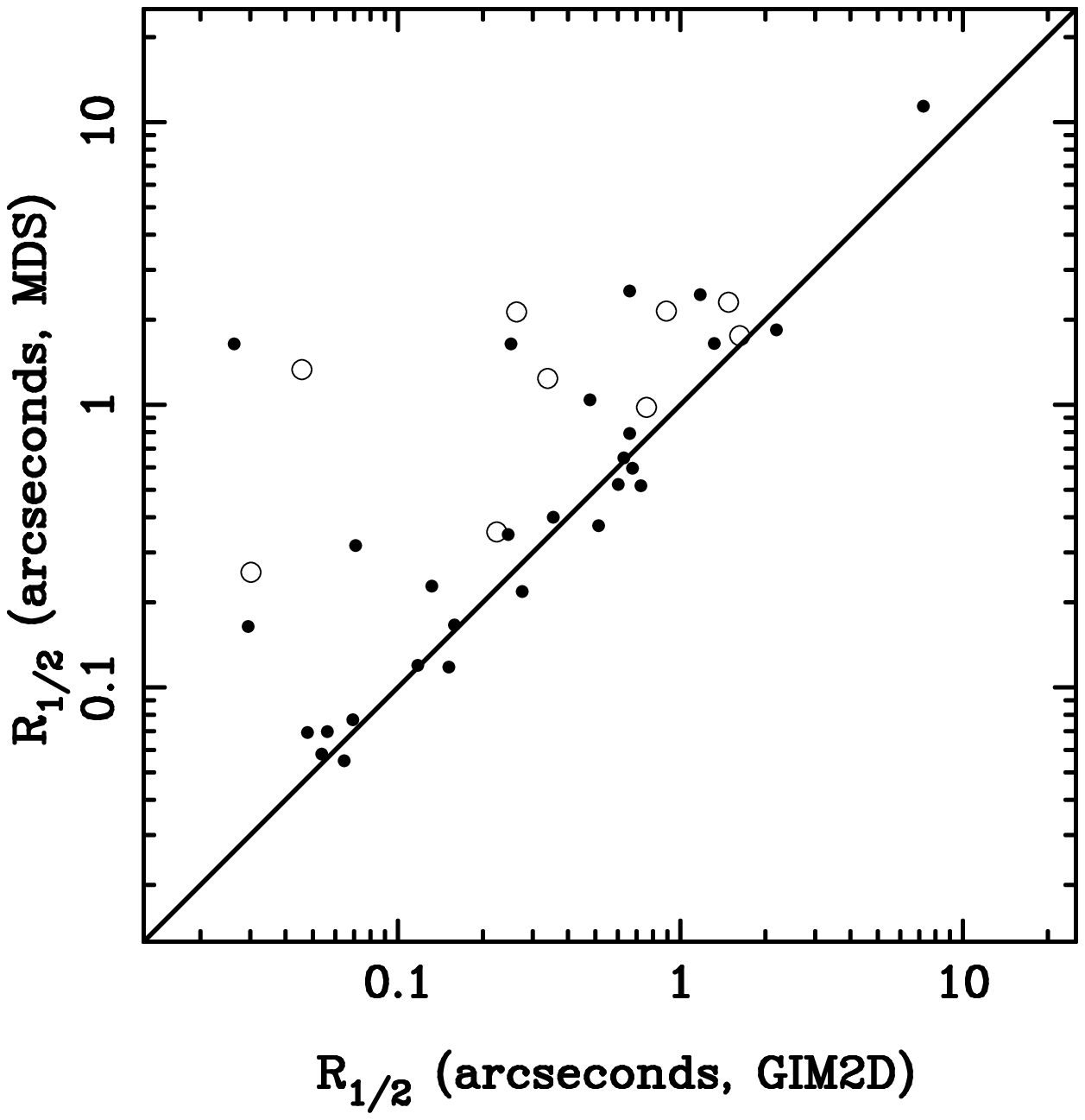}
\end{center}

\caption[holden.fig14.ps]{ The logarithmic half-light radius as measured with
GIM2D as compared with the Medium Deep Survey values for GHO~0021+0406.
Solid points represent galaxies where SExtractor found only one object
within 1\arcsec , open circles are galaxies where more than one object
from the SExtractor catalog was within 1\arcsec\ of the MDS centroid.
The line is a one to one relation, not a fit.  See \S
\ref{mds_compare} for further details.}
\label{rhalf}
\end{figure}

For GHO~0021+0406, we compare the galaxies where the MDS fit a full
bulge plus disk model to our resulting models using GIM2D.  First in
Figure \ref{rhalf} we compare the half-light radii of the two
approaches.  The solid dots are galaxies where, in the process of
matching the SExtractor catalog and the MDS catalog, there was only
one galaxy in the SExtractor catalog within 1\arcsec\ of a MDS galaxy.
The open circles represent where two or more galaxies from the
SExtractor catalog were within 1\arcsec\ of a MDS galaxy.  We then
chose the galaxy in the SExtractor catalog with the smallest
separation from the MDS galaxy.  Clearly there is a tendency for
larger half-light radii from the MDS results.  Based on the scatter in
the open circles, most, if not all, of the discrepancy between the two
sets of model fits must come from the very strict de-blending approach
we used in creating the SExtractor catalogs.  

\begin{figure}[tbp]
\begin{center}
\includegraphics[width=3.0in]{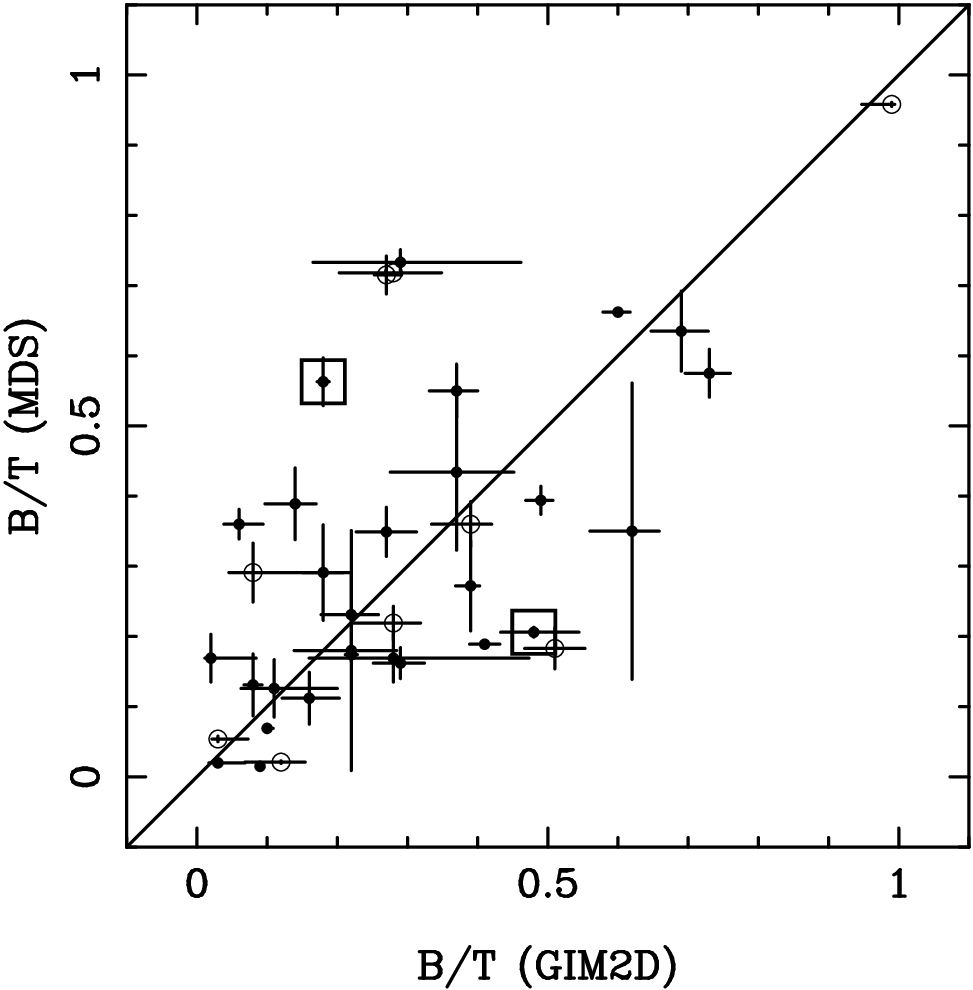}
\end{center}
\caption[holden.fig15.ps]{ The bulge to total ratio as measured with GIM2D
as compared with the Medium Deep Survey values.  We use the same
symbols as in Figure \ref{rhalf}.  The line is a one to one relation.
The two galaxies represented by boxes are shown in Figure
\ref{discrepant} and discussed in \S \ref{mds_compare}. }
\label{bulge}
\end{figure}

In Figure \ref{bulge}, we plot the ratio of the bulge flux to the
total flux, where total is defined as the bulge model flux plus disk
model flux, for the two samples.  We use the same symbols as in Figure
\ref{rhalf}.  The MDS results in general agree with the GIM2D results.
However, a few of the galaxies seem to have values of the B/T ratio that
are quite different when comparing the two algorithms.  Unlike with
the half-light radii, this does not seem to depend on the criteria
used to de-blend objects in the SExtractor catalog.

\begin{figure}[tbp]
\begin{center}
\includegraphics[width=3.0in]{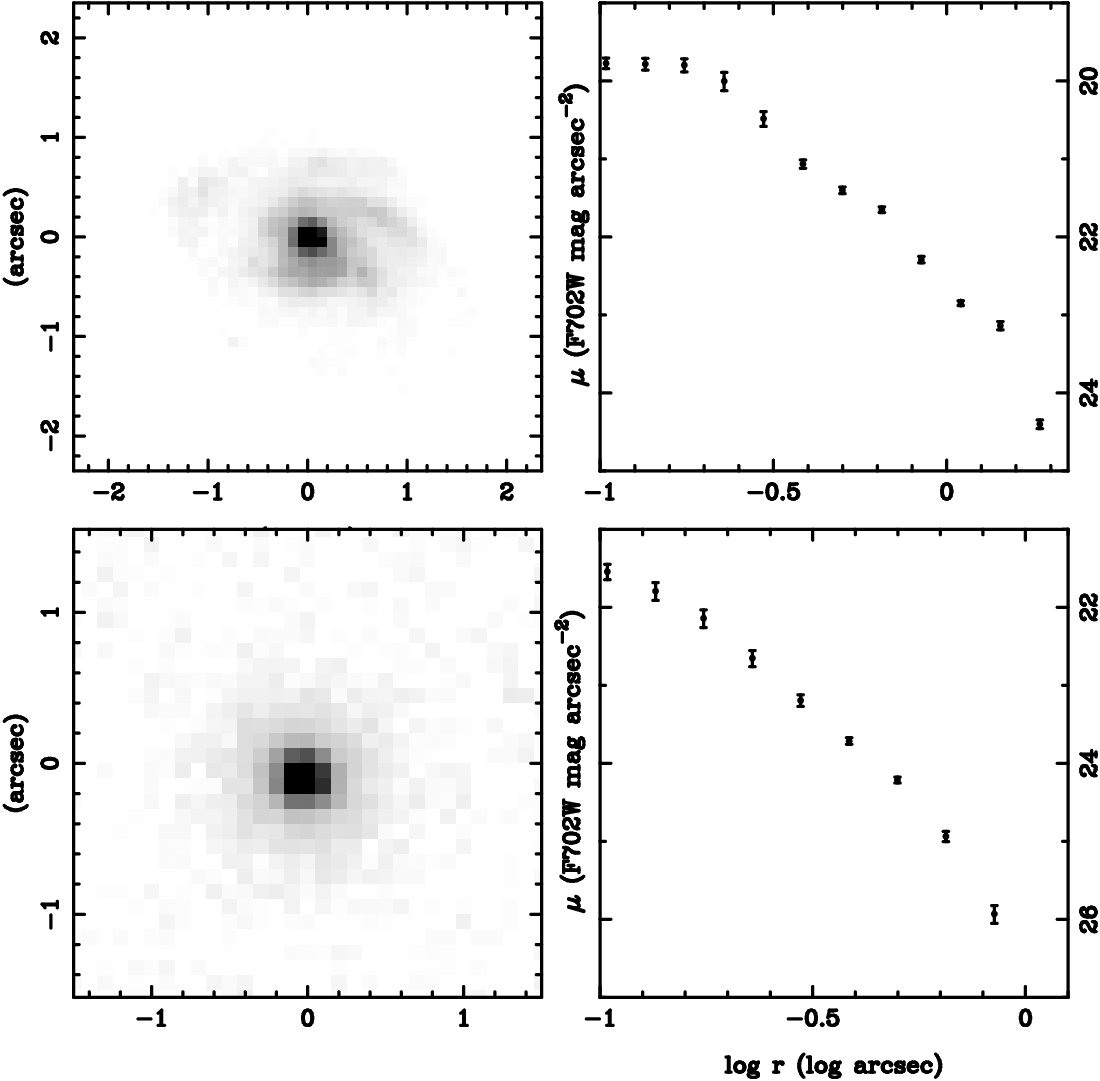}
\end{center}
\caption[holden.fig16.ps]{ The two galaxies plotted with open squares in
Figure \ref{bulge}. The top galaxy was assigned a high bulge to total
ratio by the MDS and a low value by GIM2D, thus it lies above the line
in Figure \ref{bulge}.  The bottom galaxy was assigned a low bulge
to total ratio by the MDS and high value by GIM2D.}
\label{discrepant}
\end{figure}

In Figure \ref{bulge}, we place boxes around two data points with
strongly discrepant bulge to total ratios.  In Figure \ref{discrepant}
we plot these galaxies along with their surface brightness profiles as
determined by the IRAF task {\tt ellipse}.  The object in the upper
left of Figure \ref{discrepant} was classified by GIM2D to have a
bulge to total ratio of $0.18 \pm 0.02$ while the MDS classified it to
have a bulge to total ratio of $0.56 \pm 0.03$.  Based on the surface
brightness profile in the upper right of Figure \ref{discrepant}, it
is clear that the spiral structure in the galaxy, along with an inner
ring, cause multiple breaks in the surface brightness profile.  These
breaks no doubt allow multiple models to be fit.  The second object
with a box in Figure \ref{bulge} is in the lower left of Figure
\ref{discrepant}.  GIM2D fits a bulge to total ratio of
$0.48^{+0.15}_{-0.11}$ while the MDS yields $0.206 \pm 0.007$.  This
galaxy is classified as an early--type by GIM2D, as the model residuals
are 1\% of the galaxy flux.  This object was visually classified as a
-3 in this paper and an ``E/Sa?''  by visual inspection in
\citet{lubin98}.  Clearly the extended envelope makes visual and
automated classification problematic.  We shall however, consider the
object an early--type galaxy as both GIM2D and our visual
classification yield a consistent result.

\subsection{Comparing GIM2D with Simulations}
\label{class_sim}

We used the sample of \citet{frei96} to create simulated HST galaxy
images.  For every cluster, a blank piece of sky was extracted from
the imaging data.  For RDCS~0848+4453 and 3C210, both a frame from the
WFPC2 F814W and from the NICMOS data were used.  Each galaxy from
\citet{frei96} was imaged in multiple bands.  For the simulations, we
choose the band closest to the observed HST band at the appropriate
redshift.  For example, for RDCS~0848+4453, we used the $B_j$ band or
the $g$ band images from the \citet{frei96} catalog when simulating
the WFPC2 F814W data and the $R$ or $r$ images when simulating the
NICMOS F160W data.  This has the effect of adding some morphological
changes expected when observing the same galaxies in different
filters.  Once a filter was chosen, the image size was as appropriate
for the redshift of the cluster in our assumed cosmology.  For the
local distances of the \citet{frei96} catalog, we used the HST Key
Project to Measure the Hubble Constant distances \citep{freedman2001},
when available and, additionally, assumed that all members of the
Virgo cluster were at the same distance.  In addition, we used the
results of \citet{tully2000} and the distances in \citet{giraud98}
(see references therein.)  If we did not find a distance in the
literature for a given galaxy, we instead used the redshift as a proxy
measurement for the distance to the galaxy. After rescaling, each
image was convolved with an appropriate point spread function.  The
magnitude of the galaxy was fixed at a number of different levels
starting from $m(L_{\star}) - 1$ at the redshift of the cluster in the
observed band to $m(L_{\star}) + 2$, spaced at half magnitude
intervals.  The pixel values for the fake galaxy were then used as
expectation values for Poisson random deviates.  The data were then
divided by the appropriate gain for WFPC2 and the noisy, fake galaxies
were added to the blank piece of sky from each cluster's image.  The
data reduction pipeline was run on this resulting image using the same
SExtractor and GIM2D configuration files as used for the original
cluster imaging data.

\begin{figure}[tbp]
\begin{center}
\includegraphics[width=3.0in]{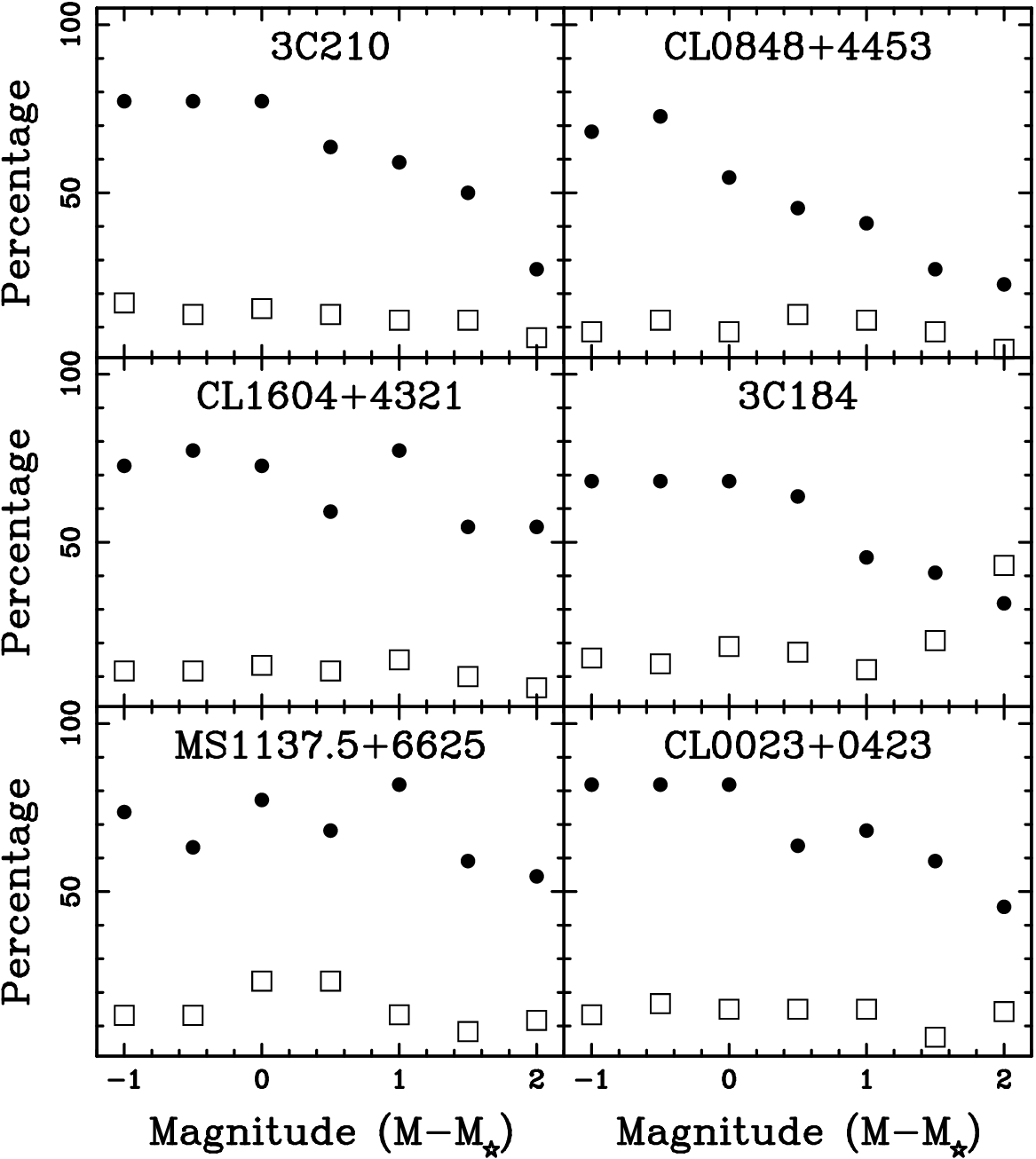}
\end{center}

\caption[holden.fig17.ps]{ We plot, for each cluster where we used
the WFPC2 imaging for automatic classification, the fraction of
correctly identified early--type galaxies in our simulations as solid
points.  As open squares, we plot the fraction of late-type galaxies
incorrectly identified as early--type galaxies from our simulations.}
\label{wfpc2_sims}
\end{figure}

The automated classification process consists of two components,
measuring the bulge-to-total ratio and measuring the flux in the
residuals.  The half-light radius is a third component, but only
affects the results in a minor way.  Overall, for the WFPC2
simulation, we found that the average error in the bulge to total
ratio was around 5\% at $m(L_{\star}) - 1$ to 30-40\%,
depending on the cluster, at $m(L_{\star}) + 2$.  Therefore, at
$m(L_{\star}) + 2$, a galaxy with a true bulge to total ratio of 50\%
would have a measured value of 10\% or less, 16\% of the time.  At
$m(L_{\star})$, the error on the bulge-to-total ratio was around 10\% or
less.  As we are most interested in measuring the color--magnitude
relation of $m(L_{\star})$ or brighter galaxies, this means that most of
them have reliable bulge to total ratios.  When we compare these
simulations with the classifications done using the bulge to total
ratio and the size of the residuals, this conclusion is borne out.  At
$m(L_{\star})$ or brighter, around 80\% of the galaxies in WFPC2 are
correctly classified as early--type galaxies.  The correct
classification rate drops to around 40\% at $m(L_{\star}) + 2$. Given the
large errors on the bulge to total ratio at these magnitudes, this
result is not surprising.  We plot our results, on a cluster by
cluster basis, in Figure \ref{wfpc2_sims}.  In that Figure, we plot
both the false positive rate, with open squares, and the fraction of
total early--type galaxies classified, using solid circles.

For RDCS~0848+4453, we found significantly worse results when using the
WFPC2 data for classification.  The best correct classification rate,
$\simeq$ 70\%, occurs at magnitudes brighter than $m(L_{\star})$ with a
rate of 55\% at $m(L_{\star})$.  The false positive rate at these
magnitudes is around 9\%.  There are five early--type galaxies brighter
than $m(L_{\star})$ according to the GIM2D classification scheme, which
means it is likely that there are seven or eight in the cluster.  At
faint magnitudes, the correct classification rate drops to 20\% at
$m(L_{\star}) + 2$.  A large part of this comes from the much worse
errors on the bulge to total ratio, usually double the size reported
for the lower redshift clusters at the same magnitude.  No doubt this
is a result of the faint magnitudes and therefore low signal-to-noise
for cluster galaxies at this redshift.

Both RDCS~0848+4453 and 3C210 were observed and classified with NICMOS
F160W imaging as well.  Similar simulations were performed for the
NICMOS data as well.  Instead of a single noisy, fake galaxy being
used for each simulation, one was constructed for each dither position
and it was added to an oversampled image.  Small random errors were
added to the true dither offsets to simulate our uncertainties in the
observed offsets between the observed images of a given galaxy.  After
a galaxy image was oversampled, it was convolved with an oversampled
point spread function made by TinyTIM.  This image was binned up to
the resolution of the instrument and then added to a noisy frame of
from the actual NICMOS imaging data from a given observation.  

\begin{figure}[tbp]
\begin{center}
\includegraphics[width=3.0in]{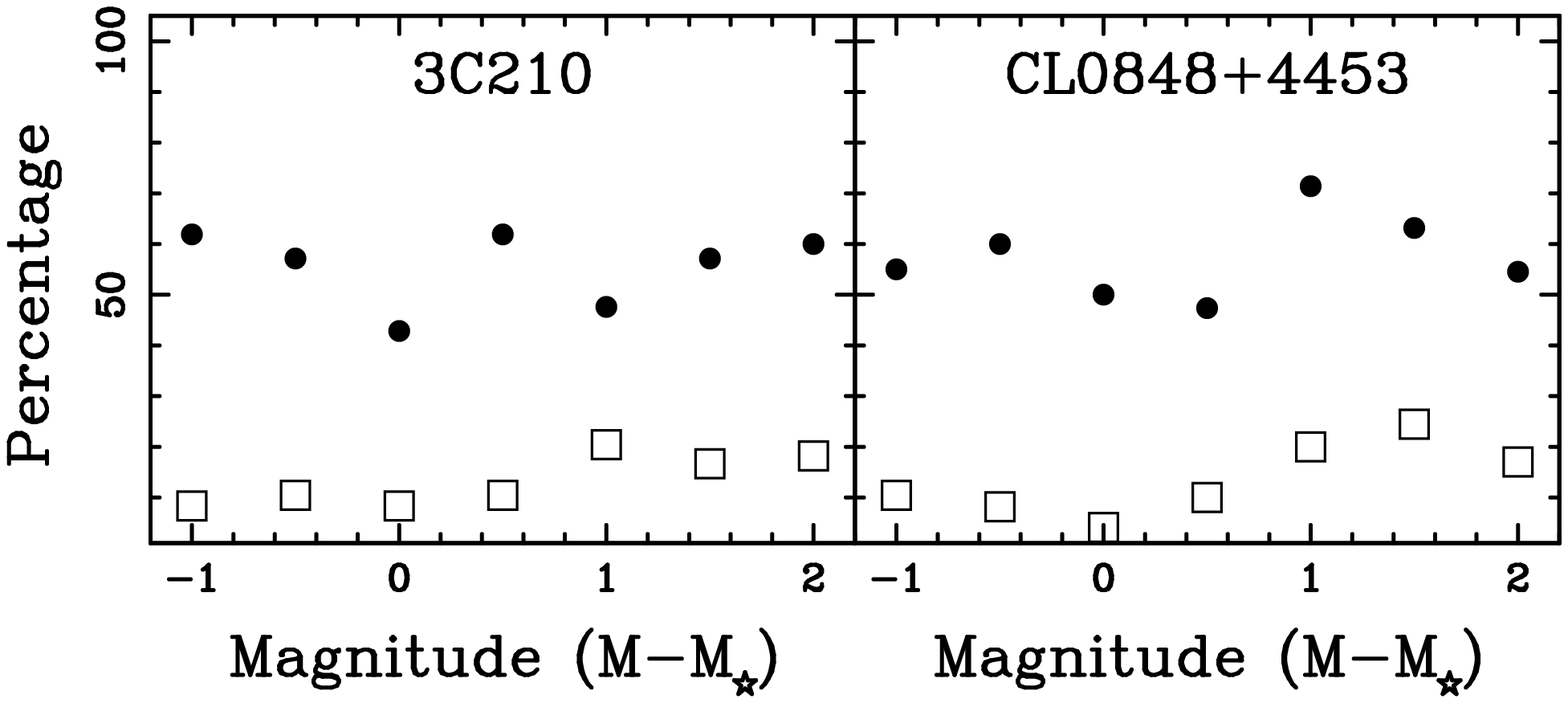}
\end{center}
\caption[holden.fig18.ps]{ We plot, for each cluster where we used
the NICMOS F160W imaging for automatic classification, the fraction of
correctly identified early--type galaxies in our simulations as solid
points.  As open squares, we plot the fraction of late-type galaxies
incorrectly identified as early--type galaxies from our simulations.}
\label{nicmos_sims}
\end{figure}

In Figure \ref{nicmos_sims} we plot the classification rate and false
positive rate for the NICMOS observations of RDCS~0848+4453 and 3C210.
We find a good rate of success for RDCS~0848+4453, roughly a $\sim$
65\% classifiction rate at most magnitudes, which is better than we found
for RDCS~0848+4453 in the F814W data.  For 3C210, we also find an
almost constant classification rate as a function of magnitude as well.
For 3C210, however, the highest classification rate is never as high as
that for the WFPC2 imaging data.  For both clusters the false positive
rate grows at fainter magnitudes, as expected.  However, we note that
the classification rate for both clusters is flat as a function of magnitude,
which we did not see in the WFPC2 simulations.  There are two reasons
for this.  First, the NICMOS data observes the rest-frame R band of
the early--type galaxies, where the light from the older stellar
populations found in bulge systems dominates.  None of the WFPC2
imaging for these two clusters sampled as red a part of the spectral
energy distribution, but rather sampled the rest frame B or U band
which is more sensitive to recent star formation.  Second, the lower
resolution of the NICMOS Camera 3 data means that, even with
dithering, the data are less sensitive to smaller bulges.  These two
effects compete, making it easy to detect bulge dominated systems in
the NICMOS imaging data, but only when the bulge size is sufficiently
large to be well resolved.  As a result of these effects, we find our
success rate for identifying pure elliptical systems is 10\% to 20\%
higher in our NICMOS simulations than our rate for S0 galaxies, where
the bulges are smaller.  We also find that our success rate for S0
galaxies falls off, as expected, with fainter magnitudes.  These
results mimic the greater success rate for pure elliptical systems
that is found in \citet{im00}.

Using the simulations of the automated identification process, as
shown in Figures \ref{wfpc2_sims} and \ref{nicmos_sims}, we only
counted galaxies with magnitudes $m(L_{\star}) + 0.6$ or brighter when
computing the early--type fraction.  For most clusters, the false
positive identification rate is around 10\% to 20\%, while the success
rate is around 70\%.  This yields a couple of galaxies per cluster
falsely identified as early--types and slightly more galaxies per
cluster missed. Therefore, our estimates of the early--type fraction
are, at most, underestimated by 20\% of the total number of galaxies.
This should not significantly bias our measures of the color
evolution of our sample.

\section{Measuring Evolution in the Color--Magnitude Relation}

\subsection{The Coma No-evolution Prediction}
\label{comapred}

We compared our high redshift cluster color--magnitude relations with
that of Coma, using the data from \citet{dePropris98} and
\citet{eisenhardt03}.  For each cluster, we transformed the Coma
photometry into the observed bands.

First, we have to take into account the different sized apertures used
for the two data sets.  The apertures used for the Coma photometry
were fixed to be 24\arcsec, or a 12\arcsec\ radius, while the
apertures used when measuring the cluster photometry were fixed
angular apertures, which are listed in \citet{stanford2002}.  We used
the color gradients of Coma from \citet{stanford98} to correct the
Coma colors to the aperture used for the high-redshift cluster
galaxies.  \citet{scodeggio2001} raised a potential bias in measuring
the color--magnitude relation.  Early--type galaxies become bluer at
larger radii so, unless measuring at the same apparent radii, we could
detect a spurious signal of evolution.  By using the color gradients
to correct all color measurements to a common physical radius, we
should compensate for this bias.

We transformed the Coma data to the observed frame of the
cluster.  We did this transformation by choosing the two Coma filters
that, when placed at the redshift of the cluster, are the best match
for the filter used for that cluster.  We then calculated the bandpass
correction, the zero-point correction and the small k-correction for
that filter to transform into the observed filter.  Below we show the
equation we fit,
\[ F_z = F_{c,1} + x*(F_{c,1}-F_{c,2}) + C \]
where $F_z$ is the filter we wish to transform the photometry to,
$F_{c,1}$ represents the filter for the Coma photometry closest, when
shifted, to the output filter, $F_{c,2}$ is the second closest filter,
$x$ represents the magnitude of the color term and $C$ represents the
small k-correction and the difference in zero-points between the two
filters.  To calculate these, we used two spectral energy
distributions, a present day early--type galaxy as represented by a 10
gigayears old model galaxy from \citet{bc96}, though based on
\citet{bruzual93}, and a slightly bluer galaxy, with colors similar to a
typical Sa galaxy, also from \citet{bc96}.  We passed these spectra
through the Coma filters and calculated the expected fluxes.  We then
red-shifted the spectra to calculate the expected flux in $F_z$ and
solved the above equation for $C$ and $x$.  Each Coma galaxy was
transformed by the above equation to have its magnitudes predicted for
the filters of the high redshift clusters.

The filters used for the cluster observations were chosen to match as
well as possible the low redshift $U$ and $B$ or $V$ filters.
Clusters observed near edges of the redshift ranges described in
\ref{grounddata}, however, may require a small extrapolation to match
the observed Coma data to the high redshift cluster observations.  The
largest extrapolation in this regard is RDCS~0848+4453 where the $I$
samples 3550 \AA , 200 \AA\ blue-ward of the central wavelength of $U$
at the redshift of the Coma cluster.  

To compute the transformed Coma galaxy colors, we use two templates
that were chosen to match the colors of either an early--type,
$m(L_{\star})$ galaxy or a red, $m(L_{\star})$ spiral, in Coma.
Nonetheless, the templates were chosen from the \citet{bc96} models.
The selection of the particular templates is a source of error in the
color transformation.  For example, for RDCS~0848+4453, we had to
extrapolate the rest-frame ultraviolet flux in order to compute the
predicted no--evolution $I-J$ color.  To estimate the size of this
error, we used the \citet{cww80} templates, specifically the
elliptical galaxy and the Sbc template.  The elliptical galaxy
template of \citet{cww80} predicts bluer colors for Coma than we
observe, possibly because it was constructed from the bulges of nearby
Spiral galaxies, rather than true elliptical galaxies..  Nonetheless,
in most cases, we find a small change, $\simeq 0.05$ magnitudes, in
predicted colors of Coma cluster galaxies when using the templates of
\citet{cww80} versus the Bruzual and Charlot model mentioned above.
This value is around the same size as the uncertainty
in the zero point of the Coma photometry, $\simeq 0.02$ in the optical
pass bands and $\simeq 0.03$ in the infrared pass bands, or $\simeq
0.04$ magnitudes in the optical - infrared colors.  These terms, along
with PSF matching errors, extinction correction errors and the
color-gradient corrections, all total an systematic error of 0.06
magnitudes, which is further described in SED95.

\subsection{High Redshift Color--Magnitude Relations}

\begin{figure}[tbp]
\begin{center}
\includegraphics[width=3.0in]{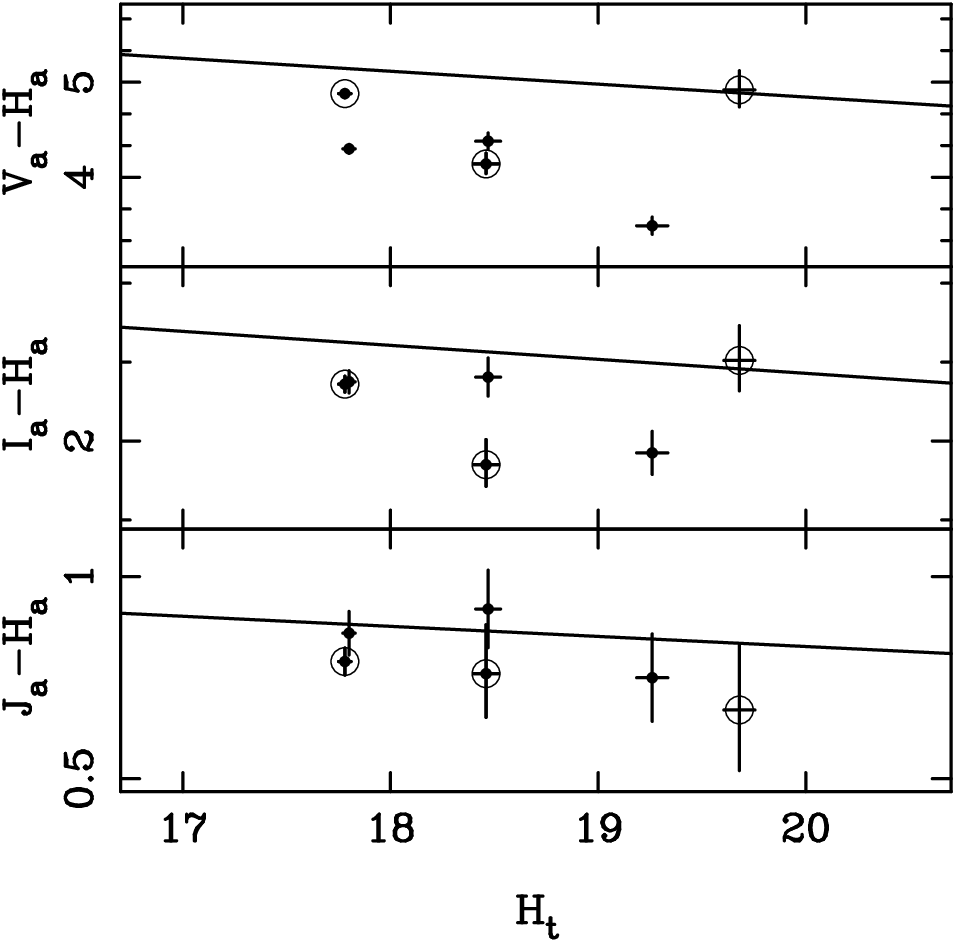}
\end{center}
\caption[holden.fig19.ps]{ Color--magnitude diagrams for the cluster
GHO~0229+0035 at z=0.607.  The colors, based on aperture magnitudes are
plotted versus the total magnitude for the early--type galaxies in
cluster.  Galaxies are represented by solid dots if visually typed or
by open circles if identified by the bulge-disk model fits.  The solid
line is the color--magnitude relation of the Coma cluster of galaxies
transformed into the observed filters.  Values with the subscript ``a''
are measured within a fixed angular aperture where all of the images
have been smoothed to the worst seeing, the aperture radius is given
in Table \ref{groundsummary}.  Values with the subscript ``t''
are the total magnitudes of the galaxies, as measured with FOCAS.  See
\ref{grounddata} for a discussion of the total magnitudes and how they
relate to ``true'' measures of the total flux from an early--type
galaxy.  }
\label{cl0231_cm}
\end{figure}

\begin{figure}[tbp]
\begin{center}
\includegraphics[width=3.0in]{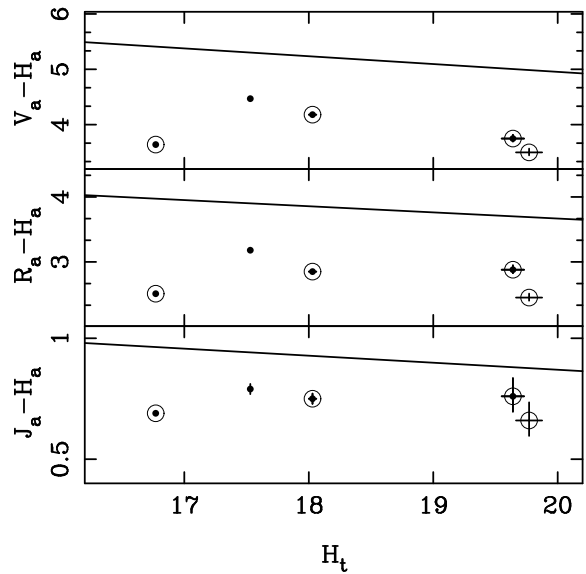}
\end{center}
\caption[holden.fig20.ps]{ Same as Figure \ref{cl0231_cm} but for
GHO~2155+0321  at z=0.7.  }
\label{cl2157_cm}
\end{figure}

\begin{figure}[tbp]
\begin{center}
\includegraphics[width=3.0in]{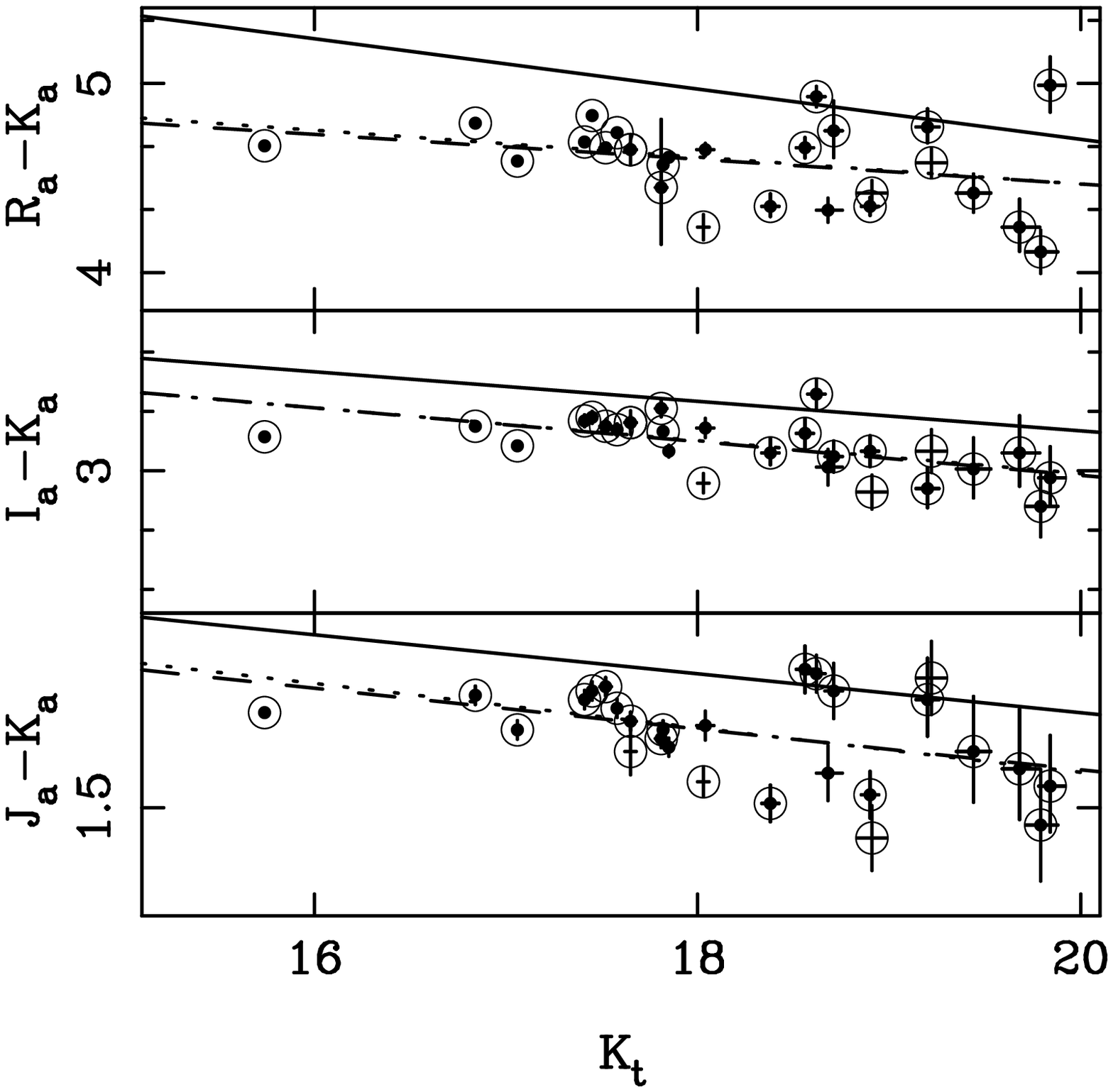}
\end{center}
\caption[holden.fig21.ps]{ 
Same as Figure \ref{cl0231_cm} but for MS~1137+6625 at z=0.782 with the
addition of a dotted line which is the fit to the visually
identified early--type galaxies and a dashed line which is the fit to
the automatically identified early--type galaxies.}
\label{ms1137_cm}
\end{figure}

\begin{figure}[tbp]
\begin{center}
\includegraphics[width=3.0in]{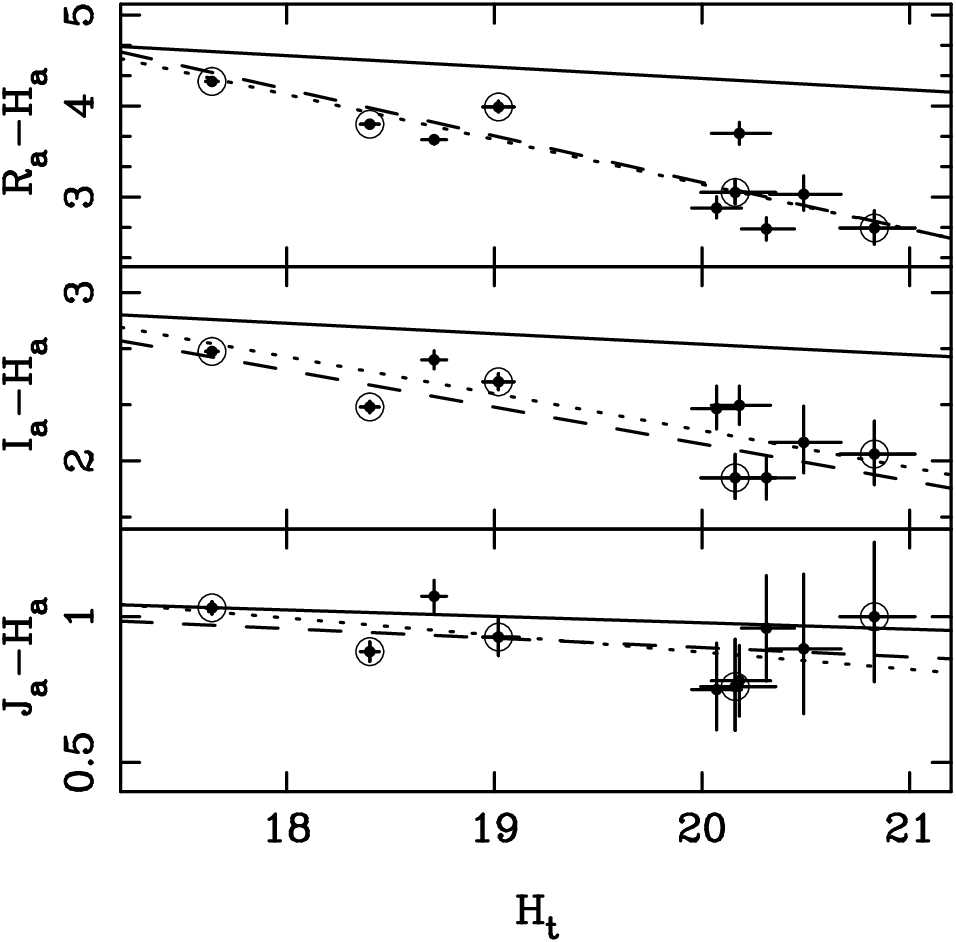}
\end{center}

\caption[holden.fig22.ps]{ 
Same as Figure \ref{cl0231_cm} but for GHO~0021+0406 at z=0.832.}
\label{cl0023_cm}
\end{figure}

\begin{figure}[tbp]
\begin{center}
\includegraphics[width=3.0in]{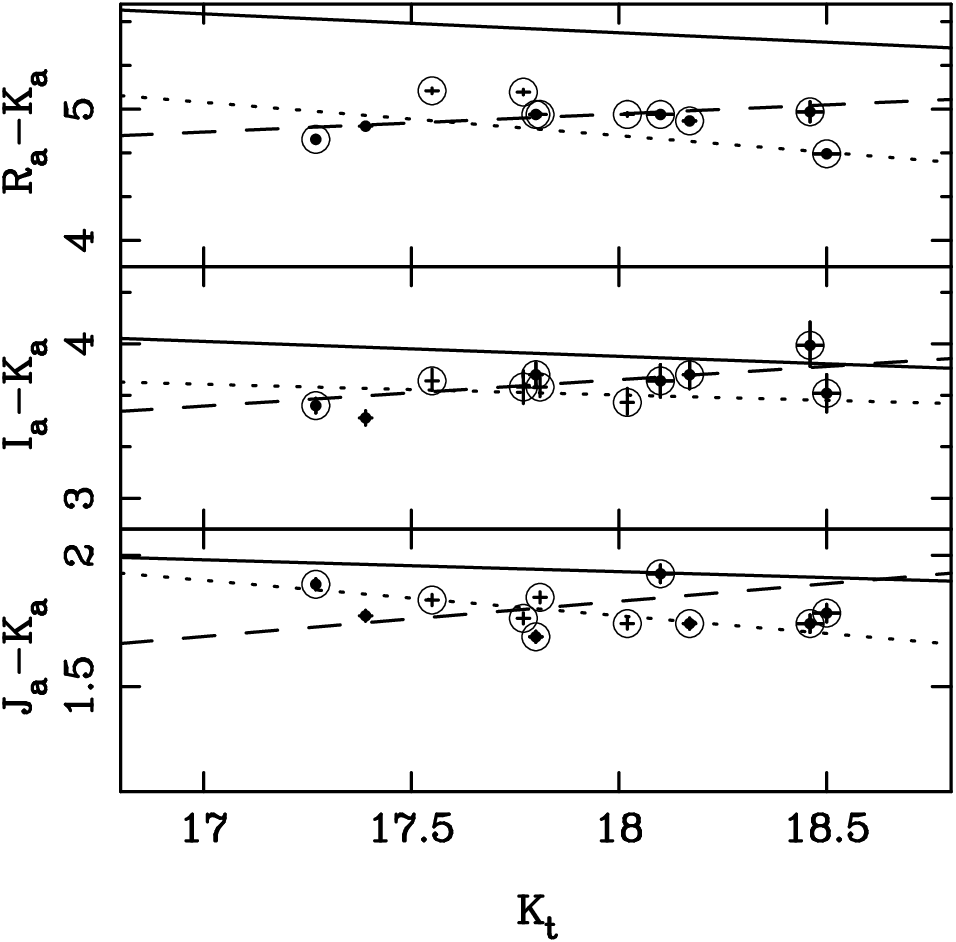}
\end{center}

\caption[holden.fig23.ps]{ 
Same as Figure \ref{cl0231_cm} but for GHO~1604+4329 at z=0.920.
}
\label{cl1604_cm}
\end{figure}

\begin{figure}[tbp]
\begin{center}
\includegraphics[width=3.0in]{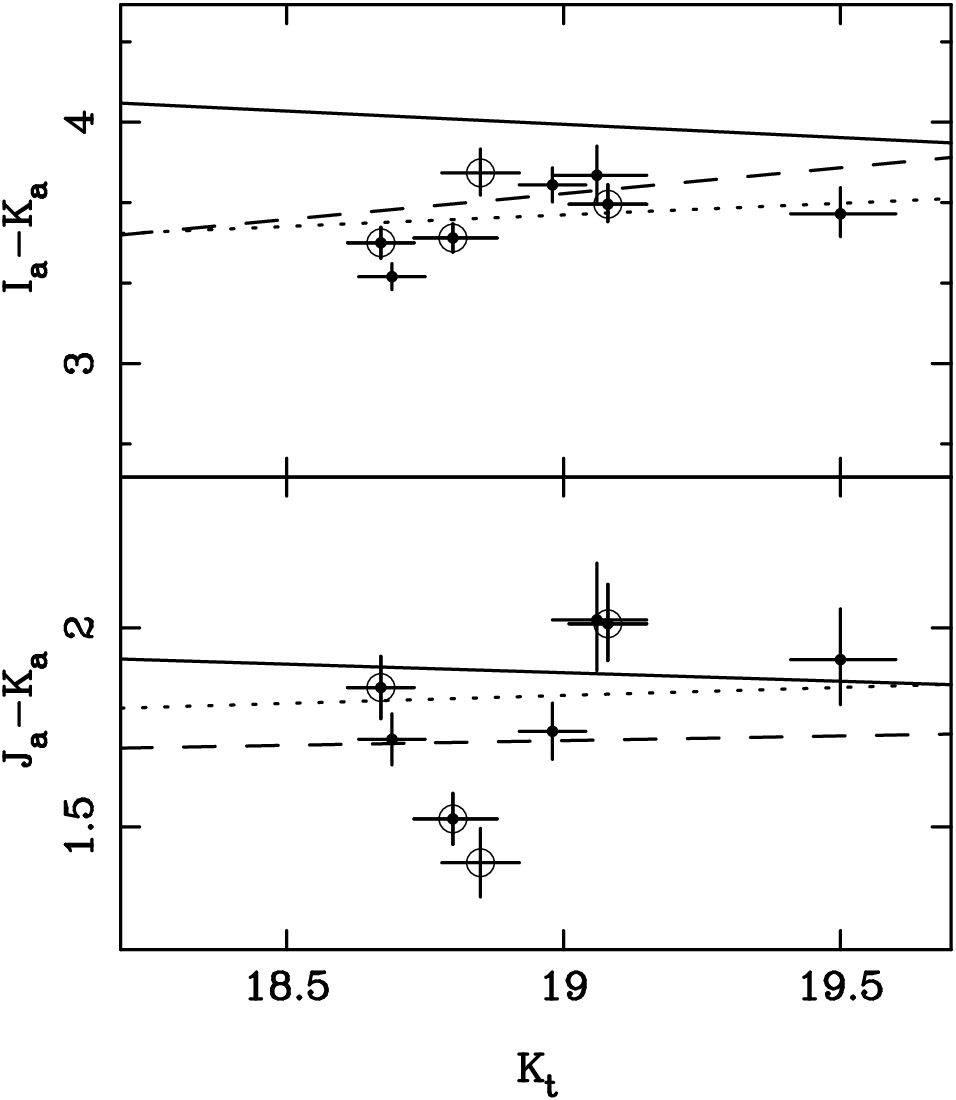}
\end{center}
\caption[holden.fig24.ps]{ 
Same as Figure \ref{cl0231_cm} but for 3C184 at z=0.996.}
\label{3c184_cm}
\end{figure}

\begin{figure}[tbp]
\begin{center}
\includegraphics[width=3.0in]{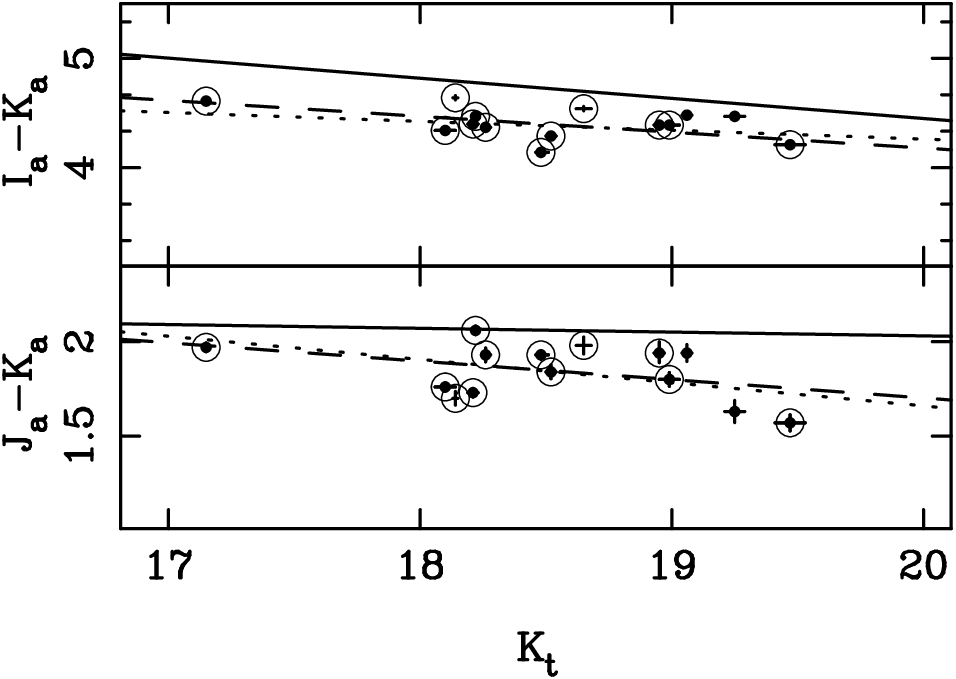}
\end{center}

\caption[holden.fig25.ps]{ 
Same as Figure \ref{cl0231_cm} but for 3C210 at z=1.169.
The open circles represent galaxies identified as early--types
using GIM2D in the WFPC2 F814W data.}
\label{3c210_cm}
\end{figure}

We plot in Figures \ref{cl0231_cm} through \ref{cl0848_nic_cm} the
color--magnitude relations for the early--type galaxies in the
clusters in our sample.  For each cluster, we plot three
color--magnitude relations, the two straddling the 4000 \AA\ break and
$J-H$ or $J-K$ relation.  We compare these colors with the total
magnitudes measured in either the $H$ or $K$ bands.  For each diagram,
we plot two sets of points.  The open circles represent early--type
galaxies as identified using GIM2D (see \S \ref{class}) while the
solid points are those visually identified (see \S \ref{class}).  We
also plot three lines for each diagram.  The solid line corresponds to
the transformed no-evolution prediction from the Coma data discussed
above.  The dashed line represents the fit to the color--magnitude
relation for the galaxies identified by GIM2D while the dotted
corresponds to the fit for the visually identified galaxies.  In two
cases, GHO 0229+0035 shown in Figure \ref{cl0231_cm} and GHO 2155+0321
shown in Figure \ref{cl2157_cm}, there were only a few early--type
galaxies identified by GIM2D or by visual inspection, so we performed
no linear fits.  After comparing the number of observed early--type
galaxies to the number of predicted early--types, see Table
\ref{numsummary}, it is entirely consistent that there is no cluster
of early--type galaxies in either case, but rather we are simply
seeing the field ellipticals population.  For the remainder of this
paper, we will not use the results from these two clusters.  Figure
\ref{cl0848_wfpc_cm} shows the GIM2D classifications for
RDCS~0848+4453 based on the WFPC2 F814W data.  These data sample the
rest frame ultra-violet at the redshift of this cluster and for 3C210.
This part of the spectrum does not sample well the older stellar
populations that dominate bulges of galaxies and, therefore, the
simulations in \S \ref{class_sim} show a low rate of correct
morphological classification.  For the rest of this paper we only use
the morphological classifications for RDCS~0848+4453 based on the
NICMOS data presented in Figure \ref{cl0848_nic_cm}.  For 3C210, we
will use both the WFPC2 and the NICMOS data because of the short
exposure times in some of the NICMOS imaging, see \S
\ref{nic_imaging}.

\begin{figure}[tbp]
\begin{center}
\includegraphics[width=3.0in]{holden.fig26.ps}
\end{center}
\caption[holden.fig26.ps]{ 
Same as Figure \ref{cl0231_cm} but for 3C210 at z=1.169.
The open circles represent galaxies identified as early--types
using GIM2D in the NICMOS F160W data.}
\label{3c210_nic_cm}
\end{figure}

\begin{figure}[tbp]
\begin{center}
\includegraphics[width=3.0in]{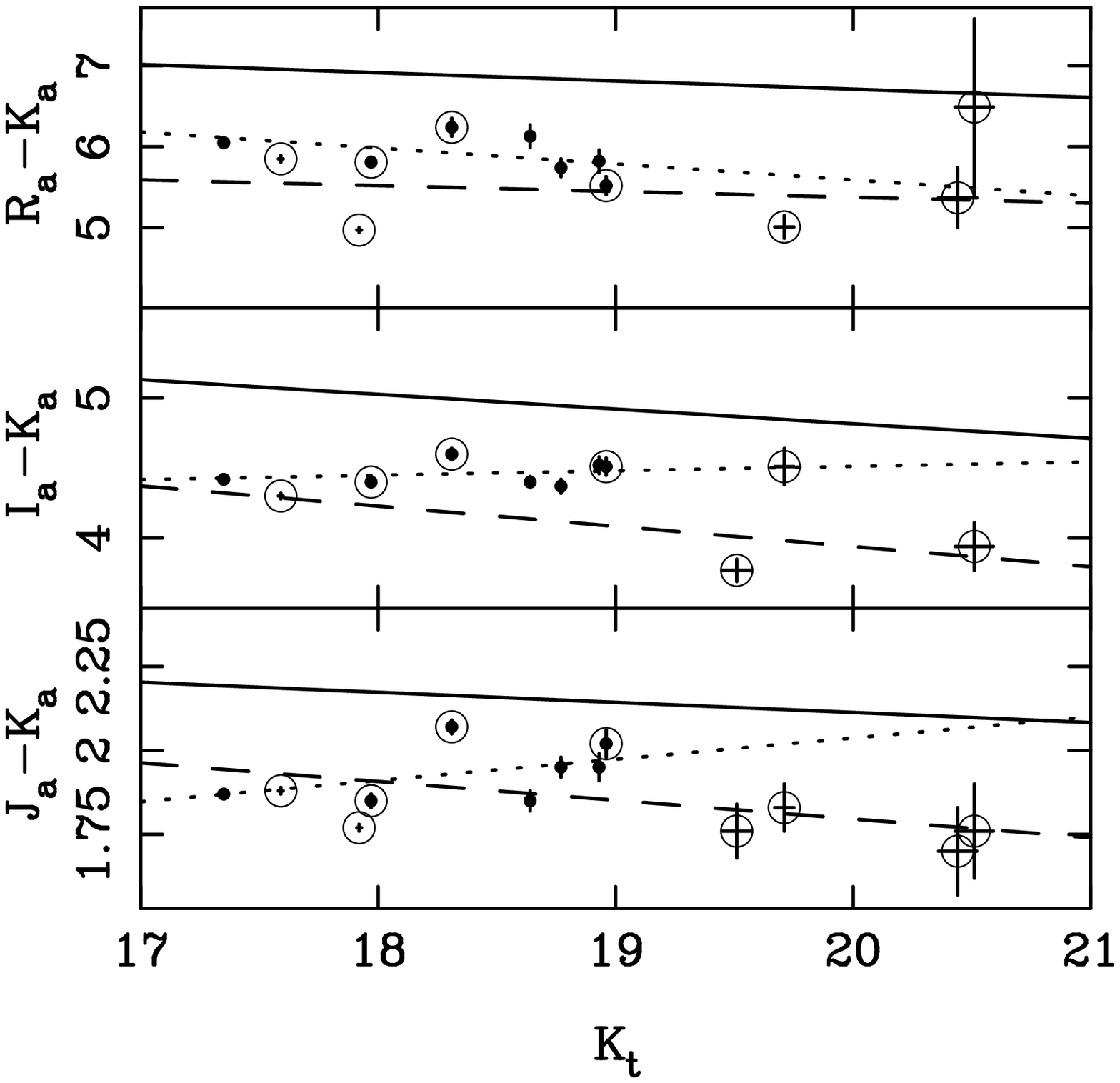}
\end{center}

\caption[holden.fig27.ps]{ 
Same as Figure \ref{cl0231_cm} but for RDCS~0848+4453 at z=1.27.
The open circles represent galaxies identified as early--types 
using GIM2D in the WFPC2 F814W data.}
\label{cl0848_wfpc_cm}
\end{figure}

\begin{figure}[tbp]
\begin{center}
\includegraphics[width=3.0in]{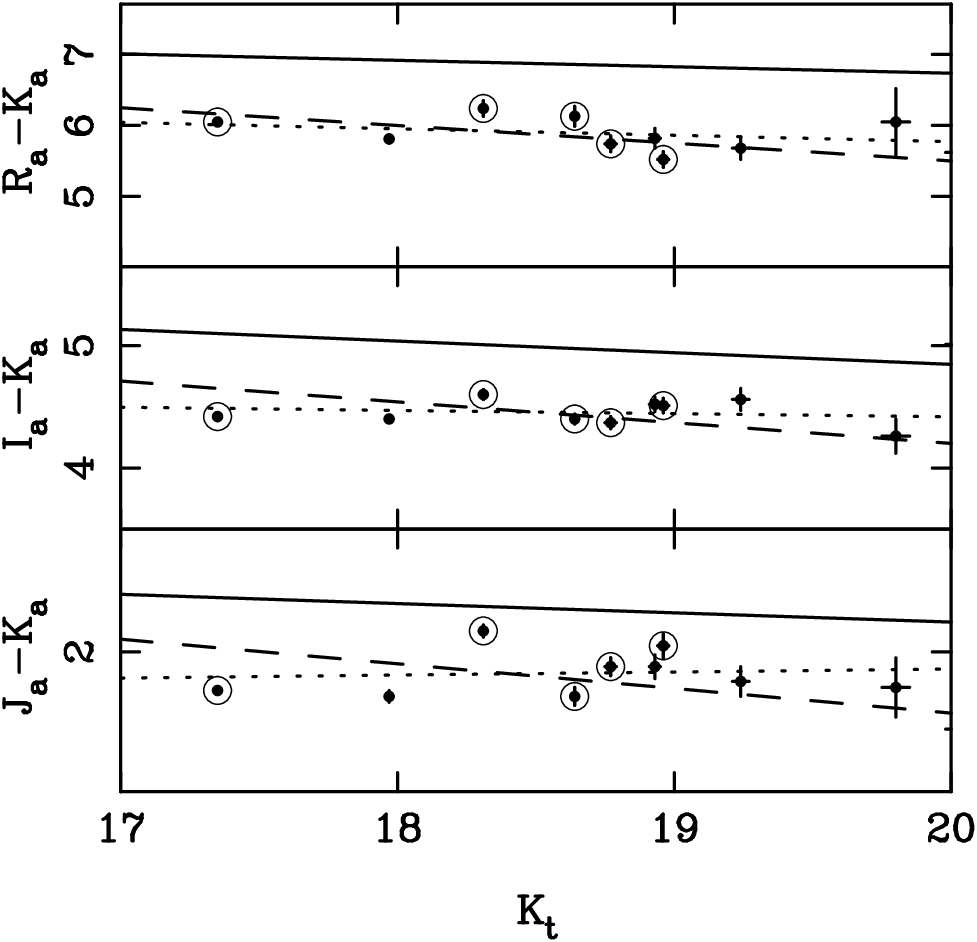}
\end{center}
\caption[holden.fig28.ps]{ 
Same as Figure \ref{cl0231_cm} but for RDCS~0848+4453 at z=1.27.
The open circles represent galaxies identified as early--types 
using GIM2D in the NICMOS F160W data.}
\label{cl0848_nic_cm}
\end{figure}

When fitting the color--magnitude relation, we first restricted the
range of total magnitudes to only those brighter than
$m(L_{\star})+2$; see Table \ref{groundsummary} for the values of
$m(L_{\star})$ we used.  We then calculated the biweight location and
scale of the colors \citep{beers90}.  If a galaxy was more than {\em
two} scale lengths, two standard deviations for a Gaussian
distribution, from either side of the peak in one of the
color--magnitude relations, it was removed from the early--type
sample.  In addition, we removed all galaxies more than half a
magnitude from the biweight peak $J-K$ or $J-H$ color and more than
one $\sigma$ redder than the expected no--evolution result.  All
galaxies used in the fit are shown in the plotted color--magnitude
diagrams in Figures \ref{ms1137_cm} through Figure
\ref{cl0848_nic_cm}.

To fit the color--magnitude relation, we used the bivariate correlated
errors and intrinsic scatter (BCES) estimator from \citet{akritas96}
as implemented in the software provided by the authors.  Specifically,
we followed the recommendation of \citet{akritas96} for fitting the
color--magnitude relation and used the BCES($X_2|X_1$) estimator.
This technique requires an estimate of the error for both the color
and magnitude as well as the covariance between the color and
magnitude.  We estimated the errors on our magnitudes by a series of
simulations described in SED98.  As the color term depends on the
magnitude, there is a strong covariance.  We simply used the errors in
the aperture magnitudes for whichever magnitude was a total magnitude,
$K$ or $H$, as our estimate of the covariance as was done in
\citet{akritas96}.  In Figure \ref{slopes}, we plot the difference in
the slopes, specifically we subtract the Coma slope from the observed
slope, such that a flatter slope will have a positive value in the
figure.

As with SED98, we will also probe the evolution in the stellar
populations of early--type galaxies using the offset of the
color--magnitude relation from the Coma no-evolution prediction.  In
Figure \ref{delta}, we plot the offset of
the color--magnitude relation for the early--type galaxies in the
clusters as a function of redshift.  This offset we calculated by
computing the y-intercept that best fit the data assuming the slope of
the Coma no-evolution relation.  We note here that, unlike in SED98,
for GHO~1603+4313 we chose the $I$ filter to represent the ``blue''
and the $J$ to represent the ``red'' for these figures only.  This is
necessary for comparing with the models described below.

Finally, we calculate the scatter in the colors around the
color--magnitude relation.  For our estimate of the scatter, we use
the biweight scale of the early--type galaxies after we have
removed the outliers.  We calculated the scatter in the colors from
the residuals in the color--magnitude relation after subtracting the
predicted Coma slope and the offset from the Coma no-evolution
relation we estimated above.  However, we change the magnitude limits,
as was done in SED98, to $m(L_{\star})$.  We plot the biweight scatter
in Figure \ref{scatter} as was done in
SED98.  In Figure \ref{br_scatter} ,
we plot the biweight scatter around the U-V color--magnitude relation.
Strictly speaking, our {\it blue} - {\it red} color is not a true
$U-V$, but is created to be as close as possible.  For comparison, we
plot the Coma result from \citet{bower92b}.  Overall, we reaffirm the
results of SED98.  First, the scatter in the color--magnitude relation
has a wavelength dependence, with a larger scatter as we compare bluer
bands with the near-infrared band.  The second result is that we measure
no change in the scatter with redshift.

\section{Evolution in Early--Type Galaxy Populations}

In this paper we have extended the range in redshift of the sample of
SED98 to $z \simeq 1.3$ and we have doubled the number of clusters of
galaxies at $z>0.8$.  Our goals, by extending the range in redshift
and number, are to examine the slope and the scatter in the
color--magnitude relation to test the ideas of elliptical galaxy
formation.  By looking at the colors of early--type galaxies, we can
only examine the history of the stars, and not the mass assembly.
Nonetheless, these photometric properties give us a measure of the
average stellar types in the early--type galaxies.

\subsection{The Slope of the Color--Magnitude Relation}

\begin{figure}[tbp]
\begin{center}
\includegraphics[width=3.0in]{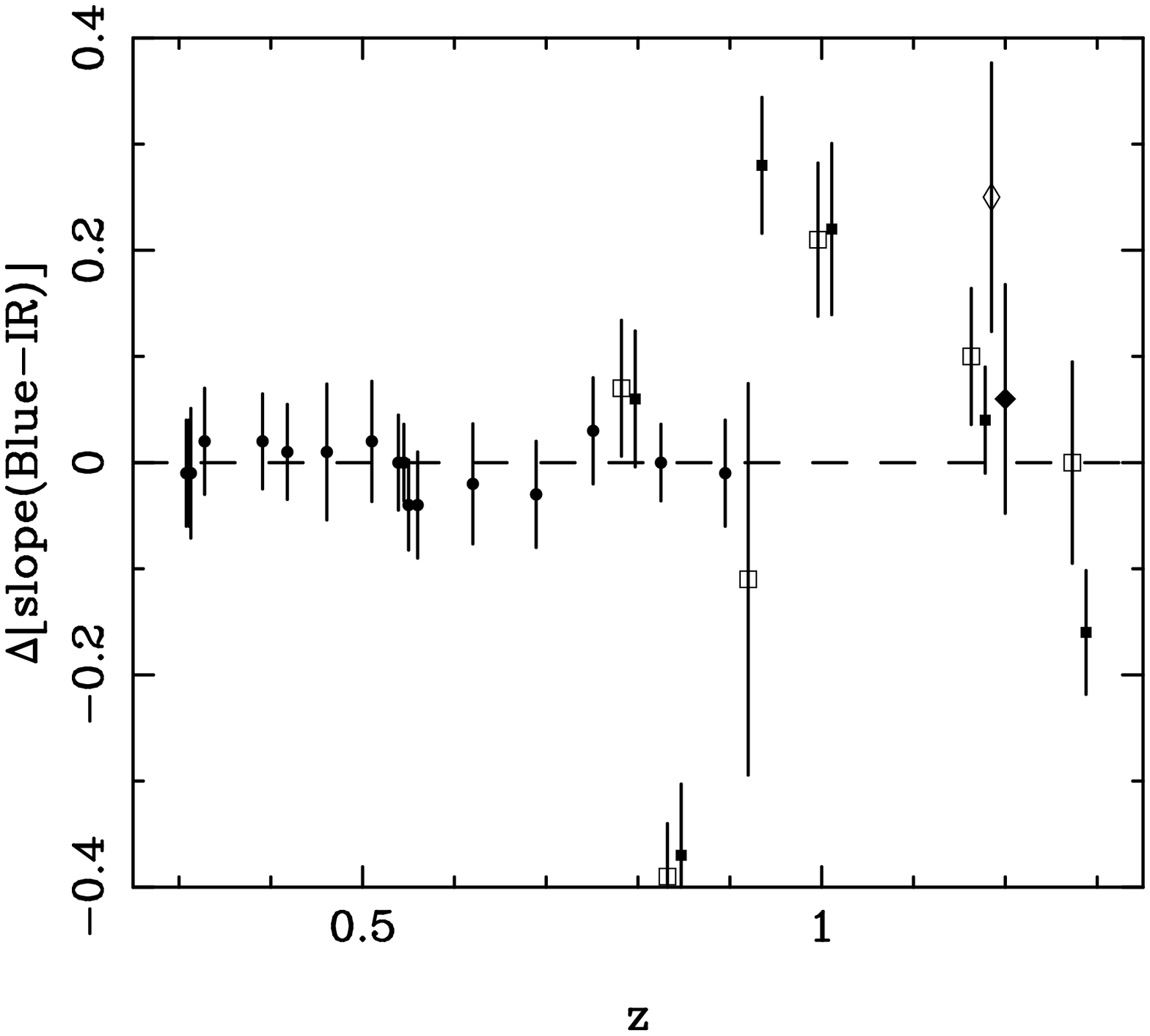}
\end{center}

\caption[holden.fig29.ps]{ The slope of the measured color--magnitude
relation minus the slope from the non-evolving Coma relation as a
function of redshift for early--type galaxies identified by visual
inspection. The solid dots are from SED98. The squares are from this
paper with the open squares representing the values from visually
classified early--type galaxies and the solid squares representing the
results of the automated classification scheme.  The solid squares are
offset from the cluster redshift for clarity. The diamonds represent
the NICMOS results for 3C210 and both the NICMOS and WFPC2 results
have been offset slight in redshift for clarity with open diamond the
visual classification result and the solid the automated
classification results.
}
\label{slopes}
\end{figure}

The slope in the color--magnitude relation is predicted to vary with
redshift.  In \citet{kauffmann98}, the authors predict a slow change
in the slope of the color--magnitude relation, with the relation
becoming flat at $z\simeq1.5$ for cluster early--type galaxies formed
by merging smaller mass systems.  In contrast, \citet{kodama97}
predict, for a purely passively evolving stellar population formed at
high redshift, that the color--magnitude relation changes only
slightly over a wide range in redshift, until very close to the epoch
of the formation of the stars in the galaxies.  Thus, we are searching
for a subtle but important effect: at what redshift does the slope of
the color--magnitude relation change?  In Figure \ref{slopes}, we find
no evidence of a change.  We plot the difference between the slope we
measure and the slope we measure for the transformed Coma photometry
we use as a no--evolution prediction.  This is the same plot as Figure
4 in SED98.  Shallower slopes, and therefore, flatter color--magnitude
relations, appear as positive values in Figure \ref{slopes}.

For some of the clusters, we find a small positive slope difference,
implying a general flattening of the color--magnitude relation.  In
general, none of these results are statistically significant.
GHO~0021+0406 shows a strong negative value, and therefore a much
steeper slope than expected from the no--evolution result, see Figure
\ref{cl0023_cm}.  We suspect, but cannot prove, that this is from
field galaxy contamination as GHO~0021+0406 has a small number of
early--type galaxies to begin with.  RDCS~0848+4453 was reported to
have a flatter than expected slope in \citet{vandokkum2001}, a result
we reproduce in Figure \ref{slopes} for the early--type
classification measurement but not when using galaxies identified with
GIM2d.  In fact, object selection appears to make a significant
difference in the measured slopes, as seen by the difference in the
slope measurements in Figure \ref{slopes} for clusters at the same
redshift.  Such sensitivity implies that, in general, our samples are
too small to make reliable slope measurements.  We note here that most
of the clusters in our sample have on the order of ten or less
early--types in the red sequence, while the data in SED98 often had
more than twice that number of early--type galaxies.  We tried two
simple tests to see if any evolution in the slopes existed, a linear
fit to the slopes as a function of redshift and power-law evolution as
a function of $1+z$.  Neither showed any statistically significant
change with redshift.  Therefore, we conclude that we cannot find any
evidence of evolution in the slope of the color--magnitude relation
but, for us to see such evolution, the trend would have to be much bigger
than that predicted in, e.g., \citet{kauffmann98}.

\subsection{The Average Color of Early--Type Galaxies}

As we have found no measurable evolution in the slope of the
color--magnitude relation, we looked for evolution in the offset and
in the scatter.  The zero-point of the color--magnitude relation must
change with time as the stellar populations that make up the cluster
early--type galaxies evolve.  By measuring the offset as a function of
redshift, and assuming a cosmology, we can estimate the average epoch
of star formation for the cluster members.  This is most easily done
by comparing models to no--evolution predictions.  

\begin{figure}[tbp]
\begin{center}
\includegraphics[width=3.0in]{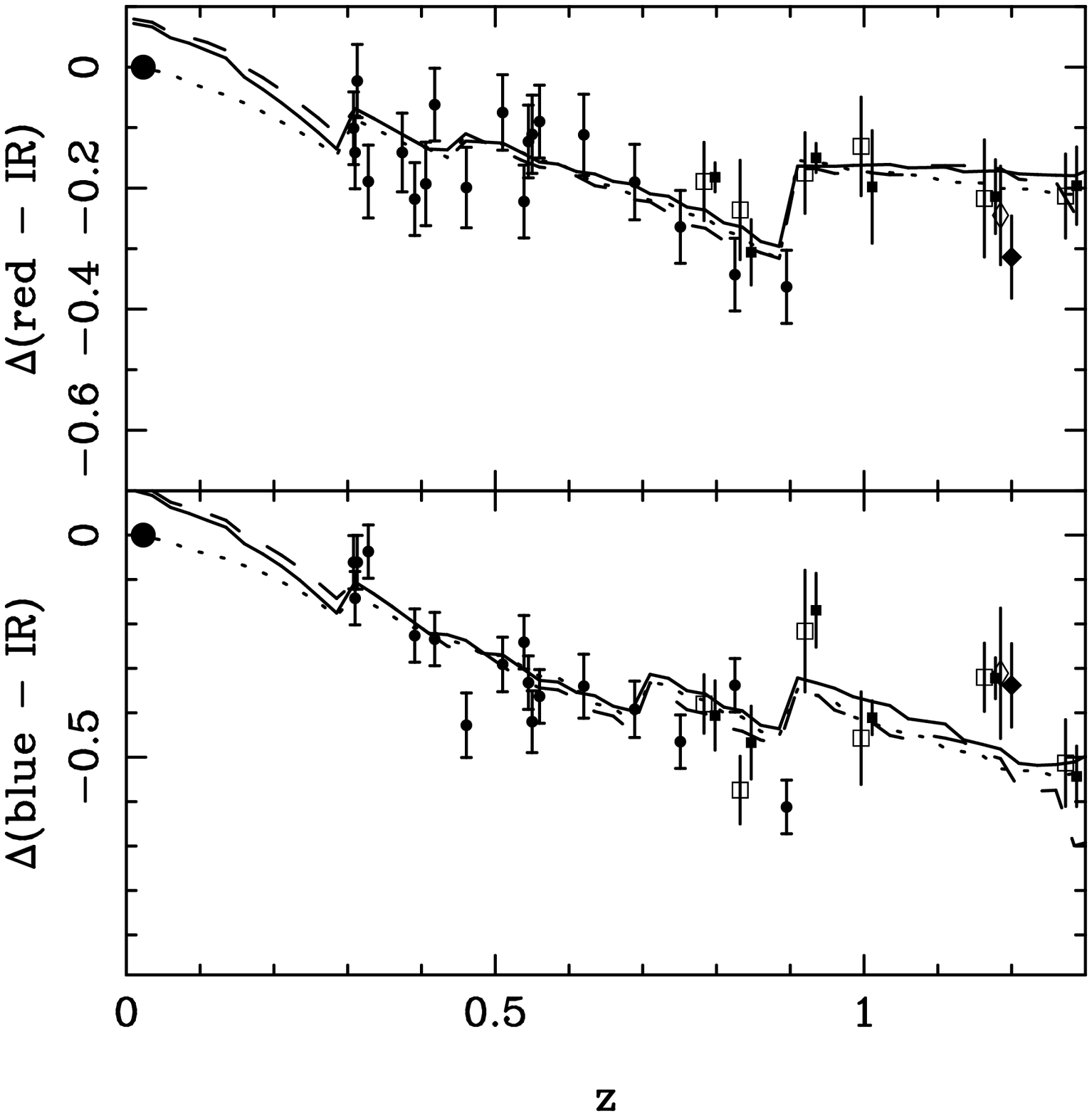}
\end{center}

\caption[holden.fig30.ps]{ The offset from the Coma no-evolution
color--magnitude relation as a function of redshift for early--type
galaxies identified by visual inspection (open squares) and GIM2D
(solid squares). The solid dots are from SED98 and the large solid
being the value for Coma, fixed to an offset of zero.  The diamonds
represent the NICMOS results for 3C210 with the open diamond the
visually identified early--type galaxies and the solid diamond the
GIM2D identified early--type galaxies.  The solid lines are
\citet{bc96} models for a single stellar population formed with a
solar metallicity at a redshift of 5 in an $\Omega= 0.1$ cosmology.
The best fitting model using a $\Omega_{m}= 0.3$,
$\Omega_{\Lambda}=0.7$ cosmology with the \citet{bc96} models and
formation of $z_f = 3$ is the dashed line, while the best fitting
\citet{bc03} model, with $z_f = 5$, is the dotted line.  All models
assume ${\rm H_{o} = 65\ km\ s^{-1}\ Mpc^{-1}}$ except for the
\citet{bc03} model which assumes ${\rm H_{o} = 71\ km\ s^{-1}\
Mpc^{-1}}$, $\Omega_{m}= 0.27$ and$\Omega_{\Lambda}=0.73$ }
\label{delta}
\end{figure}

In Figure \ref{delta}, we plot the predicted color evolution of
stellar population synthesis models along with the offsets from the
no-evolution Coma color--magnitude relation.  As discussed above,
these color offsets are measured with respect to a transformed Coma
color--magnitude relation.  We calculate the theoretical model offsets
in two different manners.  First, we use the model of
\citet{dePropris99} to describe the colors of an unevolving elliptical
galaxy, the same model we use when we compute the relative
k-corrections in \S \ref{comapred}.  This is a 10 gigayears old single
stellar population with solar metallicity and a Saltpeter mass
function where the stars were formed in a burst of 0.1 gigayears
calculated with the 1996 update of \citet{bruzual93}.  We
subtracted the predicted color of the no evolution model, in the
``blue'', ``red'' and ``IR'' pass band at the redshift for each
cluster, from the colors of the evolving models in the same filters at
the same redshift.  We will refer to this as an absolute measurement.
For our second approach, we simply subtract the expected colors of a
given model at the redshift of Coma from the expected colors at the
redshift of the cluster, which we will call the relative color
measure.  We try this second approach to test which model best
produces the relative change in color as a function of redshift, even
if the predicted colors of the model are incorrect.  In both cases, we
assume that all of the stellar populations making up the early--type
galaxies are approximately coeval, that metallicity does not evolve
with redshift and that morphological transformations have a minimal
effect.  We will discuss some of these assumptions further, below.

\subsubsection{Older Models and Previous Work}

In Figures \ref{delta}, we plot our data, the offsets, along with the
absolute model offsets.  First, with a solid line, we plot the best
model from SED98 and from \citet{nelson2001}, a single starburst with
a formation redshift, $z_f = 5$, and ${\rm H_o = 65\ km\ s^{-1}\
Mpc^{-1}}$, and $\Omega_m = 0.1$.  That model fits the high redshift
cluster colors rather well, but predicts colors far too red for the
Coma color--magnitude relation.  Updating the cosmology to
$\Omega_{\Lambda}= 0.7$ and $\Omega_m = 0.3$ does change the formation
epoch.  Using a $\chi^{2}$ test, we find the best fitting formation
epoch from the \citet{bc96} models to be $z_f = 3.0^{+1.8}_{-0.5}$ or
that the universe was $2.1^{+0.5}_{-1.0}$ gigayears old when the stars
in the galaxies formed.  This best fitting model is plotted in Figure
\ref{delta} as a dashed line.  We computed
the $\chi^{2}$ using the difference between the model and both the
``blue-IR'' and ``red-IR'' colors.  We used errors on the color offset
in the divisor but added to that value 0.06 magnitudes of systematic
error as described in \S \ref{comapred}.  The $\chi^{2} = 1.2 - 1.3$
per degree of freedom for these fits are somewhat high, the actual value
depending on the classification scheme used for the highest redshift
clusters.  However, the $\chi^{2}$ is not so high that we can reject
the best fitting model at a high level of significance.

We note here that the $z_f = 3$ model corresponds to the best fitting
formation epoch using the evolution of the mass to light ratio from
\citet{vandokkum2003}. In addition, it is a model in good agreement
with the evolution of the K-band luminosity function of
\citet{dePropris99}.  The work in \citet{dePropris99} uses the
\citet{bc96} stellar population synthesis models as well as the same
photometry as this paper, so we have achieved internal consistency for
this model comparison.  

As found in SED98, however, the colors of the \citet{bc96} model do
not match the average color of the Coma cluster members.  We find that
the best fitting model of SED98 is too red in the ``blue-IR'' color by
0.09 magnitudes and too red in the ``red-IR'' by 0.07 magnitudes at
the redshift of Coma.  The best fitting \citet{bc96} model for the
updated cosmology is too red by 0.11 and 0.08 magnitudes in the
``blue-IR'' and ``red-IR'' colors respectively.  Because of this color
mismatch, we also try to compute the relative color evolution of the
\citet{bc96} model.  We calculate how much the color should have
changed at a given redshift as compared with the $z=0.023$ color.
This mimics our actual measurement, where we compare the change in the
observed colors with respect to the Coma relation.  Here we find no
particular formation epoch to be the best fitting.  Rather, the
$\chi^{2}$ is flat for very old stellar populations, ones that formed
when the universe is 1 gigayear or less in age.  This relative
evolution test yields different results than the range of allowed
formation epochs from the absolute evolution discussed in the previous
paragraph.  This is because we are not penalizing the models for
getting the actual colors wrong, but just seeing which models get the
relative change as a function redshift correct.  Despite releasing
this additional constraint, the model of \citet{bc96} still has a
$\chi^{2} = 1.3$ per degree of freedom, and $\chi^{2} = 1.5$ for the
``blue-IR'' color, showing that the model does not predict correctly the
near ultraviolet color evolution of early--type galaxies.

\subsubsection{Bruzual and Charlot 2003 Models}

We examined the evolution of the average colors of early--type
galaxies in our sample, as well as that in SED98, using the newer
\citet{bc03} models.  Specifically, we use the Padova 1994
evolutionary tracks, as recommended, with Salpeter initial mass
function \citep{salpeter1955}.  These models produce higher formation
ages than the \citet{bc96}.  For example, we find $z_f > 3.0$, or a
maximum age of the universe at the time of formation of 2.9 gigayears,
when we use the \citet{bc03} models with solar metallicity and the
cosmological values from the Wilkinson Microwave Anisotropy Probe
\citep{bennett2003} of $\Omega_{\Lambda}= 0.73$, $\Omega_m = 0.27$ and
${\rm H_o = 71\ km\ s^{-1}\ Mpc^{-1}}$.  The distribution of
$\chi^{2}$ values stays flat to high redshifts as a function of
formation age, never crossing the 1$\sigma$ threshold. There is a
shallow minimum at $z_f = 5_{-2.7}$ or an age of $1.6^{+1.3}$
gigayears, but there is no formal upper limit.  For this comparison, we only
use the relative test mentioned above, where we compare the change of
color of the model prediction at a given clusters redshift with
respect to the predictions of the same model for the Coma cluster.
Given that the \citet{bc03} models do a much better job of reproducing
the colors of the Coma cluster, this is effectively the same as the
test where we compare with a model chosen to reproduce the colors of
Coma early--type galaxies.  The relative color evolution therefore
only rules out at high significance very recent, $z_f < 1.5$,
formation epochs, a result that is true of the \citet{bc03} models
because for those models the rate of change in the populations is too
fast to match the observed rate of change in the colors for the
highest redshift clusters.

Because of the degeneracy between age and metallicity \citep[for
example]{worthey1994}, we also compared the change in colors as a
function of redshift for both the super-solar (Z=0.05) and the Z=0.008
sub-solar metallicity populations of \citet{bc03}.  The super-solar
yielded very similar results, namely that $z_f =5_{-2}$, though the
$\chi^{2}$ = 1.77 per degree of freedom is high enough to reject the
high metallicity models at the 99.7\%, or three standard deviations,
confidence level.  Most of the mismatch occurs when comparing the
``blue-IR'' color offsets. The sub-solar models yielded a minimum
$\chi^{2}$ = 15.3 and, thus are not actually worth discussing.  We
note here, that we still assume a uniform metallicity for the
population of early--type galaxies.  Also, in \citet{bc03}, the
authors mention in Table A1 that the UV and near-infrared colors of
their models for super-solar metallicity are ``Fair/poor'', which are
exactly the colors which we are comparing against.  Thus the high
apparent significance above could be a reflection of the model's
current uncertainties.  

\subsubsection{Sources of Uncertainty}

It is important for us to note that all of these predictions are based
on one set of models, those produced by Bruzual and Charlot, with
specific underlying assumptions; a 100 megayears duration starburst,
the metal abundance of the stellar population, a Salpeter initial mass
function, a flat star formation history and that all of the galaxies
in our samples are coeval.  As shown in, for example,
\citet{vandokkum1998}, changing each of these assumptions can change
the formation epoch implied.  In \citet{trager2000}, for example, the
authors present multiple pieces of evidence that the stellar
populations of cluster elliptical galaxies are complicated, with over
a factor of two spread in ages and metallicities.  Some of this is
mitigated by the fact that we are only measuring the pace of the
evolution and not trying to actually match the colors of the galaxies
with the models directly, hence the mild independence of our results
despite a change in the metallicity of the population.  Nonetheless,
the spectroscopic techniques of \citet{trager2000} are far more
sensitive to recent bursts of star formation whereas our colors are
only measures of the overall average evolution.  A complete picture
will require a combination of the two approaches to attempt to
interpret the history of star formation in early--type galaxies.

In general, the differences between the \citet{bc96} and \citet{bc03}
models lie in the treatment of the giant branches and in how the
spectral libraries of constituent stars are assembled.  One of the key
improvements is that while the \citet{bc96} used mostly theoretical models
from R. L. Kurucz, \citet{bc03} added empirically more complete 
stellar models, affecting post main-sequence
evolution.  The changes in the giant branch stars, which dominate the
observed light in the redder bands of early--type galaxies, and the
improved stellar libraries, once again for late-type stars which
dominate early--type galaxies, explain how the difference between the
best fitting models comes about; see \citet{liu2000} for a more
detailed discussion.  Given that the colors of early--type
galaxies in the Coma cluster are much more closely matched using 
\citet{bc03}, it appears that the later models are a significant
improvement over \citet{bc96}.  Nonetheless, the large change
between the two sets of models, in the end, suggests that we only take the most
conservative results away from this analysis.  Namely, that the
average color early--type galaxies evolves as expected for a passively
evolving stellar with an average formation epoch definitely with $z_f
> 1.5$ and likely with $z_f > 2$, for our adopted cosmology.
Interestingly, our results are in excellent agreement with the results
of \citet{kelson2001} who use different models along with very
different data to yield the same results.  Our formation epoch
estimates are, of course, only an average formation epoch.  The
allowed range of values, and the lower limit, can only be addressed
using the scatter of the colors.

\subsection{The Scatter in the Color--Magnitude Relation}

Our final test for evolution, shown in Figure \ref{br_scatter}, is the
intrinsic scatter around the mean color--magnitude relation.  This
test was first proposed in \citet{bower92b}.  The power of this test
is that it is partially model independent.  Regardless of when the stars
formed, they could not have all formed at the exact same time.  As the
stellar populations making up the cluster galaxies become older, the
scatter in the observed colors becomes less.  This relies only on the
fact that the time scales for the evolution of a stellar population
become longer as the population as a whole becomes older.  Using the
models from \citet{bc96}, we examined what spread in ages would be
consistent with the observed spread in colors.  For RDCS~0848+4453,
for example, the observed biweight scale, or scatter, in the {\it
blue} - {\it red} color is $0.06 \pm 0.03$, using the visual
classifications or $0.04 \pm 0.02$ from the automated classifications.
Using stellar populations models, all with solar metallicity, such a
spread in color can be reproduced with a spread in formation times of
roughly 1 gigayear, or a spread in the formation epoch from $z_f =
2.5$ to $z_f = 3.8$, assuming a mean formation age of $z_f = 3$,
meaning that twice the standard deviation in the colors can be
contained in that redshift range.  The scatter in the {\it blue} - IR
color gives a similar result. Of course, the formation epoch can be
pushed to lower redshifts with a corresponding decrease in the spread
in formation ages.  A mean formation epoch of $z=2.3$ with a spread in
the formation redshift from $z=2.0$ to $z=2.5$, roughly 700 megayears,
also will reproduce the spread in colors.  Given the best fitting
formation age of 2.1 gigayears, a spread of 1 gigayear seems
reasonable.  The rest of the clusters have apparently larger scatters.
However, the errors on those scatter measurements are also quite
large. For example, in Figure \ref{br_scatter}, 3C 184 has an
apparently large scatter, but the errors are such that it is only two
sigma fluctuation from the no--evolution value.  This prevents us from
translating the observed scatter into a measurement of the scatter in
the age to any degree of accuracy.  Nonetheless, the other clusters do
yield measurements in agreement with the smaller results from
RDCS~0848+4453 and from SED98.

\begin{figure}[tbp]
\begin{center}
\includegraphics[width=3.0in]{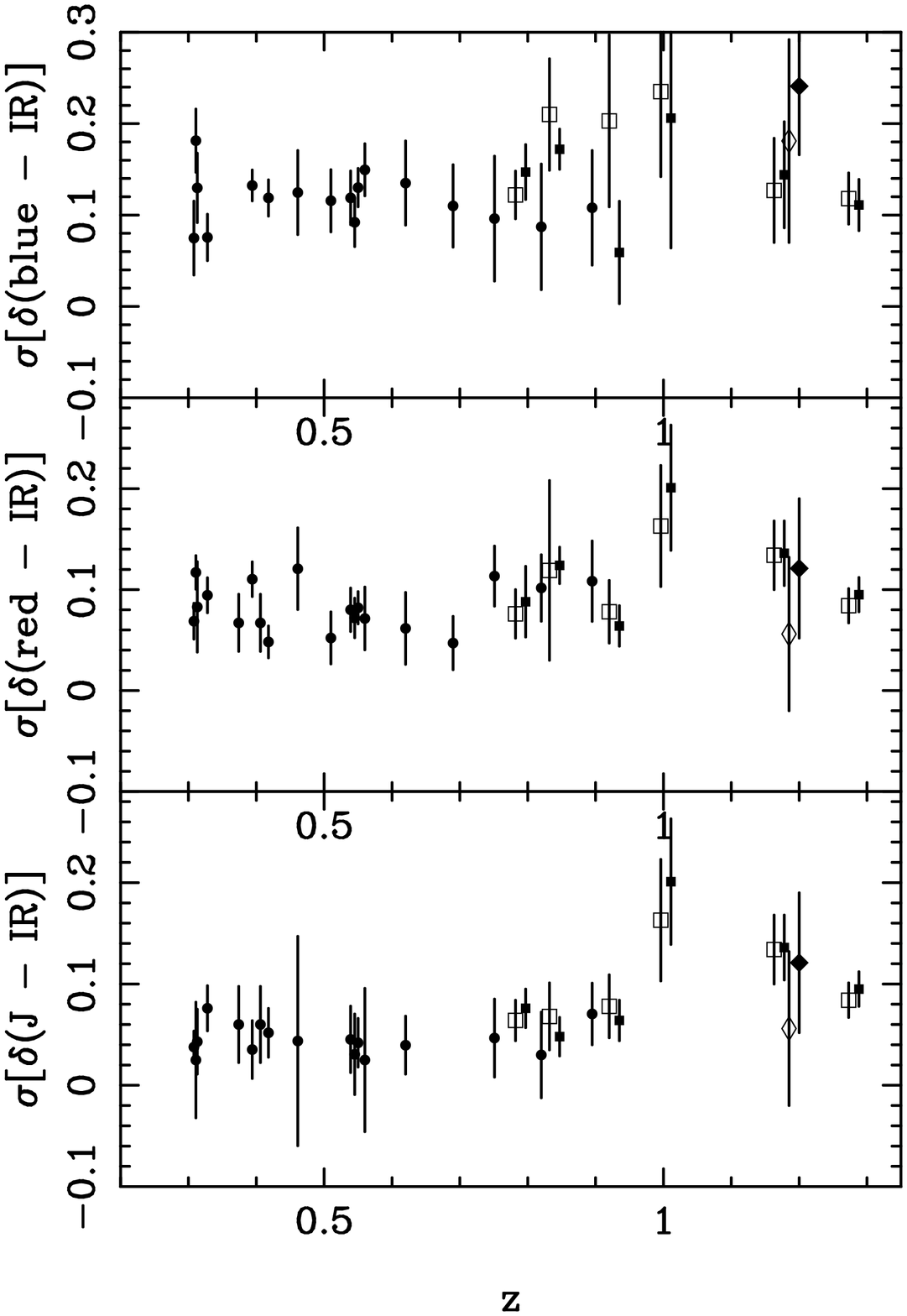}
\end{center}

\caption[holden.fig31.ps]{ The scatter around the color--magnitude
relation as a function of redshift.  The scatter was calculated for
those galaxies identified as early--types.  The symbols are the same
as in Figures 29 and 30.}
\label{scatter}
\end{figure}

Thus, we can create a self-consistent picture.  In fact, using the
\citet{bc03} models produces the same results.  The main possible source of
error is in our simple assumptions about the star formation history.
We assume above that the distribution of star formation histories
matches the distribution of the scatter.  However, the results of
\citet{pvd_mf2001} show that we could create other star formation
histories, ones where the last stars are formed much closer to the
redshift of observation and the true spread in ages is much larger
than we quote above.

\begin{figure}[tbp]
\begin{center}
\includegraphics[width=3.0in]{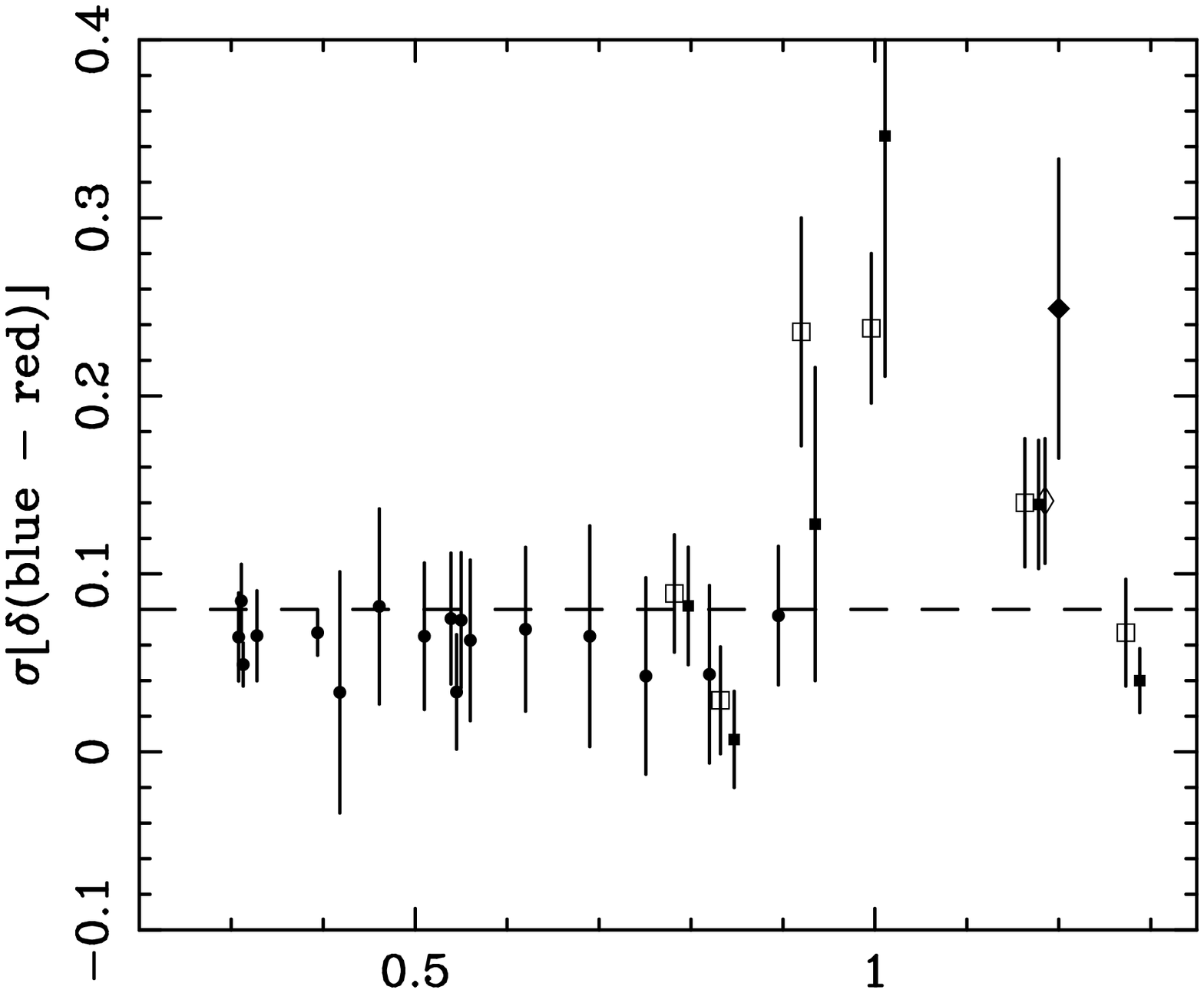}
\end{center}

\caption[holden.fig32.ps]{ The scatter around the color--magnitude
relation as a function of redshift for the ``blue - red'' color.  The
dashed line is the result from \citet{bower92b}. The symbols are the
same as in Figures 29 and 30. }
\label{br_scatter}
\end{figure}

In general, the scatter in the {\it blue} - {\it red} color, as
measured in SED98 and in this paper, is statistically consistent with
zero, and certainly with the scatter as measured by \citet{bower92b}.
We note that, in general, we have only one or two measurements in
Figure \ref{scatter} that are different from the no--evolution value
by more than one sigma, results that are entirely consistent with no
measured evolution.  In a few cases, however, the quality of the
photometry is good enough that true scatter has been measured at
greater than a 3$\sigma$ level of confidence.  We observe this in the
lowest redshift clusters in SED98.  For those clusters, the scatter
measured is in agreement with \citet{bower92b}.  For a couple of the
highest clusters, however, we also detect the scatter at $\simeq$
3$\sigma$ level of confidence.  The values we measure are only
$\simeq$ 1-2$\sigma$ away from the level as measured by
\citet{bower92b} so we cannot statistically say there is evolution in
the scatter.  However, evolution in the scatter of the colors of
cluster early--type galaxies is expected as observations approach the
formation epoch.  Given a modest investment in redshifts of cluster
members, we should be able to confirm this signal and lower the error
on the measured scatter significantly, as has been done for the
$z=1.236$ cluster, RDCS~1252-2927 using the Advanced Camera for
Surveys in \citet{blakeslee2003}

\subsection{The Fraction of Early--Type Galaxies}

\begin{figure}[tbp]
\begin{center}
\includegraphics[width=3.0in]{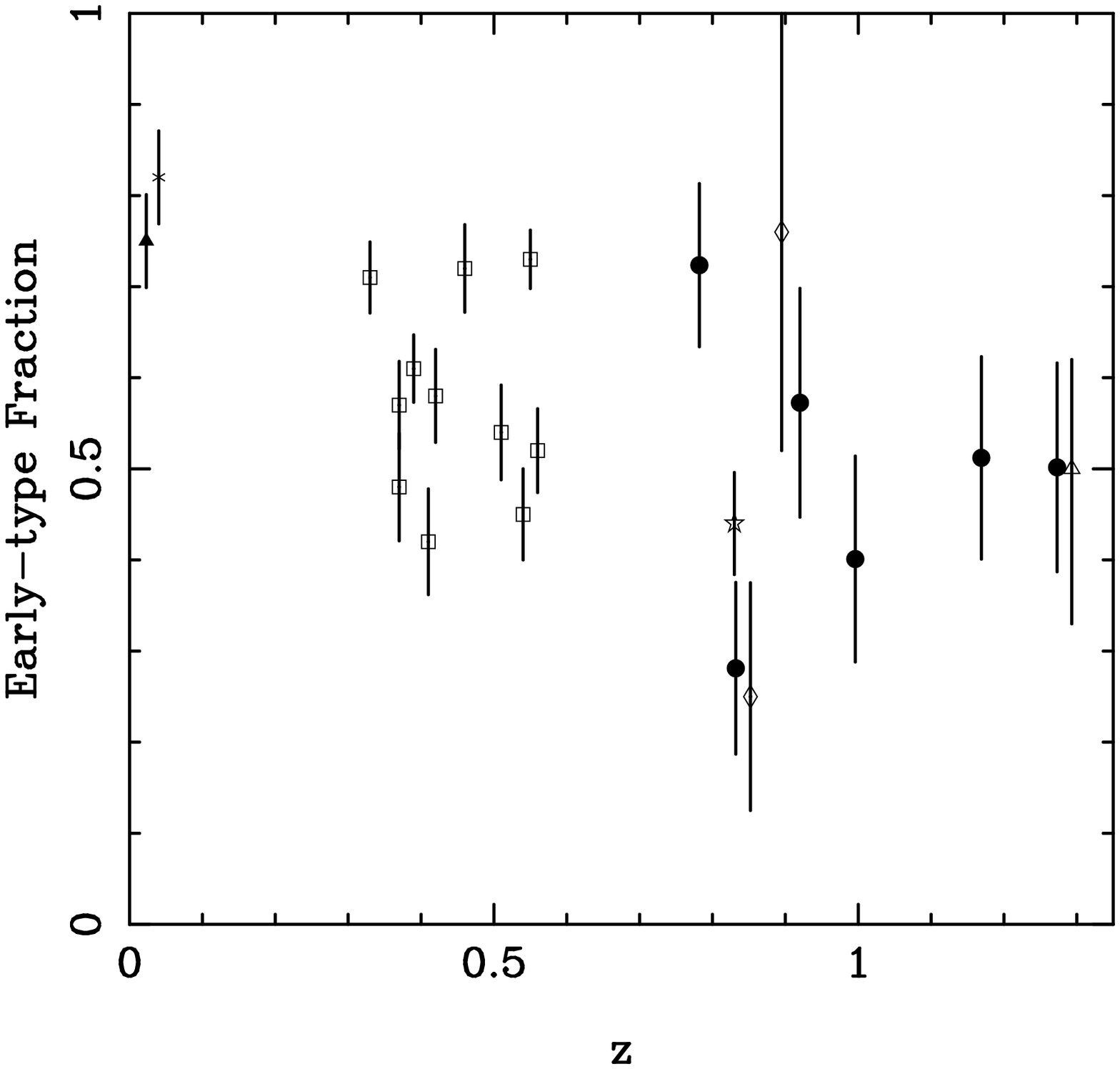}
\end{center}

\caption[holden.fig33.ps]{ The early--type galaxy fraction as a
function of redshift.  The early--type galaxy fractions for clusters in
this paper were measured by counting all of the galaxies that lie
within two standard deviations of the the mean color of the early--type
galaxies for each cluster and are symbolized by solid circles.  At low
redshifts, the solid triangle is the value for Coma from
\citet{andreon1997} while the asterisks are the average value from
\citet{dressler1980}.  The open squares are from \citet{dressler97},
and the open star is from \citet{pvd_2000}.  The two open diamonds are
from \citet{lubin98}.  One of the points is GHO~0021+0406 and is
offset from our measurement.  Finally, the open triangle is the
measurement from RDCS~0848+4453 from \citet{vandokkum2001}.
}
\label{etype_frac}
\end{figure}

A number of lines of evidence point to a smaller fraction of
early--type galaxies at high redshift
\citep{dressler97,pvd_2000,vandokkum2001}, some of the clusters in
those samples were included in SED98. In Figure \ref{etype_frac} using
solid circles, we plot the average fraction of early--type galaxies as
a function of redshift.  For comparison we plot the average results
from \citet{dressler1980} the results for the Coma cluster from
\citet{andreon1997} and the results from \citet{dressler97}.  At high
redshifts, we plot the two cluster early--type fractions for Cl
0023+0423 (GHO~0021+0406 in this paper) and Cl 1604+4304 from
\citet{lubin98}, MS 1054-0321 \citep{pvd_2000} and the measurement for
RDCS~0848+4453 in \citet{vandokkum2001}.

Our early--type galaxy fraction is calculated in an unusual manner,
and thus care should be taken when comparing it with other values.
Specifically, we define the fraction as the number of early--type
galaxies divided by the total number of galaxies within two standard
deviations of the early--type color--magnitude relation and brighter
than $m(L_{\star})+0.6$.  The magnitude limit was chosen to match
\citet{vandokkum2001}, and, therefore, we also reproduce the
early--type fraction from that paper in Figure \ref{etype_frac}.  For
each cluster, we actually have two measurements of the early--type
fraction, one from the automated classification technique and one from
visual classification, we then average those two for Figure
\ref{etype_frac}.  We perform a weighted average, assuming that each
process of morphological identification is independent and, therefore,
so are the errors.  This method stands in contrast to most measurements which
use spectra or colors to pick out cluster members and then measure the
early--type galaxy fraction.  Nonetheless, our results for clusters
with other early--type galaxy fraction measurements are in good
agreement.  \citet{lubin98} finds for GHO~0021+0406 of
$25^{+12.5}_{-12.5}$\% (using the early--type fraction from
spectroscopic redshifts, plotted as an open diamond in Figure
\ref{etype_frac}) and \citet{vandokkum2001} finds for RDCS~0848+4453
an early--type fraction of $50^{+17}_{-12}$\% (plotted as an open
triangle in Figure \ref{etype_frac}), both of which are close to our
average early--type fractions for those two clusters.  Both values
from other papers are plotted with an offset in redshift for clarity.

If we ignore the differences between the various methods and simply
average the observed early-fractions, we find that in the redshift
range of $0.78 < z< 1.273$, excluding the values for 
GHO~0021+0406, an error weighted early--type fraction of $51 \pm 9$\%.
We note here that we quote the weighted value of the error
measurements, not the error on the mean, which would be a value of
$\pm 6$\%.  We exclude GHO~0021+0406 as \citet{lubin98} finds that the
object is likely to be two groups merging rather than a cluster. 
Our measured average value is only 3.4 $\sigma$ below the
\citet{dressler1980} value of 82\%, thus our results appear strong but
are barely statistically significant.  \citet{dressler97} find an
error weighted average value of $60\pm 4$\%, a more statistically
significant result.  Taken at face value, we see no evidence for a
change in the early--type galaxy fraction from $z\simeq 0.5$ to
$z\simeq 1.0$.  Potentially there could be mild evolution, in
either direction, that we do not see.

As was originally shown in \citet{dressler1980}, there is a relation
between early--type galaxy fraction and the density of galaxies.
Therefore, a second possible explanation for our lower than expected
early--type fractions could simply be that by including all of the
galaxies, we are averaging over a lower density than in the lower
redshift comparison samples.  In addition, many of the objects in our
sample are at the low mass end of the range expected from clusters,
and, therefore have smaller central galaxy densities.  

In \S \ref{class_sim}, we discuss our simulations of the automated
classification technique.  At the magnitude limit imposed for our
early--type fraction, we should identify around 70\% of the total
population of early--type galaxies when using the automated method.
Thus, the $z=0$ value of 80\% would be observed as 56\% with the addition
of 4\% of false positive detections.  This 60\% early--type fraction is
within one standard deviation of the observed value of $51$\%, yielding
the possibility that our result could come from misidentification of
early--type galaxies at higher redshift as spirals, especially the
misidentification of S0 galaxies.  Of course, this requires that our
misidentification rate for visual classifications matches that for the
automated technique.  The fact that our overall results agree with
other published values also would imply that there is a broad tendency
to misclassify galaxies, regardless of the classifier, at high
redshift.  As we do not have evidence of this, we will simply state
that our values represent a lower bound of the true value and that,
there is no apparent evolution in the early--type galaxy fraction from
$z\simeq 0.5$ to $z\simeq 1.0$.

\subsection{Discussion}

Our results all indicate that the majority of the stars in the
cluster members are formed at $z_f = 2 - 9$, the exact value depending
on which particular set of models and results one chooses.  As noted
before, this is in good agreement with many other studies of luminous
cluster early--type galaxies such as in \citet{dePropris99},
\citet{pvd_mf2001}, \citet{nelson2001} or \citet{kelson2001}.  This
does yield the puzzling problem of how can the early--type galaxies,
that we observe, all have stars that appear to have formed at the same
range of high redshifts but the number of early--type galaxies in the
clusters appears to have shrunk at higher redshifts.

The red early--type galaxy members of RDCS~0848+4453 show residual
star formation - \citet{vandokkum2003} show spectra of the two
brightest galaxies in Figure \ref{cl0848_nic_cm} which have obvious
\ion{O}{2} emission at 3727\AA.  The two brightest members of
RDCS~0848+4453 potentially have residual star-formation (or, possibly,
a low luminosity AGN but one not detected in X-ray emission, see
\citet{stern2002}), but in general have the typical red colors of high
redshift cluster ellipticals, so it would not be surprising to find
high redshift cluster ellipticals with even bluer colors, objects that
are still forming stars or just recently stopped.  We would not
include these objects in our sample as we trim out all blue elliptical
galaxies before constructing our color--magnitude relation.  All of
this is discussed as the ``progenitor bias'' in \citet{pvd_mf2001}.
Thus, when we measure the ages of elliptical galaxy star-formation, we
are not necessarily measuring the age of formation of stars we see in
elliptical galaxies at $z=0$, but rather the oldest stars at the
observed redshifts.  The fact that most all methods yield the same
epoch of formation, regardless of the redshift range probed, has an
interesting consequence.  At all redshifts, most clusters seem to be
dominated by the light from the oldest stars.  It appears that the
morphological transformations implied by Figure \ref{etype_frac} are
both very mild and occur in objects with older stellar populations but
not in those objects still forming a majority of their stars at $z
= 1 - 2$ as is found in field galaxy surveys.  Since linear theory
predicts that the most massive objects, namely clusters, should begin
to decouple from the Hubble flow first and the smaller dark halos
inside these massive objects should collapse early, this is not
surprising.  The fact that the results of Figure \ref{etype_frac} could
represent an exaggerated version of the true amount of evolution, if
any, only further reinforces this point, namely that there is little
late time evolution in the gross properties of the stellar populations
of cluster early--type galaxies.

What would be interesting is if the reddest stars in early--type
galaxies really did form at $z_f \simeq  5$, as the \citet{bc03}
models show.  That redshift range is the same as the redshift of the
most distant quasars, radio galaxies and the allowed range ends at the
appearance of the the Gunn-Peterson trough
\citep{fan2001,becker2001,white2003}.  Given those observations point
towards a remaining neutral intergalactic medium, a formation redshift $z_f
\simeq 5$ is as high as practically can be allowed for the first
large number of stars.  In addition, given the observed correlation
between black hole mass and galaxy bulge mass, luminous quasars must
be found in galaxies that have large populations of the old stars
found in bulges, the same sorts of stars that make elliptical
galaxies.  Therefore, we have the intriguing possibility that the
oldest stellar populations we find in early--type galaxies could also
contain the first significant mass of stars in the history of the
universe, the stars that produced most of the first metals - thus
populating the intergalactic and intra-cluster medium, and the stars
that could have contributed to the re-ionization of the universe.  All
of this, of course, depends on the assumption of the models of
\citet{bc03}, and the assumptions we used in modeling the stellar
populations, with the various caveats and problems we have discussed
above.

To improve upon our measurements of the evolution of early--type
galaxy colors, future samples should be selected with either
photometric or spectroscopic redshifts.  Field contamination is a
difficult problem, as even though with the small areas covered in
this sample mean that the expected number of interloping galaxies is
small (see our estimates in Table \ref{numsummary}), the large scatter
on those small numbers of galaxies means that, for any given cluster,
there could be a large scatter in the color--magnitude relation.  In
fact, for many of our high redshift fields, we see such a high
scatter---see 3C184 for example.  However, a large scatter could also
be a natural consequence of finding cluster early--type galaxies close
to the epoch of formation.  Determining the source of this scatter
could greatly increase our understanding of the evolution and
formation epoch of early--type cluster galaxies.

A second problem is the additional scatter caused by 
contamination of the photometry from nearby galaxies.  Objects like
those shown in Figure \ref{diff_a} are separated in the HST image but
are blended together in the lower resolution ground based photometry.
This naturally causes an error in the ground based photometry that
cannot be removed except in a model dependent manner.  However, high
resolution observations using, say, the Hubble Space Telescope in
multiple bands, remove the need for this.  

Future samples of clusters of galaxies will need broad wavelength
coverage, stretching from the near-infrared to the rest-frame
ultraviolet, and will need high spatial resolution imaging.  The
natural selection will be a sample of clusters with high virial
masses, and therefore a large number of galaxies, and in conjunction
with Hubble Space Telescope ACS and NICMOS imaging.  Given the
evolution of the stars in such a sample, decomposition of the
various stellar populations should be possible. \citet{blakeslee2003}
shows the beginnings of one such program.  In that work, the authors
examine RDCS~1252-2927, at $z=1.236$, with deep ACS integrations in
two bands.  Because of the depth of the imaging data, the large number
of VLT spectra (31 cluster members to date), and the apparent richness
of the cluster, the authors have 31 elliptical galaxies within two
standard deviations of the color--magnitude sequence with which to
study the same quantities as we do in this paper.  That is more
early--type galaxies thnn in the HST imaging of 3C184, 3C210 and RDCS
0848+4453 combined.  The high quality of their data means that the slope
and scatter are well determined, though the early--type galaxy
fractions and the average color evolution cannot be studied with a
single cluster.  The second necessary component in future studies of
early--type galaxies is high resolution, and high signal to noise,
spectroscopy to examine the stellar populations.  \citet{trager2000}
has shown that, in nearby galaxies, the stellar populations show a
great deal of variation in the apparent metal abundance and the age of
the episode of the most recent star formation.  The combination of
tracking the colors, morphological distributions, and the underlying
spectral properties of galaxies will give us the opportunity to remove
some of the problems that this work has, and, thus, answer the
question of when did the stars in early--type galaxies form?

\section{Summary}

We have constructed a sample of eight clusters of galaxies and galaxy
over-densities spanning the redshift range of $0.58 < z < 1.27$ with
optical and near--infrared photometry combined with Hubble Space
Telescope (HST) morphologies.  Of these eight targets, six appear to
have a significant red sequence of early--type galaxies.  For each
cluster, we measure the colors as a function of either K or H band
magnitude.  The optical bands are all selected to measure a roughly
rest-frame U and V band at the redshift of the cluster, so we compare
the evolution of the same part of the spectral energy distribution as
a function of redshift.  The HST imaging data was generally found in
the archive, and usually sampled the rest frame B band.  In two cases,
however, NICMOS data were used that sampled the rest R band.

To morphologically identify galaxies, we chose two approaches.  First,
we visually classified the galaxies by eye.  Second, we used the
program GIM2D \citep{simard02} to fit surface brightness profiles.  We
used a combination of the residuals from the surface brightness
profile fit with the bulge to total ratio for an empirically defined
measure of the morphological type \citep{im00}.  For this second
approach, we create simulations using images of low redshift galaxies
for each observation of a cluster.  The simulations used blank
sections of the images and the image of the low redshift galaxy from
\citet{frei96} most closely matching the bandpass of the observations.
For most clusters, we found that we could successfully identify 75\%
of the early--type galaxies at $m(L_{\star})$ or brighter magnitudes.
We also compare our classifications done using the automated
classification technique of the Medium Deep Survey \citep{rog99} and
find a good agreement in most cases.  For the cluster MS1137.5+6625,
we compared our automated classifications with our visual
classifications, finding that the two approaches agree 72\% of the
time, in good agreement with our simulations.  In all of the
discrepant cases for MS1137.5+6625, objects that were classified as
early--type galaxies with our automated technique but as spiral
galaxies visually were early spiral galaxies, T=2 at the latest, and
have colors consistent with the early--type galaxies.  The later
result means that any resulting bias in the measured statistics of the
color--magnitude relation because of the inclusion of spiral galaxies
should be small.

For each cluster and galaxy over-density, we constructed
color--magnitude relations for the early--type galaxies, creating one
sequence for each of the two classification methods.  After removing
the two over-densities that had too few early--type galaxies to create
a red sequence, we measured the slope, scatter and zero-point of the
color--magnitude sequences.  Regardless of the method for
morphological identification we used, the results are similar. 

We combined the results in this work with the sample of SED98, or
\citet{stanford98}, to create a sample of 25 measurements of the
slope, scatter and average color of cluster early--type galaxies
spanning a redshift range of $0.3 < z < 1.3$, all with data acquired
in a uniform manner.  We measure no apparent evolution in the slope of
the color--magnitude relation.  However, given the small size of our
samples at the highest redshifts we cannot rule out the evolution as
predicted in such work as \citet{kauffmann98}.  We examined the shift
in the zero-point of the color--magnitude relation, assuming no
evolution in the slope of the relation, as a function of redshift.  We
compare the apparent evolution against the models of
\citet{bruzual93}, using the 1996 update, and the models of
\citet{bc03}.  In order to compare our measurement of evolution with
other results, we fit for the formation epoch of the stars in the
early--type galaxies using coeval starbursts of 0.1 gigayears for a
solar metallicity population.  We reproduce the work of earlier
groups, namely that $z_f = 5$ for a ${\rm H_o = 65\ km\ s^{-1}\
Mpc^{-1}}$, and $\Omega_m = 0.1$ universe and $z_f = 3$ for
$\Omega_{\Lambda}= 0.7$, $\Omega_m = 0.3$ and ${\rm H_o = 65\ km\
s^{-1}\ Mpc^{-1}}$.  Updating the models to the \citet{bc03} release
and using the WMAP preferred cosmology of $\Omega_{\Lambda}= 0.73$,
$\Omega_m = 0.27$ and ${\rm H_o = 71\ km\ s^{-1}\ Mpc^{-1}}$, we find
the average star formation epoch to be $z_f = 5_{-2.7}$ or an age of
the universe of $1.6^{+1.3}$ gigayears for early--type galaxies in
clusters, regardless of the metallicity of the underlying stellar
population.  Our measurements of the color scatter agree with other
work at these redshifts, such as \citet{vandokkum2001} and
\citet{blakeslee2003}, though many of our clusters have a larger error
in the scatter measurements.  Using the formalism outlined in
\citet{pvd_mf2001}, \citet{blakeslee2003} finds that the last episodes
of significant star-formation occurred at $z_L > 2.7$.  If we use the
results of \citet{pvd_mf2001}, we find, roughly, $z_L > 3$ for our
measurements of the pace of evolution of the average color.

Our results on the epoch of formation depend on the usual assumptions
of a coeval population, a single metallicity, and a simple model of
star-formation.  Results of high-resolution spectroscopy of cluster
members, such as in \citet{trager2000} show that all of these are
oversimplifications of the true picture.  While we cannot determine
how much late time star-formation there would be, we do find that
there is, at most, mild, $<20$\%, evolution in the early--type
fraction of cluster galaxies.  Thus, if there is a fair amount of late
time star-formation, it does not transform the morphologies of the
majority of cluster early--type galaxies and it does not grossly
change the colors.  Rather, we find that cluster early--type galaxies
at $z \simeq 1$ are a quiescent population and that the average
rest-frame near ultra-violet and blue colors of these galaxies can be
described by the process of passive stellar population evolution over
the observed range in redshifts.

The authors would like to thank Luc Simard for extensive help in the
use of GIM2D.  In addition, he provided the modified version of the
software we used to fit the multiple NICMOS images.  We would also
like to thank Myungshim Im for helpful suggestions on classifying
early--type galaxies using GIM2D.  BH would like to thank Pieter van
Dokkum and Marijn Franx for useful discussions on galaxy
classification and elliptical galaxy evolution.  BH would also like to
thank Marc Postman for insights on the morphology-density relation and
its impact on early--type galaxy fractions.  Finally, we would like to
thank the referee, Alfonso Arag\'on-Salamanca, for many useful
suggestions that improved this paper.  Support for SAS came from
NASA/LTSA grant NAG5-8430 and for BH from NASA/STScI AR7974.  Both BH
and SAS are supported by the Institute for Geophysics and Planetary
Physics (operated under the auspices of the US Department of Energy by
the University of California Lawrence Livermore National Laboratory
under contract W-7405-Eng-48).  Guest User, Canadian Astronomy Data
Centre, which is operated by the Dominion Astrophysical Observatory
for the National Research Council of Canada's Herzberg Institute of
Astrophysics.

\clearpage

\begin{deluxetable}{lrcccrl}
\tablecolumns{4}
\tablecaption{Summary of Ground Based Observations\label{groundsummary}}
\tablehead{
\colhead{Cluster} & \colhead{z} & \colhead{Optical Bands} &
\colhead{IR Bands} & \colhead{Aperture} &
\colhead{Other Names} & \colhead{Table No.\tablenotemark{a}} }
\startdata
GHO 0229+0035 & 0.607 & V,I & J,H & 1\farcs 27 & CL 0231+0048 & 31 \\
GHO 2155+0321 & 0.7 & V,R,I & J,H & 1\farcs 27 & CL 2157+0347 & 33 \\
MS 1137.5+6625 & 0.782 & R,I & J,H,K & 1\farcs 50 & & 37 \\
GHO 0021+0406 & 0.832 & R,I & J,H & 1\farcs 20 & CL 0023+0423 & 39 \\
GHO 1604+4329 & 0.920 & R,I & J,H,K & 1\farcs 05 & CL 1604+4321 & 42 \\
3C184 & 0.996 & I & J,K & 1\farcs 50 & & 43 \\
3C210 & 1.169 & I & J,K & 1\farcs 50 & & 44 \\
RDCS~0848+4453 & 1.273 & B,R,I & J,K & 1\farcs 20  & CL J0848+4453 & 45 \\
\enddata
\tablenotetext{a}{Refers to the table in \citet{stanford2002} where
  the photometry is tabulated.}
\end{deluxetable}  

\begin{deluxetable}{lllrcc}
\tablecolumns{5}
\tablecaption{Summary of HST Observations\label{hstsummary}}
\tablehead{
\colhead{Cluster} & \colhead{$\alpha$\tablenotemark{a}} & \colhead{$\delta$\tablenotemark{a}} & 
\colhead{Filters} & \colhead{Exposure Time} & \colhead{Table No.} \\
\colhead{} & \colhead{(J2000)} & \colhead{(J2000)} & \colhead{} &
 \colhead{(s)} & \colhead{} \\
}
\startdata
GHO 0229+0035 & 02:31:43.82  & +00:48:47.3 & F606W & 15300 & 4\\ 
GHO 2155+0321 & 21:57:52.22  & +03:48:06.3 & F702W & 18800 & 5 \\ 
MS 1137.5+6625 & 11:40:23.86  & +66:08:19.4 & F814W & 14400 & 6 \\ 
GHO 0021+0406 & 00:23:53.76 & +04:23:11.1 & F702W & 17900 & 7 \\ 
GHO 1604+4329 & 16:04:29.45  & +43:21:35.4 & F702W & 19500 & 8 \\ 
3C184 & 07:39:24.28  & +70:23:10.8 & F814W & 11000 & 9 \\ 
3C210 & 08:58:10.58  & +27:50:30.0 & F814W & 26800 & 10 \\ 
3C210 1 & 08:58:09.19  & +27:50:47.0 & F160W &  2624 & 11 \\
3C210 2 & 08:58:06.85  & +27:51:29.0 & F160W &  4416 & 11 \\
3C210 3 & 08:58:10.98  & +27:51:42.0 & F160W &  5248 & 11 \\
RDCS~0848+4453 & 08:48:33.96  & +44:53:52.2 & F814W & 27800 & 12 \\ 
RDCS~0848+4453 1 & 08:48:31.80  & +44:53:51.2 & F160W & 11200 & 13 \\ 
RDCS~0848+4453 2 & 08:48:36.85  & +44:53:55.0 & F160W & 11200 & 13 \\ 
RDCS~0848+4453 3 & 08:48:34.72  & +44:53:08.0 & F160W & 11200 & 13 \\ 
\enddata
\tablenotetext{a}{Center of HST pointing or first image in dither pattern}
\end{deluxetable}  

\begin{deluxetable}{lrcccc}
\tablecolumns{4}
\tablecaption{Summary of Merged Catalogs\label{numsummary}}
\tablehead{
\colhead{Cluster} & \colhead{z} &
\colhead{$m(L_{\star})$\tablenotemark{a}} & \colhead{$N < m(L_{\star})$} & 
\colhead{$N_{early} < m(L_{\star})$\tablenotemark{b}} & \colhead{$N_{e,pred} < m(L_{\star})$}
\\ }
\startdata
GHO 0229+0035 & 0.607 & 18.7 (18.9)\tablenotemark{c} & 12 & 4/3 & 1.4 \\
GHO 2155+0321 & 0.7 & 19.2 (19.4)\tablenotemark{c} & 19 & 3/2 & 2.1  \\
MS 1137.5+6625 & 0.782 & 18.6 (18.8) & 23 & 15/16 & 2.6 \\
GHO 0021+0406 & 0.832 & 19.7 (19.8)\tablenotemark{c} & 18 & 4/3 & 1.3 \\
GHO 1604+4329 & 0.920 & 19.0 (19.1) & 30 & 10/11 & 1.9 \\
3C184 & 0.996 & 19.2 (19.3) & 23 & 10/6  & 2.5 \\
3C210 & 1.169 & 19.6 (19.7) & 33 & 11/15 & 1.6 \\
RDCS~0848+4453 & 1.273 & 19.9 (20.0) & 27 & 10/8 & 2.4 \\
\enddata
\tablenotetext{a}{Non-evolving value of $m(L_{\star})$ for
$\Omega_m = 0.1$, $\Omega_{\Lambda} = 0.0$ and for  $\Omega_m = 0.3$,
$\Omega_{\Lambda} = 0.7$ in parentheses.  Both values use ${\rm H_o = 65\ km\ s^{-1}\
Mpc^{-1}}$ } 
\tablenotetext{b}{Two numbers of early--type galaxies are given.  The
  first is the number identified visually, the second is the number
  identified automatically}
\tablenotetext{c}{Value of $m(L_{\star})$ in $H$ band for these clusters,
not $K$ as in SED98.}
\end{deluxetable}  

\clearpage

\begin{deluxetable}{lrrcrrrrr}
\tablecolumns{9}
\tablecaption{GHO 0229+0035 HST Imaging Data\label{cl0231data}}
\tablehead{
\colhead{ID} & \colhead{$\delta$x} & \colhead{$\delta$y} & \colhead{T Type} & \colhead{B/T} & \colhead{Res.} & \colhead{${\rm R_h}$} & \colhead{Chip} & \colhead{Pos.}\\
\colhead{} & \colhead{(\arcsec )} & \colhead{(\arcsec )} & \colhead{} & \colhead{} & \colhead{} & \colhead{(\arcsec )} & \colhead{} & \colhead{}\\
}
\startdata
2 & -23.9 & -32.3 & -10 & 0.00 & -199.98 & 0.048 & wf3 & (557, 298)  \\
3 & 0.0 & 0.0 & -10 & 0.54 & -199.98 & 0.011 & wf2 & (36, 351) \\
4 & -80.5 & 81.3 & 5 & 0.00 & 0.15 & 1.784 & wf4 & (361, 692) \\
5 & 19.4 & 24.6 & 4 & 0.01 & 0.22 & 0.572 & wf2 & (318, 210) \\
6 & -71.1 & 22.0 & -1 & 0.97 & 0.03 & 0.637 & wf4 & (510, 111) \\
7 & 8.0 & 20.0 & -2 & 0.56 & 0.21 & 0.470 & wf2 & (191, 203) \\
8 & 50.0 & 63.0 & 4 & 0.11 & 0.17 & 0.847 & wf3 & (705, 408) \\
9 & -37.3 & 25.1 & 5 & 0.00 & 0.21 & 1.269 & wf4 & (190, 15) \\
11 & -56.9 & 27.3 & 3 & 0.42 & 0.15 & 0.390 & wf4 & (359, 100) \\
12 & 17.3 & -8.4 & -1 & 0.89 & 0.02 & 0.208 & wf2 & (151, 505) \\
13 & -31.0 & 10.5 & -3 & 0.29 & 0.01 & 0.389 & wf3 & (131, 186) \\
16 & 30.6 & 25.9 & 2 & 0.71 & 0.31 & 0.306 & wf2 & (420, 246) \\
17 & -66.5 & 42.0 & 3 & 0.02 & 0.21 & 0.629 & wf4 & (394, 270) \\
18 & -44.5 & 72.7 & 4 & 0.04 & 0.17 & 1.151 & wf4 & (64, 467) \\
19 & -40.0 & 11.0 & 0 & 0.28 & 0.08 & 0.305 & wf3 & (93, 266) \\
20 & -43.5 & 27.2 & 7 & 0.00 & 0.22 & 0.841 & wf4 & (240, 46) \\
22 & -33.1 & -7.2 & 3 & 0.17 & 0.09 & 0.696 & wf3 & (290, 279) \\
23 & -63.9 & 88.0 & 8 & 0.03 & 0.02 & 0.913 & wf4 & (181, 688) \\
24 & -41.9 & -13.9 & -4 & 0.11 & 0.03 & 0.009 & wf3 & (316, 380) \\
25 & -51.0 & 58.2 & 8 & 0.72 & 0.10 & 0.708 & wf4 & (177, 354) \\
26 & -49.2 & -5.0 & 8 & 0.13 & 0.12 & 0.321 & wf3 & (206, 407) \\
27 & -63.3 & 71.2 & 8 & 0.13 & 0.10 & 0.815 & wf4 & (231, 534) \\
28 & 10.8 & 36.4 & 6 & 0.05 & 0.12 & 1.053 & wf2 & (288, 67) \\
29 & -48.3 & -23.8 & 5 & 0.07 & 0.10 & 0.949 & wf3 & (378, 482) \\
30 & -58.6 & 36.7 & 6 & 0.00 & 0.23 & 1.287 & wf4 & (332, 195) \\
31 & -78.8 & 29.6 & 4 & 0.23 & 0.08 & 0.554 & wf4 & (550, 211) \\
32 & -86.1 & 12.0 & -5 & 0.00 & 0.06 & 0.034 & wf4 & (689, 82) \\
33 & 2.7 & -27.0 & -3 & 0.79 & 0.09 & 0.212 & wf3 & (620, 30) \\
36 & 40.3 & -10.1 & 5 & 0.35 & 0.00 & 0.159 & wf2 & (361, 608) \\
37 & -82.6 & 68.9 & -10 & 0.30 & -199.98 & 0.017 & wf4 & (429, 585) \\
39 & -74.5 & 56.0 & 2 & 0.07 & 0.12 & 0.691 & wf4 & (403, 435) \\
40 & -52.6 & 30.6 & 1 & 0.67 & 0.00 & 0.214 & wf4 & (304, 113) \\
41 & 15.2 & -1.0 & 8 & 0.03 & 0.11 & 0.631 & wf2 & (177, 424) \\
42 & -6.6 & -38.9 & 0 & 0.32 & 0.13 & 0.016 & wf3 & (693, 162)  \\
43 & 7.5 & 38.2 & 1 & 0.00 & 0.09 & 0.234 & wf2 & (262, 36) \\
45 & -22.0 & -0.8 & -7 & 0.81 & 0.03 & 0.211 & wf3 & (263, 155)  \\
46 & -9.0 & 21.3 & 1 & 0.49 & 0.03 & 0.113 & wf2 & (42, 120)  \\
47 & 22.6 & 47.7 & 0 & 0.37 & 0.06 & 0.150 & wf2 & (434, 15)  \\
48 & -63.6 & 21.1 & 2 & 0.36 & -0.03 & 0.399 & wf4 & (445, 65) \\
\enddata
\end{deluxetable}

\clearpage

\begin{deluxetable}{lrrcrrrrr}
\tablecolumns{9}
\tablecaption{GHO 2155+0321 HST Imaging Data\label{g2157data}}
\tablehead{
\colhead{ID} & \colhead{$\delta$x} & \colhead{$\delta$, } & \colhead{T Type} & \colhead{B/T} & \colhead{Res.} & \colhead{${\rm R_h}$} & \colhead{Chip} & \colhead{Pos.}\\
\colhead{} & \colhead{(\arcsec )} & \colhead{(\arcsec )} & \colhead{} & \colhead{} & \colhead{} & \colhead{(\arcsec )} & \colhead{} & \colhead{}\\
}
\startdata
1 & -65.7 & 58.8 & -10 & 0.40 & 0.64 & 0.557 & wf2 & (303, 239) \\
2 & -47.0 & 62.7 & -10 & 0.00 & -199.98 & 0.558 & wf2 & (117, 293) \\
3 & 18.0 & -38.7 & -10 & 0.43 & 0.15 & 0.147 & wf4 & (612, 639) \\
7 & 36.3 & 81.0 & -3 & 0.93 & 0.04 & 0.765 & wf3 & (549, 677) \\
8* & 0.0 & 0.0 & 5 & 0.00 & 0.20 & 1.368 & wf4 & (392, 275) \\
8* & 0.0 & 0.0 & -10 & 0.24 & -0.13 & 1.65 & wf4 & (394, 264) \\
9 & -18.3 & 37.1 & -10 & 0.06 & -199.98 & 0.115 & wf3 & (66, 172) \\
11 & -48.7 & 82.3 & -1 & 0.40 & 0.02 & 0.744 & wf2 & (156, 483) \\
14 & -3.7 & 49.3 & -1 & 0.48 & 0.05 & 0.440 & wf3 & (197, 305) \\
18 & -18.5 & -0.6 & -9 & 0.03 & -0.17 & 2.715 & wf4 & (212, 289) \\
24 & 20.1 & -0.9 & -9 & 0.40 & 0.01 & 0.011 & wf4 & (593, 257) \\
26 & 3.2 & 50.2 & -8 & 0.00 & 0.34 & 0.500 & wf3 & (212, 372) \\
31 & -13.1 & 85.1 & -6 & 0.00 & 0.04 & 0.018 & wf3 & (547, 180) \\
34 & 2.2 & 35.7 & 0 & 0.14 & 0.03 & 0.505 & wf3 & (69, 376) \\
36 & -1.1 & -17.7 & 4 & 0.33 & 0.18 & 1.17 & wf4 & (400, 443) \\
37 & 22.2 & 92.1 & 2 & 0.47 & 0.16 & 0.552 & wf3 & (649, 529) \\
39 & 4.2 & 59.0 & -10 & 0.03 & -199.98 & 0.025 & wf3 & (301, 377) \\
40 & 17.3 & 15.3 & 8 & 0.00 & 0.10 & 0.829 & wf4 & (558, 99) \\
41 & -42.6 & 66.2 & 3 & 0.41 & 0.32 & 1.622 & wf2 & (79, 328) \\
42 & 8.1 & -3.4 & 1 & 0.00 & 0.07 & 0.370 & wf4 & (476, 292) \\
43* & -14.8 & 71.9 & 1 & 0.00 & 0.14 & 0.455 & wf3 & (429, 188) \\
43* & -14.8 & 71.9 & 7 & 0.00 & 0.13 & 0.907 & wf3 & (408, 171) \\
47 & -27.3 & 22.9 & 1 & 0.02 & 0.10 & 0.420 & wf4 & (103, 64) \\
48 & -21.4 & 56.2 & -2 & 0.55 & 0.05 & 0.121 & wf3 & (252, 125) \\
49 & -25.2 & 53.7 & 2 & 0.07 & 0.07 & 0.391 & wf3 & (223, 90) \\
52 & -5.3 & 16.5 & 5 & 0.01 & 0.22 & 0.999 & wf4 & (327, 107) \\
53 & 31.8 & 33.2 & 0 & 0.98 & 0.08 & 0.012 & wf3 & (69, 673) \\
54 & -18.1 & 76.7 & -3 & 0.56 & 0.08 & 0.011 & wf3 & (458, 142) \\
55 & -17.8 & 94.3 & -8 & 0.03 & 0.16 & 2.476 & wf3 & (638, 127) \\
58 & 14.7 & 29.3 & 9 & 0.19 & 0.23 & 0.855 & wf3 & (17, 502) \\
62 & 25.2 & -41.5 & -2 & 0.77 & 0.01 & 0.009 & wf4 & (685, 658) \\
65 & 29.9 & 77.3 & -8 & 0.00 & -199.98 & 0.034 & wf3 & (508, 617)\\
66 & -12.3 & 45.9 & -8 & 0.06 & 0.07 & 0.752 & wf3 & (160, 229) \\
\enddata
\end{deluxetable}

\clearpage

\begin{deluxetable}{lrrcrrrrr}
\tablecolumns{9}
\tablecaption{MS 1137.5+6625\label{m1137data}}
\tablehead{
\colhead{ID} & \colhead{$\delta$x} & \colhead{$\delta$, } & \colhead{T Type} & \colhead{B/T} & \colhead{Res.} & \colhead{${\rm R_h}$} & \colhead{Chip} & \colhead{Pos.}\\
\colhead{} & \colhead{(\arcsec )} & \colhead{(\arcsec )} & \colhead{} & \colhead{} & \colhead{} & \colhead{(\arcsec )} & \colhead{} & \colhead{}\\
}
\startdata
  3 &   0.0 &   0.0 & -5 & 0.68 & 0.04 & 2.649 & wf3 & (492, 314) \\      
  8 &   4.4 &   9.7 & -5 & 0.78 & 0.07 & 0.616 & wf3 & (455, 412) \\      
 10 &   2.2 &   3.3 & -5 & 0.66 & 0.04 & 0.720 & wf3 & (471, 348) \\       
 11 &   7.4 &  -6.8 & -2 & 0.64 & 0.12 & 0.070 & wf3 & (415, 252) \\       
 12 &   8.3 &  -5.0 & -3 & 0.88 & 0.06 & 0.401 & wf3 & (411, 273) \\      
 13 & -11.3 &  12.5 & -4 & 0.24 & 0.11 & 0.442 & wf3 & (612, 429) \\      
 14 &  26.1 &  14.7 & -10 & 0.00 & 0.22 & 0.035 & wf3 & (245, 476) \\     
 15 &  -1.1 &   4.9 & -2 & 0.99 & 0.04 & 0.553 & wf3 & (508, 362) \\      
 17 &   3.8 &   2.4 & -1 & 0.82 & 0.06 & 0.422 & wf3 & (456, 338) \\      
 20 &  -4.9 &   6.2 & -2 & 0.41 & 0.02 & 0.649 & wf3 & (546, 370) \\      
 21 &   5.9 &   6.1 & -4 & 0.95 & 0.03 & 0.349 & wf3 & (438, 378) \\      
 23 &  37.7 &  33.7 & -1 & 0.36 & 0.07 & 0.754 & wf3 & (143, 673) \\      
 24 &   2.2 &  38.1 & -5 & 0.77 & 0.02 & 0.302 & wf3 & (494, 692) \\      
 25 & -26.6 &   9.1 & 2 & 0.61 & 0.07 & 0.825 & wf4 & (283, 314) \\       
 27 & -11.2 & -25.2 & 4 & 0.36 & 0.14 & 1.289 & wf3 & (589, 56) \\        
 28 &  41.4 &  31.7 & -3 & 0.51 & 0.08 & 0.321 & wf3 & (101, 655) \\      
 29 &  29.8 &  -3.9 & -1 & 0.64 & 0.08 & 0.834 & wf3 & (194, 289) \\      
 30 &  43.0 & -13.2 & -3 & 0.98 & 0.02 & 0.361 & wf3 & (57, 209) \\       
 31 &   6.5 &  -3.0 & -2 & 0.99 & 0.04 & 0.511 & wf3 & (427, 288) \\      
 32 &   7.9 &   9.4 & 1 & 0.68 & 0.07 & 0.053 & wf3 & (420, 411) \\       
 34 &  -1.4 &  -2.0 & -4 & 0.78 & 0.09 & 0.352 & wf3 & (505, 291) \\      
 39 & -10.8 & -10.4 & 0 & 0.72 & 0.03 & 0.273 & wf3 & (595, 204) \\       
 40 &  12.3 & -49.6 & -1 & 0.74 & 0.11 & 0.539 & wf2 & (156, 332) \\      
 41 &  44.7 &   3.5 & 0 & 0.88 & 0.07 & 0.37 & wf3 & (50, 375) \\         
 47 &  21.1 & -45.2 & 5 & 0.01 & 0.18 & 1.018 & wf2 & (107, 247) \\       
 49 &  32.1 &   7.9 & -3 & 0.62 & 0.06 & 0.229 & wf3 & (179, 412) \\      
 50 &  12.4 &  -5.9 & 2 & 0.27 & 0.07 & 0.379 & wf3 & (364, 262) \\       
 54 &   2.0 & -11.1 & -3 & 0.64 & 0.03 & 0.289 & wf3 & (466, 205) \\      
 55 &  29.1 &  36.2 & 5 & 0.03 & 0.12 & 0.735 & wf3 & (225, 690) \\       
 56 &  -5.3 &  11.4 & 1 & 0.84 & 0.05 & 0.204 & wf3 & (553, 422) \\       
 57 &  29.7 &  -6.5 & -1 & 0.67 & 0.07 & 0.246 & wf3 & (194, 270) \\       
 58 &  71.7 & -16.0 & 7 & 0.23 & 0.13 & 0.333 & wf4 & (208, 207) \\       
 59 & -22.4 & -64.7 & 3 & 0.51 & 0.29 & 0.199 & wf2 & (330, 670) \\       
 60 &  18.5 &  31.5 & 1 & 0.20 & 0.13 & 0.333 & wf3 & (330, 638) \\       
 61 & -22.9 & -50.6 & -3 & 0.81 & 0.09 & 0.290 & wf2 & (183, 684) \\       
 62 &  -7.1 & -11.5 & -3 & 0.84 & 0.01 & 0.234 & wf3 & (556, 195) \\      
 65 &  13.8 & -19.5 & 2 & 0.00 & 0.08 & 0.488 & wf3 & (343, 129) \\       
 71 & -11.1 & 15.6 & -4 & 0.80 & 0.03 & 0.202 & wf3 & (612, 460) \\       
 72 &  15.6 & -13.3 & 0 & 0.73 & 0.04 & 0.542 & wf3 & (330, 191) \\       
 73 &  41.5 & -25.5 & -4 & 0.50 & 0.05 & 0.158 & wf3 & (63, 83) \\        
 75 &  -2.9 & -27.6 & 5 & 0.00 & 0.26 & 0.695 & wf3 & (504, 38) \\        
 78 &  66.5 &  14.7 & 2 & 0.39 & 0.03 & 0.292 & wf4 & (510, 142) \\       
 81 &  30.7 &  28.4 & -2 & 0.64 & 0.07 & 0.211 & wf3 & (207, 616) \\      
 82 &   0.8 & -39.9 & 1 & 0.88 & -0.02 & 0.403 & wf2 & (67, 452) \\       
 83 &  17.1 &  -7.8 & 1 & 0.32 & 0.16 & 0.335 & wf3 & (320, 246) \\       
 84 &  -2.2 & -21.1 & 2 & 0.00 & 0.72 & 0.034 & wf3 & (501, 103) \\       
 85 &  13.8 & -10.2 & 6 & 0.57 & 0.21 & 1.160 & wf3 & (345, 224) \\        
 88 & -24.8 &   1.1 & -3 & 0.85 & -0.04 & 0.430 & wf3 & (740, 309) \\      
 90 & -11.9 & -43.9 & -1 & 0.48 & 0.12 & 0.300 & wf2 & (115, 571) \\        
 91 &   6.7 & -30.0 & -3 & 0.60 & 0.07 & 0.441 & wf3 & (405, 20) \\       
 93 &   5.4 & -35.2 & 1 & 0.76 & 0.02 & 0.215 & wf2 & (13, 410) \\        
 95 &  14.3 &  11.9 & 7 & 0.47 & 0.23 & 0.537 & wf3 & (360, 440) \\       
 96 &  -7.6 &  23.4 & 9 & 0.03 & 0.12 & 1.601 & wf3 & (580, 540) \\       
 98 &  15.3 &  35.8 & -3 & 0.28 & 0.28 & 0.029 & wf3 & (363, 677) \\      
102 &  43.0 & -17.1 & -10 & 0.11 & 0.28 & 0.468 & wf3 & (50, 169) \\      
103 &   5.6 &  20.8 & -1 & 0.48 & 0.07 & 0.201 & wf3 & (453, 522) \\      
107 & -12.9 &  -3.3 & -3 & 0.97 & 0.04 & 0.099 & wf3 & (616, 274) \\      
110 &  39.4 & -20.9 & -2 & 0.70 & 0.10 & 0.194 & wf3 & (86, 132) \\       
112 &  68.4 & -30.9 & 0 & 0.01 & 0.10 & 0.239 & wf4 & (55, 190) \\        
114 &  13.6 & -63.6 & -1 & 0.73 & 0.06 & 0.212 & wf2 & (298, 308) \\      
115 &  39.3 & -55.9 & 3 & 0.36 & 0.09 & 0.614 & wf2 & (202, 53) \\        
116 &  23.0 &   1.4 & 2 & 0.16 & 0.02 & 0.842 & wf3 & (265, 343) \\       
118 &  22.3 &  -0.3 & -2 & 0.68 & 0.02 & 0.253 & wf3 & (273, 324) \\       
119 &  19.0 &  -7.6 & -3 & 0.70 & 0.04 & 0.190 & wf3 & (301, 249) \\      
121 &  -4.1 &  25.0 & 5 & 0.01 & 0.14 & 0.827 & wf3 & (550, 558) \\       
123 &  26.7 & -42.6 & 2 & 0.76 & 0.05 & 0.360 & wf2 & (77, 192) \\         
125 & -17.0 &  10.2 & 1 & 0.28 & 0.05 & 0.157 & wf3 & (669, 404) \\       
126 &  13.1 & -40.9 & 6 & 0.00 & 0.10 & 0.786 & wf3 & (335, 66) \\        
128 &  14.2 & -25.7 & -2 & 0.00 & 0.13 & 0.298 & wf2 & (67, 331) \\       
129 & -14.6 & -49.2 & 9 & 0.00 & 0.15 & 0.699 & wf2 & (171, 601) \\       
133 & -10.3 &  27.7 & -7 & 0.18 & 0.21 & 0.050 & wf3 & (611, 585) \\       
\enddata
\end{deluxetable}

\clearpage

\begin{deluxetable}{lrrcrrrrr}
\tablecolumns{9}
\tablecaption{GHO 0021+0406 HST Imaging Data\label{g0023data}}
\tablehead{
\colhead{ID} & \colhead{$\delta$x} & \colhead{$\delta$, } & \colhead{T Type} & \colhead{B/T} & \colhead{Res.} & \colhead{${\rm R_h}$} & \colhead{Chip} & \colhead{Pos.}\\
\colhead{} & \colhead{(\arcsec )} & \colhead{(\arcsec )} & \colhead{} & \colhead{} & \colhead{} & \colhead{(\arcsec )} & \colhead{} & \colhead{}\\
}
\startdata
10 & -65.6 & -39.4 & 5 & 0.18 & 0.23 & 0.725 & wf3 & (493, 475) \\
11 & -25.4 & -56.1 & 7 & 0.00 & 0.26 & 1.123 & wf3 & (590, 49) \\
12 & -25.9 & -2.5 & -3 & 0.49 & 0.02 & 0.573 & wf3 & (58, 142) \\
17 & -8.1 & 12.5 & 8 & -1 & -1 & -1 & wf4 & (4, 98) \\
18 & 11.1 & -26.0 & 8 & 0.99 & 0.20 & 0.627 & wf2 & (245, 229) \\
19 & 0.9 & -42.4 & -1 & 0.60 & 0.05 & 0.217 & wf2 & (166, 399) \\
20 & 18.3 & -45.5 & 5 & 0.08 & 0.19 & 1.076 & wf2 & (345, 408) \\
21 & -51.9 & -10.0 & 10 & 0.06 & 0.17 & 0.838 & wf3 & (174, 385) \\
22 & -37.9 & -57.8 & 10 & 0.10 & 0.18 & 0.804 & wf3 & (628, 170) \\
23 & -69.8 & 1.8 & 0 & 0.41 & 0.09 & 0.287 & wf3 & (88, 583) \\
24 & -54.3 & 18.6 & 4 & 0.12 & 0.08 & 1.148 & wf4 & (477, 63) \\
25 & 51.9 & -65.3 & 6 & 0.04 & 0.00 & 1.259 & wf2 & (725, 560) \\
26 & -19.1 & 68.2 & 6 & 0.51 & 0.09 & 1.184 & wf4 & (206, 629) \\
29 & -8.7 & -6.2 & -3 & 0.48 & 0.01 & 0.267 & wf2 & (16, 58) \\
31 & -33.9 & 0.4 & 1 & 0.01 & 0.06 & 0.662 & wf3 & (36, 231) \\
32 & -76.4 & -10.1 & 7 & 0.00 & 0.09 & 0.207 & wf3 & (222, 626) \\
33 & 2.2 & -33.4 & 10 & 0.01 & 0.15 & 0.763 & wf2 & (168, 310) \\
34 & -19.8 & 15.4 & 0 & 0.42 & 0.03 & 0.211 & wf4 & (120, 109) \\
37 & -62.1 & -24.9 & -5 & 0.00 & 0.09 & 0.007 & wf3 & (342, 463) \\
39 & -66.0 & -60.0 & 0 & 0.06 & 0.15 & 0.417 & wf3 & (701, 447) \\
40 & -54.1 & -0.4 & 8 & 0.01 & 0.13 & 0.591 & wf3 & (85, 421) \\
43 & 54.0 & -13.7 & 1 & 0.62 & -0.01 & 0.872 & wf2 & (651, 32) \\
46 & 40.7 & -48.6 & 4 & 0.21 & 0.19 & 0.415 & wf2 & (574, 404) \\
49 & 51.6 & -47.3 & 6 & 0.06 & 0.01 & 0.497 & wf2 & (681, 374) \\
50 & -64.4 & 3.4 & 10 & 0.02 & 0.18 & 0.426 & wf3 & (66, 530) \\
52 & -22.1 & -55.9 & 9 & 0.08 & 0.20 & 0.875 & wf3 & (580, 16) \\
53 & -39.8 & 38.2 & 0 & 0.37 & 0.16 & 0.281 & wf4 & (358, 297) \\
54 & -70.5 & -15.8 & 1 & 0.14 & 0.08 & 0.416 & wf3 & (265, 562) \\
55 & 8.8 & -30.4 & 2 & 0.10 & 0.08 & 0.484 & wf2 & (221, 270) \\
56 & 34.0 & -57.3 & -2 & 0.28 & 0.19 & 0.218 & wf2 & (523, 498) \\
57 & 12.2 & -29.5 & -5 & 1.00 & 0.02 & 0.006 & wf2 & (261, 257) \\
60 & 42.7 & -66.1 & -4 & 0.85 & 0.04 & 0.137 & wf2 & (626, 577) \\
61 & -32.8 & -9.3 & -4 & 0.39 & 0.08 & 0.304 & wf3 & (137, 197) \\
62 & 14.8 & -59.3 & 4 & 0.27 & 0.15 & 0.832 & wf2 & (334, 551) \\
63 & -65.5 & 17.3 & -1 & 0.02 & 0.13 & 0.202 & wf4 & (575, 45) \\
65 & -23.4 & 17.7 & 1 & 0.60 & 0.20 & 0.292 & wf4 & (161, 125) \\
73 & -68.9 & -4.8 & -1 & 0.02 & 0.03 & 0.301 & wf3 & (154, 565) \\
74 & -28.1 & -49.4 & 10 & 0.03 & 0.09 & 0.820 & wf3 & (527, 83) \\
79 & -17.9 & -5.3 & 8 & 0.01 & 0.08 & 0.383 & wf3 & (72, 56) \\
80 & 5.1 & -14.9 & 10 & 0.36 & 0.06 & 0.523 & wf2 & (171, 124) \\
82 & -19.3 & 29.6 & 3 & 0.01 & 0.23 & 0.615 & wf4 & (140, 247) \\
86 & -8.2 & 25.7 & 9 & 0.11 & 0.05 & 0.541 & wf4 & (22, 230) \\
87 & -46.8 & 55.9 & 8 & 0.01 & 0.16 & 0.757 & wf4 & (457, 459) \\
88 & -12.3 & -24.7 & -3 & -1  & -1 &  -1 & wf2 & (5, 219) \\
89 & -53.7 & 40.9 & -4 & 0.75 & 0.00 & 0.015 & wf4 & (499, 300) \\
91 & -19.0 & -16.2 & 1 & 0.09 & 0.16 & 0.453 & wf3 & (178, 49) \\
\enddata
\end{deluxetable}
\clearpage

\begin{deluxetable}{lrrcrrrrr}
\tablecolumns{9}
\tablecaption{GHO 1604+4329 HST Imaging Data\label{g1603data}}
\tablehead{
\colhead{ID} & \colhead{$\delta$x} & \colhead{$\delta$, } & \colhead{T Type} & \colhead{B/T} & \colhead{Res.} & \colhead{${\rm R_h}$} & \colhead{Chip} & \colhead{Pos.}\\
\colhead{} & \colhead{(\arcsec )} & \colhead{(\arcsec )} & \colhead{} & \colhead{} & \colhead{} & \colhead{(\arcsec )} & \colhead{} & \colhead{}\\
}
\startdata
3 & -33.8 & 47.4 & -10 & 0.00 & -199.98 & 0.609 & wf2 & (258, 312) \\
5 & -80.7 & -7.0 & -10 & 0.00 & -199.98 & 0.578 & wf3 & (192, 426) \\
6 & 12.6 & 64.1 & 4 & 0.14 & 0.15 & 1.188 & wf2 & (641, 633) \\
7 & -18.7 & 46.7 & -1 & 0.43 & 0.03 & 0.752 & wf2 & (330, 448) \\
10 & -37.9 & 11.2 & -4 & 0.18 & 0.05 & 0.337 & wf3 & (465, 49) \\
13 & -21.4 & 1.5 & 1 & 0.63 & 0.04 & 0.621 & wf3 & (655, 52) \\
14 & 7.7 & 35.0 & -3 & 0.59 & 0.07 & 0.073 & wf2 & (363, 734) \\
17 & -15.0 & 0.7 & 1 & 0.63 & 0.07 & 1.043 & wf3 & (712, 29) \\
18 & -87.7 & -38.6 & 5 & 0.23 & 0.07 & 1.393 & wf3 & (289, 727) \\
20 & -24.9 & -3.8 & -2 & 0.71 & 0.04 & 0.364 & wf3 & (652, 113) \\
21 & -21.8 & 29.9 & 0 & 0.73 & 0.06 & 0.405 & wf2 & (171, 502) \\
22 & -54.5 & 19.9 & 3 & 0.03 & 0.19 & 1.577 & wf3 & (280, 58) \\
23 & -1.3 & 51.2 & 9 & 0.37 & 0.27 & 0.554 & wf2 & (458, 574) \\
27 & -19.1 & 48.0 & 4 & 0.01 & 0.17 & 0.740 & wf2 & (342, 437) \\
28 & -46.2 & -18.5 & -9 & 0.46 & 0.01 & 0.276 & wf3 & (540, 347) \\
30 & -8.6 & 31.5 & -3 & 0.82 & 0.05 & 0.406 & wf2 & (251, 611) \\
31 & -32.3 & -4.3 & 1 & 0.29 & 0.06 & 0.510 & wf3 & (591, 156) \\
32 & 6.7 & 43.9 & -3 & 0.47 & 0.02 & 0.267 & wf2 & (433, 681) \\
35 & -31.9 & 63.3 & 3 & 0.00 & 0.17 & 0.575 & wf2 & (411, 251) \\
36 & -39.9 & 85.1 & -1 & 0.02 & 0.08 & 0.367 & wf2 & (558, 74) \\
39 & -41.5 & -19.7 & -3 & 0.78 & 0.03 & 0.699 & wf3 & (588, 333) \\
40 & -53.0 & -41.1 & 4 & 0.07 & 0.06 & 0.690 & wf3 & (597, 578) \\
42 & -28.3 & 31.7 & -1 & 0.44 & 0.00 & 0.196 & wf2 & (155, 437) \\
43\tablenotemark{a} & -78.8 & -9.5 & -2 & 0.67 & 0.84 & 1.156 & wf3 & (216, 434) \\
45 & -26.5 & -15.1 & 7 & 0.00 & 0.26 & 0.496 & wf3 & (697, 220) \\
50 & -47.2 & 43.1 & 1 & 0.84 & 0.02 & 0.315 & wf2 & (161, 220) \\
52 & -93.0 & -5.8 & 6 & 0.00 & 0.15 & 0.669 & wf3 & (75, 477) \\
53 & -44.4 & -22.6 & 1 & 0.26 & 0.02 & 0.316 & wf3 & (577, 373) \\
54 & -32.4 & 71.6 & 5 & 0.04 & 0.24 & 0.945 & wf2 & (480, 206) \\
58 & 2.7 & 45.1 & 2 & 0.66 & -0.01 & 0.287 & wf2 & (423, 642) \\
62 & -38.9 & 74.1 & -3 & 0.85 & 0.02 & 0.187 & wf2 & (470, 136) \\
65 & -26.8 & 14.7 & 1 & 0.45 & 0.15 & 0.573 & wf2 & (17, 539) \\
70 & -43.9 & -41.7 & 8 & 0.00 & 0.17 & 0.444 & wf3 & (679, 537) \\
71 & -21.1 & 77.1 & 1 & 0.82 & 0.04 & 0.225 & wf2 & (583, 274) \\
73 & -55.6 & 43.4 & 7 & 0.01 & 0.16 & 0.69 & wf2 & (124, 145) \\
78 & -22.6 & 81.3 & -1 & 0.67 & 0.16 & 0.426 & wf2 & (612, 241) \\
79 & -25.4 & 4.2 & 2 & 0.38 & 0.16 & 0.659 & wf3 & (607, 48) \\
80 & -8.4 & 83.9 & -6 & 0.87 & -0.12 & 0.162 & wf2 & (708, 351) \\
85 & -10.8 & 10.8 & 3 & 0.36 & 0.09 & 0.666 & wf2 & (60, 691) \\
88 & -18.8 & 68.1 & 2 & 0.08 & 0.05 & 0.356 & wf2 & (515, 345) \\
90 & 0.8 & 33.2 & 1 & 0.01 & 0.09 & 0.303 & wf2 & (315, 684) \\
91 & -1.9 & 19.8 & 0 & 0.61 & 0.02 & 0.128 & wf2 & (180, 728) \\
94 & -40.6 & -24.7 & 4 & 0.00 & 0.21 & 0.438 & wf3 & (620, 372) \\
\enddata
\tablenotetext{a}{This object is next to the bright star, object \# 5.
The GIM2D parameters were affected by the strongly varying background.}
\end{deluxetable}
\clearpage

\begin{deluxetable}{lrrcrrrrr}
\tablecolumns{9}
\tablecaption{3C184 HST Imaging Data\label{3c184data}}
\tablehead{
\colhead{ID} & \colhead{$\delta$x} & \colhead{$\delta$, } & \colhead{T Type} & \colhead{B/T} & \colhead{Res.} & \colhead{${\rm R_h}$} & \colhead{Chip} & \colhead{Pos.}\\
\colhead{} & \colhead{(\arcsec )} & \colhead{(\arcsec )} & \colhead{} & \colhead{} & \colhead{} & \colhead{(\arcsec )} & \colhead{} & \colhead{}\\
}
\startdata
1 & 19.5 & 16.3 & 2 & 0.70 & 0.07 & 0.917 & wf3 & (645, 429) \\
2 & 8.4 & 4.9 & 3 & 0.02 & 0.12 & 1.531 & wf3 & (487, 398) \\
3 & 0.0 & 0.0 & 7 & 0.57 & 0.27 & 0.861 & wf3 & (399, 355) \\
4 & -30.2 & -13.1 & -3 & 0.94 & 0.02 & 0.528 & wf3 & (120, 172) \\
5 & -30.8 & -20.6 & 4 & 0.06 & 0.12 & 1.032 & wf3 & (51, 209) \\
6 & -52.7 & -45.8 & 5 & 0.30 & 0.19 & 1.516 & wf4 & (174, 268) \\
7 & -25.3 & 51.6 & -2 & 0.94 & 0.04 & 0.014 & wf2 & (126, 688) \\
8 & 21.6 & 1.1 & -1 & 0.58 & 0.04 & 0.385 & wf3 & (529, 530) \\
9 & -81.7 & 67.4 & 1 & 0.06 & 0.10 & 0.579 & wf2 & (701, 516) \\
10 & 15.2 & 27.3 & 3 & 0.07 & 0.15 & 0.737 & wf3 & (714, 332) \\
11 & -81.2 & 15.3 & 0 & 0.03 & 0.10 & 0.608 & wf2 & (403, 71) \\
12 & 2.4 & 2.1 & -3 & 0.74 & -0.01 & 0.419 & wf3 & (431, 365) \\
13 & 19.5 & -20.4 & 7 & 0.03 & 0.15 & 0.795 & wf3 & (333, 633) \\
14 & -42.1 & 48.0 & 4 & 0.49 & 0.13 & 0.723 & wf2 & (248, 566) \\
15 & -38.4 & -68.8 & -2 & 0.97 & 0.00 & 0.255 & wf4 & (422, 382) \\
16 & 15.4 & -20.9 & 3 & 0.02 & 0.10 & 0.567 & wf3 & (293, 601) \\
17 & -85.3 & -68.1 & -1 & 0.01 & 0.05 & 0.035 & wf4 & (20, 649) \\
18 & 7.9 & 20.1 & -3 & 0.86 & -0.01 & 0.298 & wf3 & (614, 309) \\
19 & 12.9 & 19.0 & -2 & 0.80 & 0.03 & 0.127 & wf3 & (630, 357) \\
20 & 3.9 & 7.0 & -2 & 0.19 & 0.01 & 0.250 & wf3 & (481, 346) \\
21 & -1.1 & -24.5 & 8 & 0.03 & 0.07 & 0.958 & wf3 & (187, 482) \\
22 & -63.3 & -41.3 & -1 & 0.34 & 0.02 & 0.268 & wf4 & (58, 290) \\
23 & -58.9 & -48.2 & 8 & 0.01 & 0.23 & 0.997 & wf4 & (132, 318) \\
24 & -4.7 & 7.0 & -2 & 0.57 & 0.02 & 0.180 & wf3 & (431, 277) \\
25 & 38.1 & 12.5 & 0 & 0.68 & 0.02 & 0.286 & wf3 & (717, 611) \\
26 & 15.1 & -4.4 & 1 & 0.01 & 0.15 & 0.248 & wf3 & (444, 505) \\
27 & -52.5 & 41.4 & -1 & 0.20 & 0.11 & 0.375 & wf2 & (300, 448) \\
28 & -2.1 & 10.4 & -2 & 0.22 & -0.03 & 0.163 & wf3 & (472, 280) \\
29 & 10.5 & -9.7 & -7 & 1.00 & 0.01 & 0.011 & wf3 & (374, 496) \\
30 & -18.4 & -8.3 & 7 & 0.08 & 0.20 & 0.736 & wf3 & (226, 249) \\
31 & -51.4 & 33.7 & 0 & 0.31 & 0.16 & 0.166 & wf2 & (252, 392) \\
32 & -83.5 & 71.8 & 9 & 0.05 & 0.09 & 0.651 & wf2 & (727, 544) \\
33 & -2.9 & -67.0 & 8 & 0.00 & 0.09 & 0.669 & wf4 & (718, 168) \\
34 & -48.9 & -40.7 & 1 & 0.34 & 0.02 & 0.447 & wf4 & (177, 201) \\
35 & -2.6 & 12.4 & 7 & 0.07 & -0.07 & 0.772 & wf3 & (489, 266) \\
\enddata
\end{deluxetable}
\clearpage

\begin{deluxetable}{lrrcrrrrr}
\tablecolumns{9}
\tablecaption{3C210 HST WFPC2 Imaging Data\label{3c210data}}
\tablehead{
\colhead{ID} & \colhead{$\delta$x} & \colhead{$\delta$, } &
\colhead{Class\tablenotemark{a}} & \colhead{B/T} & \colhead{Res.} & \colhead{${\rm
R_h}$} & \colhead{Chip} & \colhead{Pos.}\\ 
\colhead{} & \colhead{(\arcsec )} & \colhead{(\arcsec )} & \colhead{} & \colhead{} & \colhead{} & \colhead{(\arcsec )} & \colhead{} & \colhead{}\\
}
\startdata
2 & 21.7 & 36.7 & 1 & 0.44 & 0.06 & 0.490 & wf4 & (42, 65) \\
3 & 0.0 & 0.0 & 4 & 0.23 & 0.15 & 0.898 & wf2 & (123, 301) \\
4 & -6.5 & 27.8 & 4 & 0.31 & 0.16 & 0.467 & wf3 & (199, 125) \\
5 & 32.6 & -41.0 & 3 & 0.01 & 0.23 & 1.486 & wf2 & (648, 277) \\
6 & 23.2 & -31.1 & 1 & 0.59 & 0.06 & 0.522 & wf2 & (511, 294) \\
7 & 28.8 & -19.8 & 5 & 0.00 & 0.23 & 0.115 & wf2 & (451, 181) \\
8 & -11.6 & 12.7 & 1 & 0.19 & 0.03 & 0.361 & wf3 & (325, 34) \\
9 & -21.4 & 7.3 & 1 & 0.63 & 0.04 & 0.552 & wf3 & (438, 45) \\
10 & 21.8 & 54.8 & 2 & 0.03 & 0.07 & 0.661 & wf4 & (191, 170) \\
11 & -44.8 & 6.5 & 2 & 0.06 & 0.08 & 0.609 & wf3 & (634, 173) \\
12 & 25.8 & 46.6 & 1 & 0.01 & 0.06 & 0.300 & wf4 & (99, 157) \\
13 & 25.8 & -50.9 & 1 & 0.70 & 0.04 & 0.295 & wf2 & (689, 390) \\
14 & -16.5 & 15.8 & 2 & 0.66 & 0.03 & 0.205 & wf3 & (350, 86) \\
15 & 6.4 & -26.1 & 1 & 0.62 & 0.06 & 0.666 & wf2 & (370, 401) \\
16 & -17.5 & 55.4 & 1 & 0.45 & 0.04 & 0.637 & wf3 & (128, 413) \\
17 & -8.0 & 13.6 & 1 & 0.86 & 0.02 & 0.371 & wf3 & (292, 18) \\
18 & 43.2 & -15.0 & 3 & 0.19 & 0.21 & 1.204 & wf2 & (500, 37) \\
19 & -37.5 & 14.3 & 4 & 0.12 & 0.08 & 0.590 & wf3 & (528, 195) \\
20 & -11.0 & 24.6 & 4 & 0.01 & 0.24 & 0.308 & wf3 & (250, 125) \\
21 & -2.5 & -6.7 & 2 & 0.37 & -0.01 & 0.415 & wf2 & (82, 360) \\
22 & -3.8 & 9.5 & 1 & 0.03 & 0.09 & 0.257 & wf2 & (21, 275) \\
23 & 2.5 & 10.1 & 4 & 0.00 & 0.27 & 0.714 & wf2 & (53, 221) \\
24 & -3.1 & -27.3 & 1 & 0.85 & 0.05 & 0.323 & wf2 & (326, 484) \\
25 & -44.5 & 34.4 & 1 & 0.75 & -0.01 & 0.216 & wf3 & (468, 398) \\
26 & -2.1 & 21.6 & 1 & 0.82 & 0.03 & 0.140 & wf3 & (198, 49) \\
27 & -2.7 & -5.5 & 1 & 0.64 & -0.01 & 0.287 & wf2 & (151, 354) \\
28 & -29.5 & 38.7 & 3 & 0.12 & 0.06 & 0.915 & wf3 & (322, 346) \\
29 & 23.4 & 48.8 & 4 & 0.25 & 0.16 & 0.760 & wf4 & (132, 145) \\
30 & -26.5 & 15.4 & 4 & 0.53 & 0.01 & 0.451 & wf3 & (433, 140) \\
31 & -15.7 & 56.6 & 4 & 0.00 & 0.06 & 0.879 & wf3 & (104, 413) \\
32 & 8.3 & -4.3 & 1 & 0.63 & 0.01 & 0.167 & wf2 & (205, 259) \\
33 & -8.2 & 4.2 & 1 & 0.79 & -0.04 & 0.194 & wf2 & (39, 342) \\
34 & 12.9 & -1.8 & 4 & 0.01 & 0.15 & 0.654 & wf2 & (211, 207) \\
35 & -3.2 & 3.0 & 1 & 0.33 & 0.02 & 0.146 & wf2 & (79, 308) \\
36 & 8.5 & -1.6 & 2 & 0.00 & 0.06 & 0.353 & wf2 & (186, 242) \\
37 & -39.9 & 37.4 & 4 & 0.38 & 0.11 & 1.317 & wf3 & (410, 395) \\
38 & -24.6 & 38.7 & 1 & 0.28 & 0.02 & 0.118 & wf3 & (280, 319) \\
39 & 3.0 & 36.9 & 4 & 0.00 & 0.20 & 0.891 & wf3 & (69, 142) \\
40 & 34.3 & 43.5 & 1 & 0.44 & -0.01 & 0.370 & wf4 & (25, 206) \\
41 & 4.6 & 52.8 & 4 & 0.00 & 0.14 & 0.551 & wf4 & (267, 21) \\
42 & -2.8 & 52.7 & 3 & 0.04 & 0.12 & 0.471 & wf3 & (21, 306) \\
43 & -38.2 & 12.2 & 3 & 0.55 & 0.10 & 0.441 & wf3 & (546, 179) \\
44 & -7.3 & -11.1 & 1 & 0.95 & 0.07 & 0.310 & wf2 & (169, 424) \\
45 & -39.2 & -6.2 & 4 & 0.71 & 0.02 & 0.282 & wf3 & (663, 37) \\
46 & -41.1 & 14.0 & 4 & 0.01 & 0.07 & 0.581 & wf3 & (562, 211) \\
47 & -27.1 & 19.5 & 1 & 0.01 & 0.01 & 0.234 & wf3 & (415, 178) \\
48 & 12.2 & 6.2 & 5 & 0.64 & 0.05 & 0.022 & wf2 & (141, 166) \\
49 & 1.1 & 8.8 & 4 & 0.70 & 0.07 & 0.472 & wf2 & (57, 238) \\
50 & -6.0 & -16.9 & 1 & 0.01 & 0.04 & 0.225 & wf2 & (225, 446) \\
51 & 8.8 & -51.6 & 4 & 0.00 & 0.23 & 0.377 & wf2 & (595, 532) \\
52 & -7.4 & -21.5 & 1 & 0.54 & 0.10 & 0.145 & wf2 & (255, 488) \\
53 & 9.1 & -14.0 & 0 & 0.59 & 0.03 & 0.235 & wf2 & (291, 304) \\
54 & -23.3 & 25.8 & 3 & 0.00 & 0.32 & 0.506 & wf3 & (343, 209) \\
55 & 29.9 & -10.4 & 5 & 0.83 & 0.07 & 0.137 & wf2 & (384, 118) \\
56 & 0.1 & 4.1 & 5 & 0.93 & 0.04 & 0.390 & wf2 & (88, 275) \\
57 & -2.8 & -54.9 & 4 & 0.00 & 0.06 & 0.583 & wf2 & (555, 648) \\
58 & -18.0 & -6.4 & 2 & 0.20 & -0.11 & 0.506 & wf2 & (73, 484) \\
59 & -53.6 & 13.9 & 4 & 0.17 & -0.12 & 0.157 & wf3 & (658, 288) \\
\enddata
\tablenotetext{a}{Classification scheme from \citet{stanford2002}, 1
= E/S0, 2=Sa/Sb, 3=Sc/Sd, 4=Irr, 5=Merger and/or disturbed.}
\end{deluxetable}

\clearpage

\begin{deluxetable}{lcccrrrrr}
\tablecolumns{8}
\tablecaption{3C210 HST NICMOS Imaging Data\label{3c210nicdata}}
\tablehead{
\colhead{ID} & \colhead{R.A.} & \colhead{Dec.} &
\colhead{Class\tablenotemark{a}} & \colhead{B/T} & \colhead{Res.} &
\colhead{${\rm R_h}$} & \colhead{Pointing} &  \colhead{Pos.}\\
\colhead{} & \colhead{J2000} & \colhead{J2000} & \colhead{} &
\colhead{} & \colhead{} & \colhead{(\arcsec )} &  \colhead{} &  \colhead{}\\
}
\startdata
2 & 21.7 & 36.7 & 1 & 0.36 & 0.06 & 0.243  & 3 & (186,106) \\
3 & 0.0 & 0.0 & 4 & 0.38 & 0.15 & 0.507 & 1 & (135,70) \\
4 & -6.5 & 27.8 & 4 & 0.73 & 0.66 & 0.262 & 3 & (138,243) \\
8 & -11.6 & 12.7 & 1 & 0.04 & 0.01 & 0.135  & 1 & (51,79) \\
9 & -21.4 & 7.3 & 1 & 0.49 & 0.13 & 0.187  & 1 & (46,135) \\
10 & 21.8 & 54.8 & 2 & 0.33 & 0.07 & 0.331  & 3 & (116,52) \\
11 & -44.8 & 6.5 & 2 & 0.29 & 0.09 & 0.231  & 2 & (238,227) \\
12 & 25.8 & 46.6 & 1 & 0.54 & 0.11 & 0.176 & 3 & (161,60) \\
14 & -16.5 & 15.8 & 2 & 0.68 & 0.03 & 0.074  & 1 & (25,91) \\
17 & -8.0 & 13.6 & 1 & 0.48 & 0.05 & 0.101  & 1 & (59,63) \\
19 & -37.5 & 14.3 & 4 & 0.41 & 0.10 & 0.234  & 2 & (229,174) \\
21 & -2.5 & -6.7 & 2 & 0.91 & 0.04 & 0.205  & 1 & (116,99) \\
22 & -3.8 & 9.5 & 1 & 0.12 & 0.07 & 0.087  & 1 & (86,58) \\
23 & 2.5 & 10.1 & 4 & 0.48 & 0.17 & 0.345  & 1 & (102,32) \\
24 & -3.1 & -27.3 & 1 & 0.48 & 0.05 & 0.117  & 1 & (236,160) \\
25 & -44.5 & 34.4 & 1 & 0.82 & 0.15 & 0.061  & 2 & (132,141) \\
26 & -2.1 & 21.6 & 1 & 0.52 & 0.07 & 0.051  & 1 & (43,17) \\
27 & -2.7 & -5.5 & 1 & 0.33 & 0.03 & 0.092  & 1 & (150,96) \\
28 & -29.5 & 38.7 & 3 & 0.21 & 0.05 & 0.332  & 2 & (159,70) \\
29 & 23.4 & 48.8 & 4 & 0.21 & 0.06 & 0.341  & 3 & (144,66) \\
32 & 8.3 & -4.3 & 1 & 0.76 & 0.11 & 0.04  & 1 & (176,49) \\
33 & -8.2 & 4.2 & 1 & 0.64 & 0.06 & 0.059  & 1 & (95,91) \\
34 & 12.9 & -1.8 & 4 & 0.00 & 0.03 & 0.254  & 1 & (179,24) \\
35 & -3.2 & 3.0 & 1 & 0.97 & 0.05 & 0.051  & 1 & (115,74) \\
36 & 8.5 & -1.6 & 2 & 0.35 & 0.06 & 0.136 & 1 & (167,41) \\
37 & -39.9 & 37.4 & 4 & 0.23 & 0.13 & 0.56  & 2 & (134,113) \\
38 & -24.6 & 38.7 & 1 & 0.45 & 0.05 & 0.059  & 2 & (173,50) \\
39 & 3.0 & 36.9 & 4 & 0.16 & 0.07 & 0.395  & 3 & (132,179) \\
40 & 34.3 & 43.5 & 1 & 0.49 & 0.03 & 0.14  & 3 & (197,38) \\
42 & -2.8 & 52.7 & 3 & 0.41 & 0.07 & 0.223  & 3 & (52,154) \\
43 & -38.2 & 12.1 & 3 & 0.93 & 0.02 & 0.371  & 2 & (237,183) \\
44 & -7.3 & -11. & 1 & 0.49 & 0.05 & 0.064  & 1 & (159,130) \\
45 & -39.2 & -6.2 & 4 & 0.97 & 0.03 & 0.083  & 1 & (49,244) \\
47 & -27.1 & 19.5 & 1 & 0.42 & 0.02 & 0.117  & 2 & (240,119) \\
49 & 1.1 & 8.8 & 4 & 0.58 & 0.11 & 0.177  & 1 & (104,39) \\
50 & -6.0 & -16.9 & 1 & 0.58 & 0.04 & 0.103  & 1 & (187,141) \\
52 & -7.4 & -21.5 & 1 & 0.46 & 0.05 & 0.086  & 1 & (201,161) \\
53 & 9.1 & -14.0 & 0 & 0.79 & 0.02 & 0.072  & 1 & (218,71) \\
56 & 0.1 & 4.1 & 5 & 0.57 & 0.03 & 0.086  & 1 & (119,58) \\
59 & -53.6 & 13.9 & 4 & 0.00 & 0.27 & 0.353  & 2 & (183,238) \\
\enddata
\tablenotetext{a}{Classification scheme from \citet{stanford2002}, 1
= E/S0, 2=Sa/Sb, 3=Sc/Sd, 4=Irr, 5=Merger and/or disturbed.}
\end{deluxetable}

\clearpage

\begin{deluxetable}{lrrcrrrrr}
\tablecolumns{9}
\tablecaption{RDCS~0848+4453 HST WFPC2 Imaging Data\label{cl0848wfpc}}
\tablehead{
\colhead{ID} & \colhead{$\delta$x} & \colhead{$\delta$, } &
\colhead{Class\tablenotemark{a}} & \colhead{B/T} & \colhead{Res.} &
\colhead{${\rm R_h}$} & \colhead{Chip} & \colhead{Pos.}\\ 
\colhead{} & \colhead{(\arcsec )} & \colhead{(\arcsec )} & \colhead{} & \colhead{} & \colhead{} & \colhead{(\arcsec )} & \colhead{} & \colhead{}\\
}
\startdata
1 & 33.0 & 42.9 & 0 & 0.41 & 0.46 & 0.324 & wf4 & (319, 461) \\
2 & -71.7 & -2.4 & 0 & 0.00 & -3.42 & 0.132 & wf2 & (24, 614) \\
3 & -10.2 & -30.0 & 2 & 0.65 & 0.02 & 0.599 & wf2 & (369, 250) \\
4 & 0 & 0 & 1 & 0.27 & 0.00 & 0.999 & pc & (202, 169) \\
5 & -30.9 & -30.0 & 5 & 0.00 & 0.12 & 0.083 & wf2 & (346, 232) \\
6* & -37.5 & -0.6 & 4 & 0.36 & 0.03 & 0.547 & wf2 & (43, 271) \\
6* & -37.5 & -0.6 & 4 & 0.77 & 0.16 & 0.439 & wf2 & (44, 266) \\
6* & -37.5 & -0.6 & 4 & 0.55 & 0.06 & 0.529 & wf2 & (40, 265) \\
7 & -24.8 & -49.8 & 2 & 0.01 & 0.11 & 1.573 & wf2 & (554, 197)) \\
8 & -51.0 & 25.8 & 2 & 0.52 & 0.01 & 0.464 & wf3 & (381, 221) \\
9 & 2.4 & 41.1 & 1 & 0.70 & 0.01 & 0.602 & wf4 & (326, 150) \\
10 & 41.4 & 37.8 & 0 & 0.00 & 0.25 & 0.787 & wf4 & (262, 542) \\
11 & 3.0 & 19.2 & 1 & 0.54 & 0.07 & 0.522 & wf4 & (108, 135) \\
12 & -7.8 & -6.6 & 2 & 0.41 & 0.06 & 0.683 & pc & (13, 297) \\
13 & -6.0 & 16.2 & 1 & 0.37 & 0.03 & 0.432 & wf4 & (86, 43) \\
14 & -26.4 & -5.4 & 4 & 0.08 & 0.10 & 1.047 & wf2 & (104, 162) \\
15 & -31.5 & 10.8 & 2 & 0.23 & 0.04 & 0.217 & wf3 & (202, 49) \\
16 & 21.6 & 4.8 & 4 & 0.00 & 0.10 & 0.663 & pc & (686, 107) \\
17 & 11.4 & -2.1 & 1 & 0.34 & -0.04 & 0.485 & pc & (450, 242) \\
18 & -25.5 & -6.9 & 2 & 0.34 & 0.08 & 0.415 & wf2 & (118, 154) \\
19 & -33.6 & 69.3 & 0 & 0.23 & 0.14 & 0.778 & wf3 & (161, 641) \\
20 & 21.6 & 26.1 & 3 & 0.01 & 0.03 & 0.909 & wf4 & (156, 332) \\
21 & -32.7 & 21.6 & 1 & 0.14 & 0.05 & 0.589 & wf3 & (201, 159) \\
22 & -20.1 & -3.6 & 1 & 0.34 & 0.02 & 0.264 & wf2 & (91, 96) \\
23 & 9.0 & 25.5 & 1 & 0.95 & 0.02 & 0.327 & wf4 & (164, 204) \\
24 & -50.7 & 31.5 & 5 & 0.02 & 0.04 & 0.033 & wf3 & (373, 278) \\
25 & -41.1 & 60.6 & 0 & 0.00 & 0.75 & 0.251 & wf3 & (246, 560) \\
26 & 36.0 & 27.3 & 4 & 0.01 & 0.16 & 0.564 & wf4 & (157, 478) \\
27 & -25.2 & 8.1 & 2 & 0.24 & 0.02 & 0.478 & wf3 & (141, 15) \\
28 & -58.2 & -37.8 & 0 & 0.00 & 0.19 & 0.949 & wf2 & (402, 513) \\
29 & -45.9 & -7.5 & 3 & 0.01 & 0.09 & 0.477 & wf2 & (99, 360) \\
30 & -9.6 & -51.9 & 2 & 0.00 & 0.15 & 0.579 & wf2 & (596, 40) \\
31 & -54.9 & -0.9 & 4 & 0.00 & 0.15 & 0.457 & wf2 & (25, 441) \\
32 & -33.6 & 31.2 & 4 & 0.74 & 0.00 & 0.285 & wf3 & (202, 257) \\
33 & -44.7 & -65.4 & 0 & 0.17 & 0.00 & 0.461 & wf2 & (699, 406) \\
34 & 3.6 & 13.5 & 5 & 0.69 & 0.03 & 0.018 & wf4 & (47, 135) \\
35 & -66.6 & -45.0 & 0 & 0.32 & -0.02 & 0.233 & wf2 & (464, 609) \\
36 & -24.6 & 34.2 & 4 & 0.36 & 0.06 & 0.599 & wf3 & (109, 278) \\
37 & -70.5 & 14.4 & 2 & 0.35 & 0.07 & 0.282 & wf3 & (591, 125) \\
38 & -54.3 & 6.9 & 4 & 0.00 & 0.16 & 1.093 & wf3 & (435, 34) \\
39 & -20.1 & -7.5 & 2 & 0.41 & 0.06 & 0.683 & pc & (13, 297) \\
40 & 10.8 & 20.1 & 1 & 0.35 & -0.04 & 0.524 & wf4 & (108, 215) \\
41 & -44.4 & 66.9 & 0 & 0.86 & 0.01 & 0.210 & wf3 & (274, 627) \\
42 & 3.6 & 3.9 & 1 & 0.03 & -0.01 & 0.713 & pc & (285, 91) \\
43 & 14.7 & 39.3 & 1 & 0.28 & -0.02 & 0.261 & wf4 & (298, 274) \\
44 & 9.9 & -12.3 & 1 & 0.01 & 0.01 & 0.716 & pc & (393, 467) \\
45 & -79.8 & 9.9 & 4 & 0.20 & 0.08 & 0.824 & wf3 & (692, 89) \\
46 & -77.1 & -27.6 & 0 & 0.00 & 0.26 & 0.273 & wf2 & (273, 699) \\
47 & -54.9 & -43.8 & 0 & 0.09 & 0.11 & 0.446 & wf2 & (458, 487) \\
48 & -43.8 & -3.3 & 4 & 0.01 & 0.20 & 0.368 & wf2 & (63, 335) \\
49 & -79.8 & 18.0 & 0 & 0.01 & 0.16 & 0.671 & wf3 & (680, 172) \\
50 & -0.9 & 12.9 & 2 & 0.01 & 0.06 & 0.329 & wf4 & (45, 91) \\
51 & -21.6 & 48.6 & 0 & 0.04 & 0.21 & 0.747 & wf3 & (63, 420) \\
52 & 34.8 & 57.9 & 0 & 0.62 & 0.02 & 0.186 & wf4 & (467, 495) \\
53 & 1.8 & -10.2 & 2 & 0.76 & -0.01 & 0.399 & pc & (217, 400) \\ 
54 & -20.4 & -33.0 & 5 & 0.42 & 0.02 & 0.097 & wf2 & (389, 132) \\
55 & -39.9 & -31.8 & 4 & 0.06 & -0.09 & 0.649 & wf2 & (356, 323) \\
56 & -19.5 & 1.5 & 3 & 0.04 & 0.08 & 0.390 & wf2 & (41, 84) \\
57 & -7.2 & -1.2 & 5 & 0.23 & 0.10 & 0.064 & pc & (38, 180) \\
58 & 12.0 & 40.8 & 3 & 0.04 & 0.12 & 0.703 & wf4 & (317, 247) \\
59 & -35.4 & -61.8 & 0 & 0.01 & 0.03 & 0.236 & wf2 & (672, 310) \\
60 & -65.4 & -51.6 & 0 & 0.96 & 0.04 & 0.089 & wf2 & (529, 607) \\
61 & -35.1 & -57.9 & 0 & 0.46 & 0.08 & 0.306 & wf2 & (632, 301) \\
62 & -1.2 & -15.6 & 3 & 0.05 & 0.07 & 1.559 & pc & (137, 512) \\
63 & -31.8 & -15.0 & 2 & 0.00 & 0.04 & 0.569 & wf2 & (191, 228) \\
64 & -5.7 & 10.8 & 1 & 0.51 & 0.01 & 0.305 & wf4 & (29, 39) \\
65 & 17.7 & 47.1 & 2 & 0.01 & 0.13 & 0.196 & wf4 & (374, 310) \\
66 & -22.2 & 27.3 & 4 & 0.01 & 0.22 & 0.610 & wf3 & (91, 207) \\
67 & -44.4 & 26.4 & 1 & 0.59 & 0.00 & 0.118 & wf3 & (315, 219) \\
68 & -66.6 & -11.4 & 0 & 0.86 & -0.05 & 0.323 & wf2 & (120, 571) \\
69 & 21.6 & 16.5 & 3 & 0.28 & 0.06 & 0.493 & wf4 & (59, 322) \\
70 & -4.8 & -4.2 & 4 & 0.01 & 0.19 & 0.949 & pc & (89, 251) \\
\enddata
\tablenotetext{a}{Classification scheme from \citet{stanford2002}, 1
= E/S0, 2=Sa/Sb, 3=Sc/Sd, 4=Irr, 5=Merger and/or disturbed.}
\end{deluxetable}
\clearpage

\begin{deluxetable}{lrrcrrrrr}
\tablecolumns{8}
\tablecaption{RDCS~0848+4453 NICMOS Camera 3 Imaging Data\label{cl0848nic}}
\tablehead{
\colhead{ID} & \colhead{$\delta$x} & \colhead{$\delta$, } &
\colhead{Class\tablenotemark{a}} & \colhead{B/T} & \colhead{Res.} &
\colhead{${\rm R_h}$} & \colhead{Pointing} & \colhead{Pos.}\\
\colhead{} & \colhead{(\arcsec )} & \colhead{(\arcsec )} & \colhead{} &
\colhead{} & \colhead{} & \colhead{(\arcsec )} & \colhead{} & \colhead{}
\\
}
\startdata
3 & -10.2 & -30.0 & 2 & 0.49 & 0.05 & 0.207 & 3 & (141,124) \\
4 & 0 & 0 & 1 & 0.59 & 0.06 & 0.437 & 2 & (147,226) \\ 
5 & -30.9 & -30.0 & 5 & 0.00 & 0.10 & 0.031 & 3 & (63,190) \\
6 & -37.5 & -0.6 & 4 & 0.00 & 0.37 & 0.298 & 1 & (197,164) \\
8 & -51.0 & 25.8 & 2 & 0.54 & 0.05 & 0.159 & 1 & (61,108) \\
9 & 2.4 & 41.1 & 1 & 0.44 & 0.15 & 0.276 & 2 & (23,65) \\
11 & 3.0 & 19.2 & 1 & 0.60 & 0.04 & 0.259 & 2 & (96,145) \\
12 & -7.8 & -6.6 & 2 & 0.65 & 0.06 & 0.170 & 3 & (75,27) \\
13 & -6.0 & 16.2 & 1 & 0.71 & 0.08 & 0.241 & 2 & (72,185) \\
15 & -31.5 & 10.8 & 2 & 1.00 & 0.08 & 0.073 & 1 & (183,102) \\
16 & 21.6 & 4.8 & 4 & 0.54 & 0.24 & 0.506 & 2 & (213,139) \\
17 & 11.4 & -2.1 & 1 & 0.62 & 0.04 & 0.108 & 2 & (197,198) \\
20 & 21.6 & 26.1 & 3 & 0.00 & 0.11 & 0.401 & 2 & (144,59) \\
21 & -32.7 & 21.6 & 1 & 0.12 & 0.07 & 0.292 & 1 & (143,65) \\
22 & -20.1 & -3.6 & 1 & 0.27 & 0.04 & 0.121 & 3 & (19,55) \\
23 & 9.0 & 25.5 & 1 & 0.72 & 0.06 & 0.139 & 2 & (99,102) \\
24 & -50.7 & 31.5 & 5 & 0.39 & 0.13 & 0.001 & 1 & (44,86) \\
27 & -25.2 & 8.1 & 2 & 0.73 & 0.13 & 0.325 & 1 & (214,92) \\
29 & -45.9 & -7.5 & 3 & 0.03 & 0.18 & 0.359 & 1 & (187,217) \\
30 & -9.6 & -51.9 & 2 & 0.00 & 0.06 & 0.242 & 3 & (215,206) \\
31 & -54.9 & -0.9 & 4 & 0.17 & 0.11 & 0.198 & 1 & (133,221) \\
32 & -33.6 & 31.2 & 4 & 0.56 & 0.04 & 0.095 & 1 & (109,32) \\
34 & 3.6 & 13.5 & 5 & 0.06 & 0.14 & 0.020 & 2 & (117,165) \\
37 & -70.5 & 14.4 & 2 & 0.48 & 0.06 & 0.127 & 1 & (25,215) \\
38 & -54.3 & 6.9 & 4 & 0.46 & 0.19 & 0.658 & 1 & (110,190) \\
39 & -20.1 & -7.5 & 2 & 0.82 & 0.06 & 0.110 & 3 & (30,69) \\
40 & 10.8 & 20.1 & 1 & 0.58 & 0.02 & 0.199 & 2 & (122,117) \\
42 & 3.6 & 3.9 & 1 & 0.42 & 0.06 & 0.192 & 2 & (147,200) \\
43 & 14.7 & 39.3 & 1 & 0.44 & 0.06 & 0.115 & 2 & (76,31) \\
44 & 9.9 & -12.3 & 1 & 0.93 & 0.18 & 0.235 & 2 & (225,242) \\
48 & -43.8 & -3.3 & 4 & 0.00 & 0.16 & 0.236 & 1 & (182,195) \\
50 & -0.9 & 12.9 & 2 & 0.00 & 0.16 & 0.148 & 2 & (102,181) \\
54 & -20.4 & -33.0 & 5 & 0.95 & 0.02 & 0.023 & 3 & (112,169) \\
55 & -39.9 & -31.8 & 4 & 0.00 & 0.10 & 0.465 & 3 & (34,226) \\
58 & 12.0 & 40.8 & 3 & 0.10 & 0.05 & 0.295 & 2 & (60,34) \\
62 & -1.2 & -15.6 & 3 & 0.44 & 0.04 & 0.304 & 3 & (128,40) \\
64 & -5.7 & 10.8 & 1 & 0.48 & 0.04 & 0.134 & 2 & (91,205) \\
67 & -44.4 & 26.4 & 1 & 0.69 & 0.02 & 0.035 & 1 & (84,85) \\
69 & 21.6 & 16.5 & 3 & 0.00 & 0.28 & 0.198 & 2 & (176,95) \\
\enddata
\tablenotetext{a}{Classification scheme from \citet{stanford2002}, 1
= E/S0, 2=Sa/Sb, 3=Sc/Sd, 4=Irr, 5=Merger and/or disturbed.}
\end{deluxetable}


\clearpage
\begin{thebibliography}{75}
\expandafter\ifx\csname natexlab\endcsname\relax\def\natexlab#1{#1}\fi

\bibitem[{{Akritas} \& {Bershady}(1996)}]{akritas96}
{Akritas}, M.~G. \& {Bershady}, M.~A. 1996, \apj, 470, 706

\bibitem[{{Andreon} {et~al.}(1997){Andreon}, {Davoust}, \&
  {Heim}}]{andreon1997}
{Andreon}, S., {Davoust}, E., \& {Heim}, T. 1997, \aap, 323, 337

\bibitem[{{Arag\'on-Salamanca} {et~al.}(1993){Arag\'on-Salamanca}, {Ellis},
  {Couch}, \& {Carter}}]{aragon93}
{Arag\'on-Salamanca}, A., {Ellis}, R.~S., {Couch}, W.~J., \& {Carter}, D. 1993,
  \mnras, 262, 764

\bibitem[{{Becker} {et~al.}(2001){Becker}, {Fan}, {White}, {Strauss},
  {Narayanan}, {Lupton}, {Gunn}, {Annis}, {Bahcall}, {Brinkmann}, {Connolly},
  {Csabai}, {Czarapata}, {Doi}, {Heckman}, {Hennessy}, {Ivezi{\' c}}, {Knapp},
  {Lamb}, {McKay}, {Munn}, {Nash}, {Nichol}, {Pier}, {Richards}, {Schneider},
  {Stoughton}, {Szalay}, {Thakar}, \& {York}}]{becker2001}
{Becker}, R.~H., {Fan}, X., {White}, R.~L., {Strauss}, M.~A., {Narayanan},
  V.~K., {Lupton}, R.~H., {Gunn}, J.~E., {Annis}, J., {Bahcall}, N.~A.,
  {Brinkmann}, J., {Connolly}, A.~J., {Csabai}, I., {Czarapata}, P.~C., {Doi},
  M., {Heckman}, T.~M., {Hennessy}, G.~S., {Ivezi{\' c}}, {\v Z}., {Knapp},
  G.~R., {Lamb}, D.~Q., {McKay}, T.~A., {Munn}, J.~A., {Nash}, T., {Nichol},
  R., {Pier}, J.~R., {Richards}, G.~T., {Schneider}, D.~P., {Stoughton}, C.,
  {Szalay}, A.~S., {Thakar}, A.~R., \& {York}, D.~G. 2001, \aj, 122, 2850

\bibitem[{{Beers} {et~al.}(1990){Beers}, {Flynn}, \& {Gebhardt}}]{beers90}
{Beers}, T.~C., {Flynn}, K., \& {Gebhardt}, K. 1990, \aj, 100, 32

\bibitem[{{Bennett} {et~al.}(2003){Bennett}, {Halpern}, {Hinshaw}, {Jarosik},
  {Kogut}, {Limon}, {Meyer}, {Page}, {Spergel}, {Tucker}, {Wollack}, {Wright},
  {Barnes}, {Greason}, {Hill}, {Komatsu}, {Nolta}, {Odegard}, {Peiris},
  {Verde}, \& {Weiland}}]{bennett2003}
{Bennett}, C.~L., {Halpern}, M., {Hinshaw}, G., {Jarosik}, N., {Kogut}, A.,
  {Limon}, M., {Meyer}, S.~S., {Page}, L., {Spergel}, D.~N., {Tucker}, G.~S.,
  {Wollack}, E., {Wright}, E.~L., {Barnes}, C., {Greason}, M.~R., {Hill},
  R.~S., {Komatsu}, E., {Nolta}, M.~R., {Odegard}, N., {Peiris}, H.~V.,
  {Verde}, L., \& {Weiland}, J.~L. 2003, \apjs, 148, 1

\bibitem[{{Bertin} \& {Arnouts}(1996)}]{bertin96}
{Bertin}, E. \& {Arnouts}, S. 1996, \aaps, 117, 393

\bibitem[{{Blakeslee} {et~al.}(2003){Blakeslee}, {Franx}, {Postman}, {Rosati},
  {Holden}, {Illingworth}, {Ford}, {Cross}, {Gronwall}, {Ben{\'{\i}}tez},
  {Bouwens}, {Broadhurst}, {Clampin}, {Demarco}, {Golimowski}, {Hartig},
  {Infante}, {Martel}, {Miley}, {Menanteau}, {Meurer}, {Sirianni}, \&
  {White}}]{blakeslee2003}
{Blakeslee}, J.~P., {Franx}, M., {Postman}, M., {Rosati}, P., {Holden}, B.~P.,
  {Illingworth}, G.~D., {Ford}, H.~C., {Cross}, N.~J.~G., {Gronwall}, C.,
  {Ben{\'{\i}}tez}, N., {Bouwens}, R.~J., {Broadhurst}, T.~J., {Clampin}, M.,
  {Demarco}, R., {Golimowski}, D.~A., {Hartig}, G.~F., {Infante}, L., {Martel},
  A.~R., {Miley}, G.~K., {Menanteau}, F., {Meurer}, G.~R., {Sirianni}, M., \&
  {White}, R.~L. 2003, \apjl, 596, L143

\bibitem[{{Bower} {et~al.}(1998){Bower}, {Kodama}, \& {Terlevich}}]{bower98}
{Bower}, R.~G., {Kodama}, T., \& {Terlevich}, A. 1998, \mnras, 299, 1193

\bibitem[{{Bower} {et~al.}(1992){Bower}, {Lucey}, \& {Ellis}}]{bower92b}
{Bower}, R.~G., {Lucey}, J.~R., \& {Ellis}, R.~S. 1992, \mnras, 254, 601

\bibitem[{{Bruzual} \& {Charlot}(1993)}]{bruzual93}
{Bruzual}, A.~G. \& {Charlot}, S. 1993, \apj, 405, 538

\bibitem[{{Bruzual} \& {Charlot}(1996)}]{bc96}
---. 1996, personal communication

\bibitem[{{Bruzual} \& {Charlot}(2003)}]{bc03}
{Bruzual}, G. \& {Charlot}, S. 2003, \mnras, 344, 1000

\bibitem[{{Burstein} \& {Heiles}(1982)}]{bh82}
{Burstein}, D. \& {Heiles}, C. 1982, \aj, 87, 1165

\bibitem[{{Castander} {et~al.}(1994){Castander}, {Ellis}, {Frenk}, {Dressler},
  \& {Gunn}}]{castander94}
{Castander}, F.~J., {Ellis}, R.~S., {Frenk}, C.~S., {Dressler}, A., \& {Gunn},
  J.~E. 1994, \apjl, 424, L79

\bibitem[{{Clowe} {et~al.}(1998){Clowe}, {Luppino}, {Kaiser}, {Henry}, \&
  {Gioia}}]{clowe98}
{Clowe}, D., {Luppino}, G.~A., {Kaiser}, N., {Henry}, J.~P., \& {Gioia}, I.~M.
  1998, \apjl, 497, L61

\bibitem[{{Coleman} {et~al.}(1980){Coleman}, {Wu}, \& {Weedman}}]{cww80}
{Coleman}, G.~D., {Wu}, C.-C., \& {Weedman}, D.~W. 1980, \apjs, 43, 393

\bibitem[{{de Propris} {et~al.}(1998){de Propris}, {Eisenhardt}, {Stanford}, \&
  {Dickinson}}]{dePropris98}
{de Propris}, R., {Eisenhardt}, P.~R., {Stanford}, S.~A., \& {Dickinson}, M.
  1998, \apjl, 503, L45

\bibitem[{{de Propris} {et~al.}(1999){de Propris}, {Stanford}, {Eisenhardt},
  {Dickinson}, \& {Elston}}]{dePropris99}
{de Propris}, R., {Stanford}, S.~A., {Eisenhardt}, P.~R., {Dickinson}, M., \&
  {Elston}, R. 1999, \aj, 118, 729

\bibitem[{{de Vaucouleurs} {et~al.}(1991){de Vaucouleurs}, {de Vaucouleurs},
  {Corwin}, {Buta}, {Paturel}, \& {Fouque}}]{devauc1991}
{de Vaucouleurs}, G., {de Vaucouleurs}, A., {Corwin}, H.~G., {Buta}, R.~J.,
  {Paturel}, G., \& {Fouque}, P. 1991, {Third Reference Catalogue of Bright
  Galaxies} (Volume 1-3, XII, 2069 pp.~7 figs..~ Springer-Verlag Berlin
  Heidelberg New York)

\bibitem[{{Deltorn} {et~al.}(1997){Deltorn}, {Le Fevre}, {Crampton}, \&
  {Dickinson}}]{deltorn97}
{Deltorn}, J.-M., {Le Fevre}, O., {Crampton}, D., \& {Dickinson}, M. 1997,
  \apjl, 483, L21

\bibitem[{{Donahue} {et~al.}(1999){Donahue}, {Voit}, {Scharf}, {Gioia},
  {Mullis}, {Hughes}, \& {Stocke}}]{donahue99b}
{Donahue}, M., {Voit}, G.~M., {Scharf}, C.~A., {Gioia}, I.~M., {Mullis}, C.~R.,
  {Hughes}, J.~P., \& {Stocke}, J.~T. 1999, \apj, 527, 525

\bibitem[{{Dressler}(1980)}]{dressler1980}
{Dressler}, A. 1980, \apj, 236, 351

\bibitem[{{Dressler} {et~al.}(1997){Dressler}, {Oemler}, {Couch}, {Smail},
  {Ellis}, {Barger}, {Butcher}, {Poggianti}, \& {Sharples}}]{dressler97}
{Dressler}, A., {Oemler}, A.~J., {Couch}, W.~J., {Smail}, I., {Ellis}, R.~S.,
  {Barger}, A., {Butcher}, H., {Poggianti}, B.~M., \& {Sharples}, R.~M. 1997,
  \apj, 490, 577

\bibitem[{{Eggen} {et~al.}(1962){Eggen}, {Lynden-Bell}, \& {Sandage}}]{els62}
{Eggen}, O.~J., {Lynden-Bell}, \& {Sandage}, A. 1962, \apj, 136, 748

\bibitem[{{Eisenhardt} {et~al.}(2004){Eisenhardt}, {de Propris}, {Gonzalez},
  {Stanford}, {Wang}, \& {Dickinson}}]{eisenhardt03}
{Eisenhardt}, P., {de Propris}, R., {Gonzalez}, A., {Stanford}, S.~A., {Wang},
  M., \& {Dickinson}, M. 2004, \apj, in preparation

\bibitem[{{Eisenhardt} {et~al.}(2001){Eisenhardt}, {Elston}, {Dickinson},
  {Stanford}, \& {Stern}}]{eisenhardt_spices}
{Eisenhardt}, P., {Elston}, R., {Dickinson}, M., {Stanford}, S.~A., \& {Stern},
  D. 2001, \apjs, in prep

\bibitem[{{Ellis} {et~al.}(1997){Ellis}, {Smail}, {Dressler}, {Couch},
  {Oemler}, {Butcher}, \& {Sharples}}]{ellis97}
{Ellis}, R.~S., {Smail}, I., {Dressler}, A., {Couch}, W.~J., {Oemler}, A.~J.,
  {Butcher}, H., \& {Sharples}, R.~M. 1997, \apj, 483, 582+

\bibitem[{{Faber}(1973)}]{faber73}
{Faber}, S.~M. 1973, \apj, 179, 731

\bibitem[{{Fan} {et~al.}(2001){Fan}, {Narayanan}, {Lupton}, {Strauss}, {Knapp},
  {Becker}, {White}, {Pentericci}, {Leggett}, {Haiman}, {Gunn}, {Ivezi{\' c}},
  {Schneider}, {Anderson}, {Brinkmann}, {Bahcall}, {Connolly}, {Csabai}, {Doi},
  {Fukugita}, {Geballe}, {Grebel}, {Harbeck}, {Hennessy}, {Lamb}, {Miknaitis},
  {Munn}, {Nichol}, {Okamura}, {Pier}, {Prada}, {Richards}, {Szalay}, \&
  {York}}]{fan2001}
{Fan}, X., {Narayanan}, V.~K., {Lupton}, R.~H., {Strauss}, M.~A., {Knapp},
  G.~R., {Becker}, R.~H., {White}, R.~L., {Pentericci}, L., {Leggett}, S.~K.,
  {Haiman}, Z., {Gunn}, J.~E., {Ivezi{\' c}}, {\v Z}., {Schneider}, D.~P.,
  {Anderson}, S.~F., {Brinkmann}, J., {Bahcall}, N.~A., {Connolly}, A.~J.,
  {Csabai}, I., {Doi}, M., {Fukugita}, M., {Geballe}, T., {Grebel}, E.~K.,
  {Harbeck}, D., {Hennessy}, G., {Lamb}, D.~Q., {Miknaitis}, G., {Munn}, J.~A.,
  {Nichol}, R., {Okamura}, S., {Pier}, J.~R., {Prada}, F., {Richards}, G.~T.,
  {Szalay}, A., \& {York}, D.~G. 2001, \aj, 122, 2833

\bibitem[{{Freedman} {et~al.}(2001){Freedman}, {Madore}, {Gibson}, {Ferrarese},
  {Kelson}, {Sakai}, {Mould}, {Kennicutt}, {Ford}, {Graham}, {Huchra},
  {Hughes}, {Illingworth}, {Macri}, \& {Stetson}}]{freedman2001}
{Freedman}, W.~L., {Madore}, B.~F., {Gibson}, B.~K., {Ferrarese}, L., {Kelson},
  D.~D., {Sakai}, S., {Mould}, J.~R., {Kennicutt}, R.~C., {Ford}, H.~C.,
  {Graham}, J.~A., {Huchra}, J.~P., {Hughes}, S.~M.~G., {Illingworth}, G.~D.,
  {Macri}, L.~M., \& {Stetson}, P.~B. 2001, \apj, 553, 47

\bibitem[{{Frei} {et~al.}(1996){Frei}, {Guhathakurta}, {Gunn}, \&
  {Tyson}}]{frei96}
{Frei}, Z., {Guhathakurta}, P., {Gunn}, J.~E., \& {Tyson}, J.~A. 1996, \aj,
  111, 174

\bibitem[{{Fruchter} \& {Hook}(2002)}]{fruchter2002}
{Fruchter}, A.~S. \& {Hook}, R.~N. 2002, \pasp, 114, 144

\bibitem[{{Gioia} \& {Luppino}(1994)}]{gioia94}
{Gioia}, I.~M. \& {Luppino}, G.~A. 1994, \apjs, 94, 583

\bibitem[{{Gioia} {et~al.}(1990){Gioia}, {Maccacaro}, {Schild}, {Wolter},
  {Stocke}, {Morris}, \& {Henry}}]{gioia90a}
{Gioia}, I.~M., {Maccacaro}, T., {Schild}, R.~E., {Wolter}, A., {Stocke},
  J.~T., {Morris}, S.~L., \& {Henry}, J.~P. 1990, \apjs, 72, 567

\bibitem[{{Giraud}(1998)}]{giraud98}
{Giraud}, E. 1998, \aj, 116, 1125

\bibitem[{{Gunn} {et~al.}(1986){Gunn}, {Hoessel}, \& {Oke}}]{gunn86}
{Gunn}, J.~E., {Hoessel}, J.~G., \& {Oke}, J.~B. 1986, \apj, 306, 30

\bibitem[{{Henry} {et~al.}(1992){Henry}, {Gioia}, {Maccacaro}, {Morris},
  {Stocke}, \& {Wolter}}]{henry92}
{Henry}, J.~P., {Gioia}, I., {Maccacaro}, T., {Morris}, S.~L., {Stocke}, J., \&
  {Wolter}, A. 1992, \apj, 386, 408

\bibitem[{{Holtzman} {et~al.}(1995){Holtzman}, {Burrows}, {Casertano},
  {Hester}, {Trauger}, {Watson}, \& {Worthey}}]{holtzman95}
{Holtzman}, J.~A., {Burrows}, C.~J., {Casertano}, S., {Hester}, J.~J.,
  {Trauger}, J.~T., {Watson}, A.~M., \& {Worthey}, G. 1995, \pasp, 107, 1065

\bibitem[{{Im} {et~al.}(2002){Im}, {Simard}, {Faber}, {Koo}, {Gebhardt},
  {Willmer}, {Phillips}, {Illingworth}, {Vogt}, \& {Sarajedini}}]{im00}
{Im}, M., {Simard}, L., {Faber}, S.~M., {Koo}, D.~C., {Gebhardt}, K.,
  {Willmer}, C.~N.~A., {Phillips}, A., {Illingworth}, G., {Vogt}, N.~P., \&
  {Sarajedini}, V.~L. 2002, \apj, 571, 136

\bibitem[{{Kauffmann} \& {Charlot}(1998)}]{kauffmann98}
{Kauffmann}, G. \& {Charlot}, S. 1998, \mnras, 294, 705

\bibitem[{{Kelson} {et~al.}(2001){Kelson}, {Illingworth}, {Franx}, \& {van
  Dokkum}}]{kelson2001}
{Kelson}, D.~D., {Illingworth}, G.~D., {Franx}, M., \& {van Dokkum}, P.~G.
  2001, \apjl, 552, L17

\bibitem[{{Kodama} \& {Arimoto}(1997)}]{kodama97}
{Kodama}, T. \& {Arimoto}, N. 1997, \aap, 320, 41

\bibitem[{{Kron}(1980)}]{kron80}
{Kron}, R.~G. 1980, \apjs, 43, 305

\bibitem[{{Liu} {et~al.}(2000){Liu}, {Charlot}, \& {Graham}}]{liu2000}
{Liu}, M.~C., {Charlot}, S., \& {Graham}, J.~R. 2000, \apj, 543, 644

\bibitem[{{Lubin} {et~al.}(1998){Lubin}, {Postman}, {Oke}, {Ratnatunga},
  {Gunn}, {Hoessel}, \& {Schneider}}]{lubin98}
{Lubin}, L.~M., {Postman}, M.~P., {Oke}, J.~B., {Ratnatunga}, K.~I., {Gunn},
  J.~E., {Hoessel}, J.~G., \& {Schneider}, D.~P. 1998, \aj, 116, 586

\bibitem[{{Mathis}(1990)}]{mathis90}
{Mathis}, J.~S. 1990, \araa, 28, 37

\bibitem[{{Nelson} {et~al.}(2001){Nelson}, {Gonzalez}, {Zaritsky}, \&
  {Dalcanton}}]{nelson2001}
{Nelson}, A.~E., {Gonzalez}, A.~H., {Zaritsky}, D., \& {Dalcanton}, J.~J. 2001,
  \apj, 563, 629

\bibitem[{{Oke} {et~al.}(1998){Oke}, {Postman}, \& {Lubin}}]{oke98}
{Oke}, J.~B., {Postman}, M., \& {Lubin}, L.~M. 1998, \aj, 116, 549

\bibitem[{{Postman} {et~al.}(2001){Postman}, {Lubin}, \& {Oke}}]{postman2001}
{Postman}, M., {Lubin}, L.~M., \& {Oke}, J.~B. 2001, \aj, 122, 1125

\bibitem[{{Ratnatunga} {et~al.}(1999){Ratnatunga}, {Griffiths}, \&
  {Ostrander}}]{rog99}
{Ratnatunga}, K.~U., {Griffiths}, R.~E., \& {Ostrander}, E.~J. 1999, \aj, 118,
  86

\bibitem[{{Rosati} {et~al.}(1998){Rosati}, {Della Ceca}, {Norman}, \&
  {Giacconi}}]{rosati98}
{Rosati}, P., {Della Ceca}, R., {Norman}, C., \& {Giacconi}, R. 1998, \apjl,
  492, L21

\bibitem[{{Salpeter}(1955)}]{salpeter1955}
{Salpeter}, E.~E. 1955, \apj, 121, 161

\bibitem[{{Sandage} \& {Visvanathan}(1978)}]{sandage78}
{Sandage}, A. \& {Visvanathan}, N. 1978, \apj, 225, 742

\bibitem[{{Schade} {et~al.}(1995){Schade}, {Lilly}, {Crampton}, {Le Fevre}, \&
  {Tresse}}]{schade95}
{Schade}, D., {Lilly}, S.~J., {Crampton}, D., H.~F., {Le Fevre}, O., \&
  {Tresse}, L. 1995, \apj, 451, 1

\bibitem[{{Scodeggio}(2001)}]{scodeggio2001}
{Scodeggio}, M. 2001, \aj, 121, 2413

\bibitem[{{Simard} {et~al.}(2002){Simard}, {Willmer}, {Vogt}, {Sarajedini},
  {Phillips}, {Weiner}, {Koo}, {Im}, {Illingworth}, \& {Faber}}]{simard02}
{Simard}, L., {Willmer}, C.~N.~A., {Vogt}, N.~P., {Sarajedini}, V.~L.,
  {Phillips}, A.~C., {Weiner}, B.~J., {Koo}, D.~C., {Im}, M., {Illingworth},
  G.~D.~I., \& {Faber}, S.~M. 2002, \apjs, 142, 1

\bibitem[{{Spinrad} {et~al.}(1985){Spinrad}, {Marr}, {Aguilar}, \&
  {Djorgovski}}]{spinrad85}
{Spinrad}, H., {Marr}, J., {Aguilar}, L., \& {Djorgovski}, S. 1985, \pasp, 97,
  932

\bibitem[{{Stanford} {et~al.}(1998){Stanford}, {Eisenhardt}, \&
  {Dickinson}}]{stanford98}
{Stanford}, S.~A., {Eisenhardt}, P.~R., \& {Dickinson}, M. 1998, \apj, 492, 461

\bibitem[{{Stanford} {et~al.}(2002){Stanford}, {Eisenhardt}, {Dickinson},
  {Holden}, \& {De Propris}}]{stanford2002}
{Stanford}, S.~A., {Eisenhardt}, P.~R., {Dickinson}, M., {Holden}, B.~P., \&
  {De Propris}, R. 2002, \apjs, 142, 153

\bibitem[{{Stanford} {et~al.}(1995){Stanford}, {Eisenhardt}, \&
  {Dickinson}}]{stanford95}
{Stanford}, S.~A., {Eisenhardt}, P.~R.~M., \& {Dickinson}, M. 1995, \apj, 450,
  512

\bibitem[{{Stanford} {et~al.}(1997){Stanford}, {Elston}, {Eisenhardt},
  {Spinrad}, {Stern}, \& {Dey}}]{stanford97}
{Stanford}, S.~A., {Elston}, R., {Eisenhardt}, P.~R., {Spinrad}, H., {Stern},
  D., \& {Dey}, A. 1997, \aj, 114, 2232

\bibitem[{{Stanford} {et~al.}(2000){Stanford}, {Holden}, {Rosati}, {Tozzi},
  {Borgani}, {Eisenhardt}, \& {Spinrad}}]{stanford00}
{Stanford}, S.~A., {Holden}, B., {Rosati}, P., {Tozzi}, P., {Borgani}, S.,
  {Eisenhardt}, P.~R., \& {Spinrad}, H. 2000, \apjl, submitted

\bibitem[{{Stern} {et~al.}(2002){Stern}, {Tozzi}, {Stanford}, {Rosati},
  {Holden}, {Eisenhardt}, {Elston}, {Wu}, {Connolly}, {Spinrad}, {Dawson},
  {Dey}, \& {Chaffee}}]{stern2002}
{Stern}, D., {Tozzi}, P., {Stanford}, S.~A., {Rosati}, P., {Holden}, B.,
  {Eisenhardt}, P., {Elston}, R., {Wu}, K.~L., {Connolly}, A., {Spinrad}, H.,
  {Dawson}, S., {Dey}, A., \& {Chaffee}, F.~H. 2002, \aj, 123, 2223

\bibitem[{{Terlevich} {et~al.}(2001){Terlevich}, {Caldwell}, \&
  {Bower}}]{terlevich2001}
{Terlevich}, A.~I., {Caldwell}, N., \& {Bower}, R.~G. 2001, \mnras, 326, 1547

\bibitem[{{Trager} {et~al.}(2000){Trager}, {Faber}, {Worthey}, \& {Gonz{\'
  a}lez}}]{trager2000}
{Trager}, S.~C., {Faber}, S.~M., {Worthey}, G., \& {Gonz{\' a}lez}, J.~J. 2000,
  \aj, 120, 165

\bibitem[{{Tully} \& {Pierce}(2000)}]{tully2000}
{Tully}, R.~B. \& {Pierce}, M.~J. 2000, \apj, 533, 744

\bibitem[{{van Dokkum} \& {Franx}(2001)}]{pvd_mf2001}
{van Dokkum}, P.~G. \& {Franx}, M. 2001, \apj, 553, 90

\bibitem[{{van Dokkum} {et~al.}(2000){van Dokkum}, {Franx}, {Fabricant},
  {Illingworth}, \& {Kelson}}]{pvd_2000}
{van Dokkum}, P.~G., {Franx}, M., {Fabricant}, D., {Illingworth}, G.~D., \&
  {Kelson}, D.~D. 2000, \apj, 541, 95

\bibitem[{{van Dokkum} {et~al.}(1998){van Dokkum}, {Franx}, {Kelson}, \&
  {Illingworth}}]{vandokkum1998}
{van Dokkum}, P.~G., {Franx}, M., {Kelson}, D.~D., \& {Illingworth}, G.~D.
  1998, \apjl, 504, L17+

\bibitem[{{van Dokkum} \& {Stanford}(2003)}]{vandokkum2003}
{van Dokkum}, P.~G. \& {Stanford}, S.~A. 2003, \apj, 585, 78

\bibitem[{{van Dokkum} {et~al.}(2001){van Dokkum}, {Stanford}, {Holden},
  {Eisenhardt}, {Dickinson}, \& {Elston}}]{vandokkum2001}
{van Dokkum}, P.~G., {Stanford}, S.~A., {Holden}, B.~P., {Eisenhardt}, P.~R.,
  {Dickinson}, M., \& {Elston}, R. 2001, \apjl, 552, L101

\bibitem[{{Visvanathan} \& {Sandage}(1977)}]{visvanathan77}
{Visvanathan}, N. \& {Sandage}, A. 1977, \apj, 216, 214

\bibitem[{{White} {et~al.}(2003){White}, {Becker}, {Fan}, \&
  {Strauss}}]{white2003}
{White}, R.~L., {Becker}, R.~H., {Fan}, X., \& {Strauss}, M. 2003, \aj,
  accepted

\bibitem[{{Worthey}(1994)}]{worthey1994}
{Worthey}, G. 1994, \apjs, 95, 107

\end{thebibliography}
\end{document}